\documentclass[msom,nonblindrev]{informs3} 

\OneAndAHalfSpacedXI
\AtBeginDocument{\textheight610pt}  

\usepackage{graphicx}
\usepackage{amsfonts}
\usepackage{subfigure}
\usepackage{longtable}
\usepackage{rotating}
\usepackage{epstopdf}
\usepackage{color}
\usepackage{multirow}
\usepackage{booktabs}
\usepackage{bbm}
\usepackage{setspace}
\usepackage{enumerate}
\usepackage{diagbox}
\usepackage{threeparttable} 

\usepackage{xurl}
\usepackage{natbib}
\usepackage{hyperref}
\hypersetup{
    colorlinks=true,
    linkcolor=black,
    filecolor=black,
    urlcolor=black,
    citecolor=blue,
}

\usepackage{tabularx,tikz,microtype,enumitem,caption}
\usetikzlibrary{arrows.meta,positioning}

\makeatletter
\providecommand\normalsizeXI{\normalsize}
\providecommand\normalsizeX{\normalsize}
\providecommand\smallX{\small}
\providecommand\footnotesizeIX{\footnotesize}
\makeatother

\usepackage{endnotes}
\makeatletter
\def\enoteheading{\section*{\notesname
		\@mkboth{\MakeUppercase{\notesname}}{\MakeUppercase{\notesname}}}%
	\noindent
}
\makeatother

\usepackage{lipsum}
\makeatletter
\renewcommand\footnoterule{%
  \kern-3\p@
  \hrule\@width 6.5in
  \kern2.6\p@}
\makeatother

 \bibpunct[, ]{(}{)}{,}{a}{}{,}%
 \def\bibfont{\fontsize{10}{12}\selectfont}
 \def\bibsep{\smallskipamount}
 \def\bibhang{24pt}%

 \def\argmax{{\rm argmax}\,}

 \def\do{{\rm dom }\,}

 \def\today{\ifcase\month\or January\or February\or  March\or  April\or
 May\or June\or July\or August\or  September\or October\or November\or
 December\fi  \space\number\day, \number\year}

 \def\qed{\hspace*{10pt}\hfill{$\square$}\hfilneg\par}
 \def\sqr#1#2{{\vcenter{\hrule height .#2pt
       \hbox{\vrule width .#2pt height#1pt \kern#1pt\vrule width.#2pt}
                        \hrule height.#2pt}}}
 \def\square{\mathchoice\sqr54\sqr54\sqr{2.1}3\sqr{1.5}3}
 \def\bm{\boldsymbol}

 \theoremstyle{TH}
 \newtheorem{thm}{Theorem}
 \newtheorem{lem}{Lemma}
 \newtheorem{cor}{Corollary}
 \newtheorem{prop}{Proposition}
 \newtheorem{defi}{Definition}
 \newtheorem{exap}{Example}

 \newcommand{\enu}{\begin{enumerate}}
 \newcommand{\eenu}{\end{enumerate}}
 \newcommand{\des}{\begin{description}}
 \newcommand{\edes}{\end{description}}
 \newcommand{\iit}{\begin{itemize}}
 \newcommand{\eiit}{\end{itemize}}
 \newcommand{\tab}{\begin{tabular}}
 \newcommand{\etab}{\end{tabular}}
 \newcommand{\ben}{\begin{equation}}   \newcommand{\een}{\end{equation}}
 \newcommand{\bea}{\begin{eqnarray}}   \newcommand{\eea}{\end{eqnarray}}
 \newcommand{\beq}{\begin{eqnarray*}}  \newcommand{\eeq}{\end{eqnarray*}}

\TheoremsNumberedThrough     

\EquationsNumberedThrough    

\newcommand{\PaperTitle}{Payment Design under Belief Disagreement: Rent Placement, Bid Resolution, and Auction Choice}
\begin{document}
\RUNTITLE{\PaperTitle}
\TITLE{\PaperTitle}
\ARTICLEAUTHORS{%
    \AUTHOR{Lijian Lu\thanks{The authors used generative AI tools (ChatGPT and Claude Code) to assist with editing, derivations, proof validation, and numerical verification. Authors retain sole responsibility for the paper's content.}}
    \AFF{HKUST Business School, Hong Kong University of Science and Technology, \EMAIL{lijianlu@ust.hk}}
    \AUTHOR{Ce ZHANG}
    \AFF{HKUST Business School, Hong Kong University of Science and Technology, \EMAIL{czhangdp@connect.ust.hk}}
}
\ABSTRACT{%
\noindent\textbf{Problem definition:} When auction bidders misperceive their rivals, a seller can change not only the amount of information rent but also the states in which it is delivered. We ask when this payment-design opportunity survives a finite bidding interface, nonnegative-payment restrictions, and comparison with conventional auctions under the same beliefs.
\noindent\textbf{Methodology/results:} We formulate payment design as a linear program in nonnegative bidder rents. A single state can safely carry a report's rent when it is both cheapest for its intended type and least attractive to potential mimics. We characterize this alignment, bound the cost of a restricted payment architecture when alignment is imperfect, and use shadow prices when the rankings conflict. Own-type projection yields a unique optimal expected-payment rule for a fixed allocation. Contiguous bid bands implement truthful reporting at every value, but create pooling losses and within-band surplus. Under a separate smooth-belief model, feasible truthful rules and revenue upper bounds support comparisons with a first-price equilibrium under the same beliefs. A calibration using 1,000 eBay proof-set auction records illustrates a partition-dependent reversal: first price dominates all exactly truthful, no-subsidy rules on ten equal-probability bands in one scenario, while a richer contingent rule on ten equal-dollar bands outperforms it.
\noindent\textbf{Managerial implications:} An operator should choose interface resolution, payment dependence, and integrity tolerances jointly. Validate the probabilities that support incentives, compare formats under the same beliefs, and distinguish redesign from deployment. Any receipt advantage must cover implementation costs before it can imply a profit advantage.
}
\KEYWORDS{marketplace design; auctions; subjective beliefs; platform operations; revenue management; incentive compatibility}
\maketitle

\section{Introduction}\label{sec:intro}
An auction operator chooses how buyers communicate their values, how reports determine allocation, and how the winner pays. These choices are linked. A finite bidding interface may make a contingent payment rule easier to specify, yet pool buyers with different values and limit the prices compatible with participation. Even a payment rule that exploits bidders' mistaken assessments of competition can therefore earn less than a conventional auction. The question is not simply whether belief disagreement can raise receipts, but which part of that opportunity survives the interface and payment restrictions used to deliver it.

Collectible-item auctions provide the motivating setting. Buyers can value the same item differently and must assess how strongly others value it. \emph{Taste projection}---excessive reliance on one's own preferences when predicting others' values---provides one reason those assessments may differ systematically from the seller's population model \citep{Ross1977,Krueger1994,gagnon-bartsch2021projection}. Prior auction research shows that projection can increase first-price bids while leaving truthful bidding in private-value second-price auctions unchanged \citep{gagnon-bartsch2021projection}. Our question concerns the design of payments and the interface through which those payments are implemented, rather than projection's effect within a given format alone.

The theoretical difficulty is that a cheap payment contingency need not provide cheap incentives. Suppose a buyer overweights a particular configuration of rivals. Leaving rent in that configuration is inexpensive under the seller's objective distribution. However, rent attached to one report also enters the payoffs of every type considering that report. A configuration that is cheap for the intended recipient may be especially attractive to a mimic, increasing the rents required elsewhere. Payment design must account jointly for objective rent cost and these cross-type incentive effects. Our central contribution is to characterize when this state-location problem can be separated from the determination of rent levels.

We study an owner-seller of one object, symmetric buyers with independent private values, and known type-dependent subjective kernels. The seller maximizes objective expected receipts at a truthful subjective equilibrium. In the discrete model, efficient allocation is fixed to isolate payment design. With continuous values, a published partition replaces exact-value messages: the winner is selected uniformly from the highest occupied band, and payments depend on band reports. This changes allocation as well as payment. We compare candidate partitions; we do not derive an optimal communication constraint or assume that the proposed interface is already used in the motivating market.

Our contribution connects three parts of this design problem. \textit{First, we characterize when rent location can be determined before rent levels.} Universal incentive-safe compression is possible precisely when one state minimizes both the intended type's rent cost and every mimic's perceived benefit. This characterizes a replacement operation applicable to every nonnegative schedule, rather than a necessary condition for a particular optimum to use one state. Under strict alignment, the state-contingent program reduces to a unique fixed point in perceived rent levels. A contraction bracket quantifies the cost of restricting the architecture when alignment is approximate; shadow prices characterize the cost--leakage tradeoff otherwise. The contribution is this integrated treatment of interacting incentives, not linear programming or state-contingent transfers alone.

\textit{Second, we translate rent placement into an exact finite-message implementation.} Own-type projection yields a unique diagonal expected-payment rule. For continuous values, the recursion evaluated at band floors is optimal for every contiguous partition under common positive projection intensity. Incentives hold at every value, rather than only at a representative point. The accounting separates receipts, floor rent, within-band buyer surplus, and pooling loss. Competition removes the behavioral premium for fixed bands but not the resolution cost. Refinement under the atomic projection model approaches full extraction, while nonnegative payments limit the preferred state's rent capacity and require overflow rather than arbitrary discount clipping.

\textit{Third, we show why interface and auction comparisons must use the same belief environment.} Under a separate smooth, atomless kernel, the simple diagonal rule is no longer the general solution. We construct a richer class of exactly truthful rules, bound the receipts of all exactly truthful no-subsidy rules on a fixed partition, and derive a symmetric increasing first-price equilibrium with global best responses. Separated feasible revenues and upper bounds rank the alternatives in the specified economy; overlapping bounds leave them unresolved. Message count and the number of contingencies entering payments are different dimensions of design.

The proof-set records of \citet{wegmann2011} provide a monetary scale for these comparisons, not an estimate of bidders' belief distortion. With four buyers and ten equal-probability bands, the projection rule earns \$20.91 at truthful reporting, against \$16.54 for the rational band benchmark and \$16.84 for second price. Its \$4.37 behavioral increment is not its \$4.07 receipt advantage; with two bands, a positive increment coexists with a \$3.80 loss. Under concentrated smooth beliefs, first price exceeds the \$20.82 upper bound for ten equal-probability bands. A feasible ten-band equal-dollar rule instead earns \$21.59, above the \$21.19 first-price upper bound. This positive reversal uses richer payments and has a \$0.40 certified margin in the evaluated scenario. 

These results are conditional design benchmarks. Truthful implementation establishes an equilibrium, not uniqueness or selection of that equilibrium. The calibration does not identify beliefs, validate strategy conjectures, or establish the performance of a frozen mechanism when beliefs change. Keeping these distinctions explicit gives the results a concrete role: identifying promising payment structures, measuring the costs of a finite interface, and screening candidate designs against conventional auctions before deployment-specific validation.

\subsection{Related Literature and Positioning}\label{sec:lit}

Classical auction design links allocation, incentive rents, and seller revenue, which typically assumes a common and known probabilistic model
\citep{Myerson1981optimal,che1998standard,armstrong1996multiproduct,Vohra2012}.
We retain its optimization approach but allow the probability law in the incentive constraints to differ from the law in the objective. Conditional on a fixed allocation and unrestricted transfers, rent schedules delivering the same perceived utility can then have different objective costs. Our contribution is to characterize when a state-cost ranking survives all report deviations and how that characterization supports finite-message payment design. We do not claim a new revelation principle or optimality among every possible allocation and communication rule.

\citet{JACQUES1988} are the closest antecedent for using transfers contingent on rivals' reports. Their surplus-extraction results use objective information dependence and interim participation. Here values are objectively independent, while bidders' subjective conditional distributions are type-dependent and need not agree with the objective law. We impose nonnegative truthful utility in every rival configuration. The question is therefore where to place the rents that statewise participation and incentive compatibility require, not how to append unrestricted side bets to eliminate every type's interim surplus. This participation protection should also be distinguished from robustness of incentives to uncertain beliefs
\citep{BergemannMorris2005,ChungEly2007,Roberto2012}.

That distinction also separates our nominal problem from mechanisms designed for sets of possible beliefs. \citet{Lopomo2022} characterize surplus extraction when each type's beliefs belong to a set; \citet{Li2024Robust} study locally uncertain beliefs and mechanisms that remain truthful throughout the specified uncertainty. Our approximation bound instead compares reoptimized payment architectures at a given kernel. It neither constructs a belief-robust posted contract nor selects among equilibria. Distributionally robust selling \citep{Kocyigit2020,allouah2020prior,Wang2024,Anunrojwong2025} and approximate incentive compatibility \citep{Balseiro2024} likewise address different restrictions on the design problem.

Behavioral auction theory studies how distorted beliefs and strategic reasoning affect outcomes under given formats \citep{crawford2007level,breitmoser2019knowing,gagnon-bartsch2021projection}. In \citet{gagnon-bartsch2021projection}, taste projection changes both perceptions of values and the strategies that bidders conjecture within their perceived games. We maintain correct conjectures about actual strategy maps under subjective value distributions. Neither our kernel nor that convention follows empirically from taste projection. Relative to \citet{morita2024optimal}, who study interpersonal projection in multi-agent moral hazard, our adverse-selection problem couples report deviations and adds a finite bidding interface. Behavioral contracting and operations further motivate attention to contract form and participants' assessments \citep{eliaz2006contracting,grubb2009selling,katok2009contracting,Bordalo2020,Madar2012,Donohue2020,cui2020simplicity,Gao2025}. 

Operational market design emphasizes how institutional constraints change the mechanisms a designer can offer \citep{bhargava2020optimal,Saban2021procurement,choi2023design,Chen2022}. We contribute a belief-sensitive payment framework in which a finite interface changes both the achievable receipts and the allocation, while a conventional auction remains a relevant alternative. Unlike a field implementation study, the calibration does not establish that bidders understand these payments or that the measured belief environment supports them. It identifies the operating tradeoffs and the scope of the receipt comparison. The acceptability of earning receipts from mistaken beliefs is also separate from nonnegative realized utility \citep{Cappelen2025}.

\textbf{Roadmap.} The paper proceeds from the model and equilibrium convention in Section~\ref{sec:model}, to general rent placement in Section~\ref{sec:optimal}, the projection interface in Section~\ref{sec:projection}, payment restrictions and smooth beliefs in Section~\ref{sec:extensions}, and calibrated comparisons in Section~\ref{sec:evidence}. Section~\ref{sec:conclusion} draws bounded operating implications. All proofs and additional calibration details appear in the online appendix.

\textbf{Notation.}
Throughout, boldface denotes vectors (e.g., $\mathbf{s}$), which are taken to be row vectors; a prime denotes transpose (e.g., $\mathbf{s}^{\prime}$).
Vector inequalities are interpreted componentwise.
Let $\mathbb{Z}_+$ denote the set of nonnegative integers, $\mathbf{e}_k$ the $k$th standard unit vector, with a $1$ in the $k$-th position and $0$ elsewhere, and $\mathbf{1}(A)$ the indicator of event $A$.
The set of buyers is $\mathcal{M}:=\{1,\ldots,m\}$, and the type index set is $\mathcal{N}:=\{1,\ldots,n\}$ with the ordered type space
$\mathcal{T}:=\{T_1>T_2>\cdots>T_n>0\}$.
For buyer $i\in\mathcal{M}$, $\mathbf{s}_{-i}:=(s_1,\ldots,s_{i-1},s_{i+1},\ldots,s_m)$ denotes the profile of all other buyers' reports.

\section{Environment, Equilibrium, and Design Framework}\label{sec:model}
An operator (owner seller or mechanism designer) sells a single indivisible object to $m\ge 2$ risk-neutral buyers indexed by $i\in\mathcal M:=\{1,\ldots,m\}$. Buyer $i$ has a private valuation $\widetilde S_i$, drawn independently from 
$
\mathcal T:=\{T_1>T_2>\cdots>T_n>0\}, n\ge2,
$
with \emph{objective} distribution $\mathbf q=(q_1,\ldots,q_n)$, 
$
q_j=\Pr(\widetilde S_i=T_j)>0$ and $\sum_{j=1}^n q_j=1.
$
Write $\widetilde{\mathbf S}=(\widetilde S_1,\ldots,\widetilde S_m)$ for the random valuation profile and $\mathbf s=(s_1,\ldots,s_m)\in\mathcal T^m$ for its realization. The operator's value for the object is normalized to zero. The mechanism and available reports are publicly specified. The operator knows the objective law and the subjective kernels used in design; buyers need not assess rival values according to that law.

\noindent\textbf{Beliefs.}
A type-$T_j$ buyer assigns each rival a \emph{subjective distribution} $\boldsymbol\pi_j=(\pi_{j1},\ldots,\pi_{jn})$, where
$
\pi_{jh}:=\Pr(\widetilde S_\ell=T_h\mid \widetilde S_i=T_j)
$
for $\ell\neq i$; the \emph{rational benchmark} is $\pi_{jh}=q_h$ for all $j,h$. Throughout the finite-type analysis, $\pi_{jh}>0$ and $\sum_h\pi_{jh}=1$; this is the full-support condition used in the probability ratios and proofs. Conditional on her type, the buyer regards rivals as independent draws from $\boldsymbol \pi_j$. Thus $\Pi_j(\mathbf s_{-i})=\prod_{\ell\ne i}\pi_{j,o(s_\ell)}$, where $s_\ell=T_{o(s_\ell)}$. The rational benchmark is $\pi_{jh}=q_h$. These are subjective probabilities, not objective conditional probabilities under independence. We collect these beliefs in the matrix $\boldsymbol\pi=(\pi_{jh})_{j,h\in\mathcal N}$ with $\mathcal N:=\{1,\ldots,n\}$.

Each row of $\boldsymbol \pi$ defines a valid conditional law. The model does not require a common joint law consistent with every buyer's assessment. Buyers are naive: they use their kernels without correcting them, but correctly conjecture the actual reporting strategy in the equilibrium defined below. This is the maintained solution concept, not a claim that buyers solve a common-prior game or are naive about every aspect of others' behavior. The operator is sophisticated and unbiased: it anticipates this behavior and evaluates receipts under $\boldsymbol q$.

\noindent\textbf{Mechanisms and configurations.}
A direct mechanism $(\widetilde{\bm p},\widetilde{\bm x})$ specifies allocation probabilities and expected transfers for every report profile $\bm s \in \mathcal T^m$, with $\sum_i\widetilde p_i(\bm s)\le1$. We study symmetric rules and summarize a buyer's rivals by a \emph{configuration} $\mathbf k=(k_1,\ldots,k_n)$, where $k_r \in\mathbb{Z}_+$ counts reports $T_r$ and $\sum_{r=1}^n k_r = m-1$. The set of all feasible configurations is
$\mathcal K:=\left\{\mathbf k\in\mathbb Z_+^n:\ \sum_{r=1}^n k_r=m-1\right\}.$
Under truthful rival reporting, the objective and type-$j$ subjective probabilities and their ratio are:
\bea
\Phi(\mathbf{k})=\frac{(m-1)!}{\prod_{r=1}^n k_r!}\prod_{r=1}^n q_r^{\,k_r},\ 
\phi_j(\mathbf{k})=\frac{(m-1)!}{\prod_{r=1}^n k_r!}\prod_{r=1}^n \pi_{jr}^{k_r},\ 
\Lambda_j(\mathbf{k}) = \frac{\phi_j(\mathbf{k})}{\Phi(\mathbf{k})}=\prod_{r=1}^{n}\left(\frac{\pi_{jr}}{q_r}\right)^{k_r},\label{eq:weights}
\eea
Let $\mathbf D_j:=(m-1)\mathbf e_j$ be the $j$-th \emph{diagonal} configuration in which every rival reports $T_j$. Define
\bea
\Lambda_j = \max_{\mathbf{k}\in\mathcal{K}} \Lambda_j(\mathbf{k}), \quad
r_j^* = \argmax_{r\in\mathcal{N}} \left\{\pi_{jr}/q_r\right\}, \quad \mathbf{k}_j^* = \mathbf{D}_{r_j^*}. \label{equ:peak-ratio}
\eea Choose one maximizer when there is a tie. Multiplicativity of $\Lambda_j(\mathbf{k})$ implies that its maximum $\Lambda_j$ is attained at $\bm k_j^*$, which we call type $T_j$'s \emph{peak diagonal}; the strict conditions below make the maximizer unique. Under \emph{projection bias}, studied below, this peak occurs at the buyer's own type, so $r_j^*=j$ and $\mathbf{k}_j^* = \mathbf{D}_j$.  A large subjective-to-objective ratio means that type $j$ overweights a configuration relative to its objective frequency.

Let $p_j(\mathbf k)$ and $x_j(\mathbf k)$ be the allocation probability and expected payment\endnote{We note that any mechanism on configuration spaces uniquely derives the direct reporting mechanism. Specifically, given a reporting profile $\mathbf{s}=(s_1,\ldots,s_m)$, each buyer $i\in\mathcal{M}$ is uniquely assigned type $o(s_i)$ and configuration $\mathbf{k}(\mathbf{s}_{-i})\in\mathcal{K}$, where $k_r(\mathbf{s}_{-i}) = \sum_{j\in\mathcal{M},j\neq i}\mathbf{1}(s_j = T_r)$ for all $r\in\mathcal{N}$. Specifically, we set $\tilde{p}_i(\mathbf{s}) = p_{o(s_i)}\left(\mathbf{k}(\mathbf{s}_{-i})\right)$ and $\tilde{x}_i(\mathbf{s}) = x_{o(s_i)}\left(\mathbf{k}(\mathbf{s}_{-i})\right)$. Consequently, characterizing the optimal mechanism on configuration spaces is sufficient.} of a buyer who reports $T_j$ when her rivals generate configuration $\mathbf k$. For a type-$T_j$ buyer, reporting $T_h$ yields an expected utility of $u_{jh}(\mathbf{k})=T_j \, p_h(\mathbf{k})-x_h(\mathbf{k})$, with truthful rent $u_j(\mathbf{k})=T_jp_j(\mathbf k)-x_j(\mathbf k)$.
The corresponding subjective interim utilities, as \emph{perceived} before observing the rivals' configuration, are given by:
\bea\label{eq:two-aggregates}
\hat U_j={\sf E}_{\phi_j}\left[u_j\left(\mathbf{k}\right)\right] =\sum_{\mathbf{k}\in\mathcal{K}}\phi_j\left(\mathbf{k}\right)u_j\left(\mathbf{k}\right), \hat U_{jh}={\sf E}_{\phi_j}\left[u_{jh}\left(\mathbf{k}\right)\right] =\sum_{\mathbf{k}\in\mathcal{K}}\phi_j\left(\mathbf{k}\right)u_h\left(\mathbf{k}\right)+(T_j-T_h) P_{jh},
\eea
where $P_{jh} = \sum_{\mathbf{k}}\phi_j\left(\mathbf{k}\right)\,p_h(\mathbf{k})$ is the subjective probability that a type $T_j$ reporting $T_h$ wins. The last equality holds because $u_{jh}(\mathbf{k}) = u_h(\mathbf{k}) + (T_j-T_h)p_h(\mathbf{k})$.  The first term in $\hat U_{jh}$, $\sum_{\mathbf{k}\in\mathcal{K}}\phi_j\left(\mathbf{k}\right)u_h\left(\mathbf{k}\right)$, reflects the \emph{perceived rent leakage} to a type-$j$ buyer imitating report $h$, while the second, $(T_j-T_h) P_{jh}$, is the \emph{perceived gain from deviating}, weighted by her perceived winning probability.

\noindent\textbf{Allocation and participation.} The discrete analysis centers on an \emph{efficient} allocation: 
\begin{equation}\label{eq:efficient-allocation}
p_j(\mathbf{k})=
\mathbf{1}\left(\sum_{r=1}^{j-1}k_r=0\right)/(k_j+1), \quad \mbox{for each type $j\in\mathcal{N}$ and configuration $\mathbf{k}\in\mathcal{K}$}. \tag{EF}
\end{equation}
The highest report wins, regardless of beliefs, with uniform tie-breaking. This restriction isolates payment design. When reports are bands rather than exact values, the same formula implements a highest-band lottery and is not generally efficient in raw values. Statewise participation requires
\begin{equation}
\label{eq:robust-ir}
u_j(\mathbf k)\ge0,
\qquad \forall j\in\mathcal N,\ \forall \mathbf k\in\mathcal K.
\tag{Robust-IR}
\end{equation}
Because $p_j$ and $x_j$ are expectations over the allocation lottery, this guarantees nonnegative truthful utility for every rival configuration prior to the lottery's outcome. The winner-only implementations detailed below also secure nonnegative utility post-selection. This protection is independent of the probabilities associated with configurations, making it robust against misspecified beliefs and providing a limited-liability interpretation for the mechanism.

\noindent\textbf{Subjective equilibrium (SE).}
Consider a symmetric mixed reporting strategy $\mathbf B=(b_{rh})_{r,h\in\mathcal N}$, where $b_{rh}$ is the
probability that a type $T_r$ reports $T_h$, with every row summing to one. A type-$T_j$ buyer who correctly conjectures $\mathbf B$ assigns rival report $T_h$ probability $\sum_r\pi_{jr}b_{rh}$. Independent rival types and private randomizations induce
$\phi_j^{\mathbf B}(\mathbf k)
=
\frac{(m-1)!}{\prod_{h\in\mathcal N}k_h!}
\prod_{h\in\mathcal N}
\left(\sum_{r\in\mathcal N}\pi_{jr}b_{rh}\right)^{k_h}$,
where $\mathbf k\in\mathcal K$.

\begin{defi}[Subjective equilibrium (SE)]
\label{defi-sbi}
A reporting strategy $\mathbf B$ is a \emph{subjective equilibrium with correct strategy conjectures}\endnote{The subjective equilibrium in a continuous value mechanism is defined as follows. Let $A_i$ denote buyer $i$'s action space and write $p_i(\mathbf a)$ and $x_i(\mathbf a)$ for her allocation probability and expected payment under action profile $\mathbf a$. Analogous to \eqref{eq:configurationEquilibrium}, a strategy profile $\boldsymbol{\mathsf{b}}=(\mathsf{b}_i)_{i\in\mathcal M}$ with $\mathsf{b}_i:[\underline s,\bar s]\to A_i$ is a SE if, for every buyer $i$ and valuation $s_i$,
\[
\mathsf{b}_i(s_i)\in
\argmax_{a_i\in A_i}
\int_{[\underline s,\bar s]^{m-1}}
\Bigl[
s_i p_i\bigl(a_i,\boldsymbol{\mathsf{b}}_{-i}(\mathbf s_{-i})\bigr)
-
x_i\bigl(a_i,\boldsymbol{\mathsf{b}}_{-i}(\mathbf s_{-i})\bigr)
\Bigr]
\prod_{\ell\ne i}\widehat g(s_\ell\mid s_i)\,
\mathrm d\mathbf s_{-i}.
\]
In this framework, each buyer accurately conjectures the strategies of rivals while assessing them based on her own conditional value distribution. For example, in a first-price auction with reserve $s_0$ with a symmetric strictly increasing strategy $\mathsf b$, the condition for an active value $s\ge s_0$ simplifies to
$
s\in\argmax_{y\in[s_0,\bar s]}
\bigl[s-\mathsf b(y)\bigr]\widehat G(y\mid s)^{m-1},
$
with buyers whose valuations fall below $s_0$ choosing to abstain, yielding zero utility.} if,
for $j,h$,
\begin{equation}\label{eq:configurationEquilibrium}
b_{jh}>0
\quad\Longrightarrow\quad
\sum_{\mathbf k\in\mathcal K}
\phi_j^{\mathbf B}(\mathbf k)
\bigl[T_jp_h(\mathbf k)-x_h(\mathbf k)\bigr]
\ge
\sum_{\mathbf k\in\mathcal K}
\phi_j^{\mathbf B}(\mathbf k)
\bigl[T_jp_\ell(\mathbf k)-x_\ell(\mathbf k)\bigr],
\qquad
\forall\,\ell\in\mathcal N.
\end{equation}
\emph{Truthful implementation} means that $b_{jh}=\mathbf{1}(j=h)$ for all $j,h\in\mathcal N$ is such an equilibrium.
\end{defi}

A unilateral deviation changes neither the buyer's true type nor her subjective kernel. Linearity in a report lottery makes pure deviations sufficient. At truthful reporting, $\mathbf B = I_n$ and $\phi_j^{\mathbf B}(\mathbf k)=\phi_j(\mathbf k)$, the truthful implementation becomes
\begin{equation}
\label{eq:biased-ic}
\sum_{\mathbf k\in\mathcal K}\phi_j(\mathbf k)u_j(\mathbf k)
\ge
\sum_{\mathbf k\in\mathcal K}\phi_j(\mathbf k)u_h(\mathbf k)
+(T_j-T_h)P_{jh},
\qquad \forall j\ne h.
\tag{IC}
\end{equation}
The key asymmetry is that buyers evaluate \eqref{eq:configurationEquilibrium} and \eqref{eq:biased-ic} using
their subjective probabilities, whereas the operator evaluates revenue using the objective frequencies $\Phi$.

\noindent\textbf{Objective and implementation.} The operator chooses a mechanism in the class
$\mathcal A_{EF}:=\{(\mathbf p,\mathbf x): \eqref{eq:efficient-allocation},\ \eqref{eq:robust-ir},\ \eqref{eq:biased-ic}\ \text{hold}\}$ to maximize the expected revenue:
\begin{equation}
\label{eq:rev-max}
U_0^*
:=
\max_{(\mathbf p,\mathbf x)\in\mathcal A_{EF}}
 m\sum_{j\in\mathcal N}\sum_{\mathbf k\in\mathcal K}q_j\Phi(\mathbf k)x_j(\mathbf k),
\tag{REV-MAX}
\end{equation}
where the term $\sum_{j\in\mathcal{N}}\sum_{\mathbf{k}\in\mathcal{K}}q_j\,\Phi(\mathbf{k})\,x_j\left(\mathbf{k}\right)$ represents the expected payment from each buyer, calculated under the objective distribution. Since efficiency fixes the allocation rule, the operator's problem reduces to choosing payments subject to robust-IR and IC. Let $U_0^{S}$ denote optimal revenue at the rational benchmark ($\pi_{jh}=q_h$), $\overline U_j:=\sum_{\mathbf k\in\mathcal K}\Phi(\mathbf k)u_j(\mathbf k)$ the expected rent of a type-$T_j$ buyer, and $
\overline T
:=
m\sum_{j\in\mathcal N}\sum_{\mathbf k\in\mathcal K}
q_jT_j\,\Phi(\mathbf k)\,p_j(\mathbf k)$
the expected total \emph{efficient surplus}. Finally, let
$E(\boldsymbol\pi):=U_0^{*}(\boldsymbol\pi)-U_0^{S}$ denote the \emph{behavioral increment} attributable to probability disagreement. 

\begin{lem}[Revenue–rent equivalence]\label{prop-revenue-rent-eq}
\eqref{eq:rev-max} is equivalent to
\begin{equation}
\label{eq:rent-min}
U_{\min}^*
:=
\min_{\mathbf u\in\mathcal U_{EF}}
 m\sum_{j\in\mathcal N}q_j\overline U_j,
\ \mbox{where $\mathcal U_{EF}:= \left\{\mathbf{u}\ge 0: \  \eqref{eq:efficient-allocation}, \eqref{eq:biased-ic} \ \text{hold}\right\}$.}
\tag{RENT-MIN}
\end{equation}
Any optimal utility schedule $\mathbf u^*$ induces payments
$
x_j^*(\mathbf k)=T_jp_j(\mathbf k)-u_j^*(\mathbf k)
$, and $U_0^*(\boldsymbol{\pi}) + U_{\min}^*(\boldsymbol{\pi})=\overline T$.
\end{lem}

\begin{figure}[ht]
\vspace{-0.1in}
\centering
\caption{Economic flow from beliefs to payments and receipts}\label{fig:notations}
\begin{tikzpicture}[node distance=7mm and 7mm,>=Latex,every node/.style={font=\footnotesize},box/.style={draw,rounded corners,align=center,minimum height=8mm,minimum width=22mm}]
\node[box] (prim) {Primitives\\$T_j,q_j$}; \node[box,right=of prim] (obj) {Objective\\$\Phi(\mathbf{k})$};
\node[box,below=of prim] (bel) {Belief\\$\boldsymbol{\pi}_j\Rightarrow\phi_j(\mathbf{k})$}; \node[box,right=of bel] (lr) {Disagreement\\$\Lambda_j(\mathbf{k})$};
\node[box,right=of lr] (loc) {Rent location\\cheap + low leakage}; \node[box,above=of loc] (ic) {Interacting IC\\$\widehat U_j$};
\node[box,right=of loc] (rent) {Rent\\$u_j(\mathbf{k})$}; \node[box,right=of rent] (pay) {Payment\\$x_j=T_jp_j-u_j$};
\node[box,above=of pay] (rev) {Revenue\\surplus $-$ rent};
\draw[->](prim)--(obj);\draw[->](prim)--(bel);\draw[->](obj)--(lr);\draw[->](bel)--(lr);\draw[->](lr)--(loc);\draw[->](loc)--(rent);\draw[->](ic)--(rent);\draw[->](rent)--(pay);\draw[->](pay)--(rev);\draw[->](bel.east) to[bend left=18] (ic.west);
\end{tikzpicture}
\vspace{-0.2in}
\end{figure}
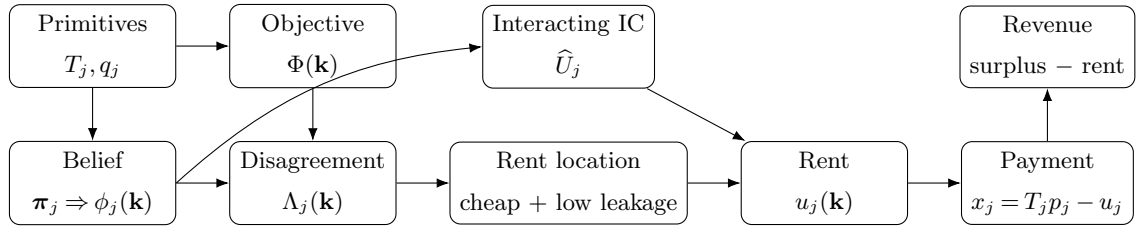

Given efficient allocation, payment design does not change total surplus; instead, it minimizes the objective cost of the rents necessary for incentive compatibility, thereby determining how the surplus is divided between buyers and the operator. Our analysis identifies a truthful subjective equilibrium, establishing the uniqueness of an optimal expected-payment rule within efficient allocation mechanisms.
Figure~\ref{fig:notations} summarizes the role of objective frequencies, subjective beliefs, and interacting incentives.

The model isolates four linked design decisions. First, when buyers and the operator assign different probabilities to the same competitive state, where should incentive rent be placed? Second, how can the resulting state-contingent payment rule be implemented when bids are continuous but the interface is finite? Third, how much of the resulting revenue advantage survives comparison with conventional auction formats under the \emph{same} beliefs and with the same market primitives? Fourth, how does the advantage change when payments must be nonnegative or beliefs vary smoothly within a band?

These questions separate three objects that are often conflated: the behavioral distortion, the mechanism used to monetize it, and the operational environment in which that mechanism is deployed. The theoretical analysis first characterizes when rent location can be separated from interacting incentive constraints. The subsequent sections ask whether that structural advantage survives finite message resolution, a nonnegative-payment constraint, smooth within-band beliefs, and comparison with simpler auction formats. Thus, the paper evaluates belief disagreement as a \emph{design opportunity}, not as a sufficient condition for higher revenue.

\noindent\textbf{Scope and maintained assumptions.}
The baseline discrete model permits unrestricted transfers and imposes statewise individual rationality. Discreteness gives positive probability to the rival configurations on which optimal rent may be concentrated and therefore makes the rent-location logic transparent. For continuous values, exact matches disappear; Section~\ref{sec:bands} replaces them with published bid bands and establishes truthfulness for every value within a band. Section~\ref{sec:robust} then imposes nonnegative payments: the preferred rent state has finite capacity, so the mechanism reallocates overflow rather than truncating the unconstrained discount. Section~\ref{sec:smooth} replaces band-constant projection with one specified smooth belief family.


\subsection{Projection: Behavioral Foundation and Auction Mapping}
\label{subsec:projection-evidence}

Our auction belief distortion is motivated by \emph{taste projection}: individuals tend to overestimate the similarity between others' preferences and their own. A broad false-consensus literature documents excessive reliance on one's own tastes when predicting those of others \citep{Ross1977,Marks1987,Krueger1994,Ambuehl2021}. Most directly for our setting, \citet{gagnon-bartsch2021projection} model auction bidders as perceiving rivals' valuations to be excessively concentrated around their own values. We capture this directional distortion in discrete form as
\begin{equation}\label{eq:projection-beliefs}
\pi_{jj}=A_j:=q_j+\alpha(1-q_j),
\qquad
\pi_{jr}=(1-\alpha)q_r,\quad r\neq j,
\end{equation}
where $\alpha\in[0,1)$ measures the extent of own-type overweighting and $\alpha=0$ is the rational benchmark.

Equation~\eqref{eq:projection-beliefs} is a tractable behavioral specification rather than an empirically identified law for the marketplace we study. Its role is to provide a transparent specialization of the general belief-disagreement model. The main rent-placement results do not require this functional form; they require alignment between the state that is objectively cheapest for providing perceived rent and the state that creates the least leakage to potential mimics. Projection is important because it satisfies this alignment and therefore yields a particularly simple optimal payment architecture.

Related evidence on \emph{information projection} concerns a different mistake: individuals overestimate how much of their private information others possess or can infer \citep{Madar2012,busse2015psychological,gagnon-bartsch2022quality,bushong2024failures}. This literature documents systematic underappreciation of informational differences, but it does not imply that a rival's valuation is perceived to be close to one's own. We therefore use taste projection, rather than information projection, to motivate the auction-specific mapping in \eqref{eq:projection-beliefs}. Information-projection evidence provides broader support for probability disagreement across agents, not identification of our particular valuation-belief distortion.

The distinction matters for interpretation. The data used below do not identify $\alpha$ or validate strategy conjectures. Our nominal results show which rules are truthful when the specified kernel is correct. Belief sensitivity exercises that redesign the rule are not evidence that a fixed rule is incentive-robust. We neither estimate such robustness nor infer projection from a simulated revenue increase. Section~\ref{sec:smooth} separately studies one smooth kernel to examine how the preferred payment architecture changes when beliefs vary within a band.

\section{Incentive-Compatible Rent Placement}\label{sec:optimal}
This section supplies the structural part of the design framework: an exact compression condition, a bound on the cost of a restricted architecture, and a shadow-price characterization when rent cost and imitation incentives conflict. Its results are nominal statements for a given belief matrix.

\subsection{Incentive-Safe Compression}\label{sec:compat}
We begin with the rent-minimization problem \eqref{eq:rent-min} under general beliefs. Its first ingredient is that the objective cost of perceived rent is governed by a single likelihood ratio. Since $\phi_j(\mathbf{k})\leq\Lambda_j\Phi(\mathbf{k})$ for every $\mathbf{k}\in\mathcal{K}$ by \eqref{equ:peak-ratio}, any nonnegative schedule $\mathbf{u}$ delivering perceived interim rent $\hat U_j$ costs the operator at least
\begin{equation}\label{eq:exchange}
\overline{U}_j=\sum_{\mathbf{k}\in\mathcal{K}}\Phi(\mathbf{k})\,u_j(\mathbf{k})
\ \geq\ \frac{1}{\Lambda_j}\sum_{\mathbf{k}\in\mathcal{K}}\phi_j(\mathbf{k})\,u_j(\mathbf{k})=\frac{\hat U_j}{\Lambda_j},
\end{equation}
with equality exactly when the schedule is supported where $\Lambda_j(\mathbf{k})$ attains its maximum. The subjective-to-objective likelihood ratio is therefore an \emph{exchange rate} between perceived and objective rent: one unit of perceived interim rent costs the operator only $1/\Lambda_j$ in objective terms, so sharper disagreement makes incentives cheaper. For a fixed perceived rent the operator favors states with a high ratio: the crucial state is not the one the buyer deems most probable, but the one she \emph{overweights} most relative to its actual occurrence.

Inequality \eqref{eq:exchange} is only half of the design problem. The target rent $\hat U_j$ is not exogenous; it enters every other type's payoff through that type's own subjective weights. Conceding rent in a state relaxes one constraint while tightening others, because a mimic evaluates the same rent with different weights. A state that is cheapest for its intended recipient may still be the wrong place to concede rent if it is conspicuous to a mimic. To capture the effect on mimics, for each report $h$ and each potential mimic $j\neq h$, define the \emph{leakage ratio}
\begin{equation}\label{eq:leak-ratio}
\varrho_{jh}(\mathbf{k}):=\frac{\phi_j(\mathbf{k})}{\phi_h(\mathbf{k})}=\frac{\Lambda_j(\mathbf{k})}{\Lambda_h(\mathbf{k})},
\qquad
\varrho_{jh}:=\min_{\mathbf{k}\in\mathcal{K}}\varrho_{jh}(\mathbf{k}),
\end{equation}
the perceived rent that type $T_j$ collects, per unit of perceived rent delivered to report $T_h$, when that rent sits in state $\mathbf{k}$. The pair $\bigl(1/\Lambda_h(\mathbf{k}),\,(\varrho_{jh}(\mathbf{k}))_{j\neq h}\bigr)$ is the state's \emph{cost--leakage signature}: its first coordinate is what the operator pays, and the remaining $n-1$ coordinates are what each mimic collects. For $\hat U_h>0$ the weights $\phi_h(\mathbf{k})u_h(\mathbf{k})/\hat U_h$ form a probability distribution over $\mathcal{K}$, and both the objective cost of report $h$'s rent and every mimic's leakage from it, each divided by $\hat U_h$, are the corresponding averages of the signature coordinates. Rent placement is thus a choice among convex combinations of signatures, and optimizing the first coordinate alone can worsen the others.

A state that can replace every report-$h$ rent schedule safely must minimize all these coordinates at once. This universal requirement is stronger than the possibility that a particular optimum happens to use one state. We use the following strict alignment conditions on the belief system $\boldsymbol\pi$, for every report $h$ and mimic $j\ne h$: 

\noindent\textbf{(C1)} \emph{Unique peak.} The peak diagonal $\mathbf{k}_h^{\ast}(=\mathbf{D}_{r_h^{\ast}})$ uniquely maximizes $\Lambda_h(\mathbf{k})$ over $\mathbf{k}\in\mathcal{K}$.\\
\noindent\textbf{(C2)} \emph{Minimal leakage.} The same $\mathbf{k}_h^{\ast}$ uniquely minimizes $\varrho_{jh}(\mathbf{k})$ over $\mathbf{k}\in\mathcal{K}$.

By \eqref{eq:weights} both ratios are products of per-rival ratios across the $m-1$ rivals, so each is extremized at a diagonal configuration that assigns all rivals to a single type; \eqref{equ:peak-ratio} identifies the maximizer of $\Lambda_h(\cdot)$ as $\mathbf{k}_h^{\ast}$, and {\rm (C1)} adds only that it is the sole maximizer. Condition~{\rm (C2)} is the substantive requirement: the state cheapest for report $h$ must also be the least visible to every type that might imitate $h$. The next result shows that this pair is not a convenient sufficient condition but the exact boundary of the compression architecture, and that within the boundary the whole design problem collapses onto $n$ rent levels.

\begin{thm}[Incentive-safe compression]\label{thm:compat}
{\rm (i)} For report $h$, a state $\mathbf{k}_h^*\in\mathcal{K}$ permits \emph{universal incentive-safe compression}---replacing any nonnegative schedule by the schedule concentrated at $\mathbf{k}_h^*$, while preserving intended perceived rent and weakly reducing both objective cost and every mimic's payoff---if and only if
\begin{equation}\label{eq:compatibility}
\mathbf{k}_h^*\ \in\ \arg\max_{\mathbf{k}\in\mathcal{K}}\Lambda_h(\mathbf{k})\ \cap \bigcap_{j\neq h}\ \arg\min_{\mathbf{k}\in\mathcal{K}}\varrho_{jh}(\mathbf{k}) .
\end{equation}

\noindent{\rm (ii)} Under {\rm (C1)--(C2)}, these extrema in \eqref{eq:compatibility} are unique for every report, $\bar\varrho:=\max_{j\neq h}\varrho_{jh}<1$, and \eqref{eq:rent-min} is equivalent to the \emph{rent-level program}
\begin{equation}\label{eq:rlp}
\min_{\hat{\boldsymbol U}\,\geq\,0}\ \ m\sum_{j\in\mathcal N}\frac{q_j\hat U_j}{\Lambda_j}
\quad\text{s.t.}\quad
\hat U_j\ \geq\ \varrho_{jh}\,\hat U_h+(T_j-T_h)\,P_{jh},\quad \forall j\neq h.
\tag{RLP}
\end{equation}
The two programs share their optimal value and their unique optima are paired by
\begin{equation}\label{eq:compat-support}
u_j^{\ast}(\mathbf{k})=\frac{\hat U_j^{\ast}}{\phi_j(\mathbf{k}_j^{\ast})}\,\mathbf{1}(\mathbf{k}=\mathbf{k}_j^{\ast}),\qquad \forall j\in\mathcal{N}.
\end{equation}

\noindent{\rm (iii)} The unique optimal perceived rents $\hat{\boldsymbol U}^{\ast}$ are the fixed point of 
\begin{equation}\label{eq:fixed}
\mathcal{R}(\hat{\boldsymbol U})_j:=\max\Bigl\{0,\ \max_{r\neq j}\bigl[\varrho_{jr}\hat U_r+(T_j-T_r)P_{jr}\bigr]\Bigr\},\qquad j\in\mathcal{N},
\end{equation}
iteration from zero increases to $\hat{\boldsymbol U}^{\ast}$.
\end{thm}

Theorem~\ref{thm:compat} separates the \emph{location} of rent from its endogenous \emph{level}. By part~(i), the two roles a state must play---objectively cheap for its intended recipient and perceptually unattractive to every mimic---are ranked by two different ratios, and compression into one state is safe precisely where the two rankings agree; necessity makes \eqref{eq:compatibility} a boundary rather than a hypothesis. Part~(ii) pins the support at the peak diagonals and leaves only the levels: a report of $T_h$ then leaks exactly $\varrho_{jh}\hat U_h$ to a type-$T_j$ deviator and costs the operator $\hat U_h/\Lambda_h$, so \eqref{eq:biased-ic} becomes linear in perceived rents and $n|\mathcal{K}|$ state-contingent variables collapse to $n$. Uniqueness in {\rm (C2)} makes $\mathcal{R}$ a contraction and uniqueness in {\rm (C1)} leaves a single optimal schedule, so the operator settles where every discount sits from the belief matrix alone and, by part~(iii), solves for its size without knowing which constraints bind.


\subsection{The Cost of Approximate Compatibility}\label{sec:approx}
An operator may prefer the operational simplicity of a single contingent discount per report, as outlined in Theorem~\ref{thm:compat}, even when the cost and leakage rankings (i.e., Conditions~{\rm (C1)--(C2)}) do not perfectly align. We evaluate the cost of this restriction at a specified kernel, with payments in each program reoptimized accordingly. This comparison does not imply that a posted rule remains fixed while users' beliefs are altered.

Restrict report $h$'s rent to its peak diagonal $\mathbf{k}_h^{\ast}$. Let 
\[
\varrho^{\rm D}_{jh}:=\varrho_{jh}(\mathbf{k}_h^{\ast}), \ 
\bar\varrho^{\rm D}:=\max_{j\neq h}\varrho^{\rm D}_{jh}, \ 
\Delta\varrho:=\max_{j\neq h}\bigl(\varrho^{\rm D}_{jh}-\varrho_{jh}\bigr)\ \geq\ 0.
\]
Condition~{\rm (C2)} is exactly $\Delta\varrho=0$ together with uniqueness. The restricted program is \eqref{eq:rlp} with $\varrho^{\rm D}$ in place of $\varrho$: rent costs $\hat U_h/\Lambda_h$ and leaks $\varrho^{\rm D}_{jh}\hat U_h$ to mimic $j$. Let $\mathcal{R}^{\rm D}$ be \eqref{eq:fixed} with these coefficients, $\hat{\boldsymbol U}^{\rm D}$ its fixed point, and $U_{\min}^{\rm D}:=m\sum_hq_h\hat U_h^{\rm D}/\Lambda_h$ the optimal value of that restricted program. On the other side, \eqref{eq:rlp} remains well defined without {\rm (C2)}: the map $\mathcal{R}$ contracts whenever $\bar\varrho^{\rm D}<1$ since $\bar\varrho\leq\bar\varrho^{\rm D}$, so $\hat{\boldsymbol U}^{\ast}$ is still its componentwise-minimal solution, and we write $U_{\min}^{\rm L}:=m\sum_hq_h\hat U_h^{\ast}/\Lambda_h$. Without {\rm (C2)}, \eqref{eq:rlp} is only a relaxation, and its value need not be attained.

\begin{thm}[Bracket and loss bound]\label{thm:approx}
Suppose $\bar\varrho^{\rm D}<1$. Then $\mathcal{R}^{\rm D}$ is a contraction of modulus $\bar\varrho^{\rm D}$, its fixed point $\hat{\boldsymbol U}^{\rm D}$ solves the restricted program, and $\hat{\boldsymbol U}^{\rm D}\geq\hat{\boldsymbol U}^{\ast}\geq0$.

\noindent{\rm (i)} \emph{Bracket.} The unrestricted optimum satisfies
$U_{\min}^{\rm L}\ \leq\ U_{\min}^{\ast}\ \leq\ U_{\min}^{\rm D}$.

\noindent{\rm (ii)} \emph{Loss bound.} The cost of restricting the architecture satisfies
\begin{equation}\label{eq:approx}
0\leq U_{\min}^{\rm D}-U_{\min}^{\ast}
 \leq U_{\min}^{\rm D}-U_{\min}^{\rm L}
 \leq 
m\left(\sum_{h\in\mathcal N}\frac{q_h}{\Lambda_h}\right)
\cdot\frac{\Delta\varrho}{1-\bar\varrho^{\rm D}}\cdot\bigl\|\hat{\boldsymbol U}^{\ast}\bigr\|_\infty,
\ 
\bigl\|\hat{\boldsymbol U}^{\ast}\bigr\|_\infty \leq \frac{\max_{j<h}(T_j-T_h)P_{jh}}{1-\bar\varrho}.
\end{equation}

\noindent{\rm (iii)} Under {\rm (C1)--(C2)}, $\Delta\varrho=0$ and the bracket collapses to Theorem~\ref{thm:compat}.
\end{thm}

The bound factors into three terms an operator can interpret separately: the aggregate exchange rate $m\sum_hq_h/\Lambda_h$ of Section~\ref{sec:compat}; the misalignment relative to the slack left by incentive feedback, $\Delta\varrho/(1-\bar\varrho^{\rm D})$, small when beliefs are sharply distinguishable across types; and the scale $\|\hat{\boldsymbol U}^{\ast}\|_\infty$ of the rent at stake, controlled by the largest downward deviation wedge. Exact own-type overweighting is therefore not what justifies a single contingent discount; an operator need only check that $\Delta\varrho$ is small relative to $1-\bar\varrho^{\rm D}$, and both numbers come from the belief matrix directly.

Part~{\rm (i)} is the operational content. Iterating the two contractions from $\mathbf 0$ brackets the unrestricted optimum from both sides without solving the $n|\mathcal{K}|$-variable program \eqref{eq:rent-min}, and the width of the bracket is itself the certified loss. Its lower end bounds \emph{every} feasible schedule, because it relaxes each leakage coefficient to its minimum; its upper end prices one specific architecture. The two need not coincide, as Example~\ref{exap:leakage} below shows, a different single-state architecture may perform better by accepting a worse objective exchange rate in return for lower leakage.

\subsection{Shadow Prices When the Rankings Conflict}\label{sec:lp}
When the intersection in \eqref{eq:compatibility} is empty, the two rankings a state must satisfy point in different directions, thus no state uniformly minimizes all signature coordinates. Which state is preferable now can depend on \emph{which} incentive constraints are binding, and that is endogenous to the mechanism. Duality supplies exactly the missing information. Let $\eta_{jh}\geq0$ be the multiplier on the constraint in \eqref{eq:biased-ic} that deters type $T_j$ from reporting $T_h$, collected in $\boldsymbol\eta$, and define the \emph{incentive-adjusted cost} of one unit of perceived rent delivered to report $h$ in state $\mathbf{k}$,
\begin{equation}\label{eq:adjprice}
c_h(\mathbf{k};\boldsymbol\eta)
:=\underbrace{\frac{mq_h}{\Lambda_h(\mathbf{k})}}_{\text{objective outlay}}
+\underbrace{\sum_{j\neq h}\eta_{jh}\,\varrho_{jh}(\mathbf{k})}_{\text{shadow value of leakage}} .
\end{equation}
The first term is the exchange rate of \eqref{eq:exchange}, weighted by how often report $h$ is made; the second charges the rent for the incentive constraints it loosens on the way out. Program \eqref{eq:rent-min} is a finite linear program in $\mathbf{u}$, so its optimum is characterized by these prices.

\begin{prop}[Shadow-priced placement and sparse support]\label{prop:dual}
Suppose \eqref{eq:rent-min} is feasible, as it is whenever $\bar\varrho^{\rm D}<1$.

\noindent{\rm (i)} \emph{Certificate.} A feasible schedule $\mathbf{u}$ is optimal if and only if there exists $\boldsymbol\eta\geq0$ such that
\begin{equation}\label{eq:dual}
c_h(\mathbf{k};\boldsymbol\eta)\ \geq\ \sum_{r\neq h}\eta_{hr}
\qquad\text{for every }h\in\mathcal{N},\ \mathbf{k}\in\mathcal{K},
\end{equation}
with equality wherever $u_h(\mathbf{k})>0$, and every constraint carrying $\eta_{jh}>0$ binding. The common optimal value is $U_{\min}^{\ast}=\sum_{j\neq h}\eta_{jh}(T_j-T_h)P_{jh}$.

\noindent{\rm (ii)} \emph{Sparse support.} Some optimal schedule places each report's rent on at most $n$ states.

\noindent{\rm (iii)} A state minimizes $c_h(\cdot;\boldsymbol\eta)$ over $\mathcal{K}$ for \emph{every} $\boldsymbol\eta\geq0$ if and only if it satisfies \eqref{eq:compatibility}.
\end{prop}

Rent appears only where its incentive-adjusted cost equals the value of the constraints it relaxes. The left side of \eqref{eq:dual} is what a unit of perceived rent costs the operator in state $\mathbf{k}$ once the rent leaked to every mimic is charged at its shadow value; its right side is what that unit buys, namely relaxation of the constraints report $h$ itself must satisfy. Rent is placed only where the two are equal, and a state can be preferred over one with a higher likelihood ratio because it is less visible to the types whose constraints are expensive. Part~{\rm (ii)} bounds the complexity of the resulting rule: the operator never needs more than $n$ contingent transfers per report, even without compatibility, so the rule stays finite and auditable.

Part~{\rm (iii)} answers the question in the heading. What replaces the likelihood-ratio ranking is the same ranking shifted by the shadow value of leakage, and conditions~{\rm (C1)--(C2)} are precisely the case in which the shift changes nothing: the compatible state is cheapest at every price vector, so the operator can rank states before knowing which constraints bind. Outside that case the ranking is jointly determined with $\boldsymbol\eta$, which is why Theorem~\ref{thm:compat} has no belief-only analogue and why Theorem~\ref{thm:approx} prices the simple architecture rather than repairing it.


\begin{exap}[Cost and leakage need not align]\label{exap:leakage}
Take $m=2$ and $n=3$, with $\mathbf q=(0.40,0.40,0.20)$, $(T_1,T_2,T_3)=(30,20,10)$, and
\[
\boldsymbol\pi=(\pi_{jr})=\begin{pmatrix}
0.68&0.30&0.02\\
0.15&0.60&0.25\\
0.10&0.10&0.80
\end{pmatrix}.
\]
Every type's own diagonal is its unique likelihood-ratio peak, so {\rm (C1)} holds, but rent for report $T_2$ leaks least at $\mathbf{D}_3$ rather than at $\mathbf{D}_2$, so {\rm (C2)} fails there. The optimal schedule places $u_1^{\ast}(\mathbf{D}_1)=45/17$ and $u_2^{\ast}(\mathbf{D}_3)=5$, all other rents being zero, at objective cost $28/17$; the multipliers $\eta_{12}=8/17$ and $\eta_{23}=288/425$, all others zero, certify it through \eqref{eq:dual}. Confining each report to its own likelihood-ratio peak instead costs $449/255$, an excess of $29/255$, or $6.90\%$ of minimum rent cost. Report $T_2$'s optimal one-state location is therefore \emph{not} its cheapest state in isolation. Appendix~\ref{app:proofs} verifies every constraint and the dual certificate.
\end{exap}
Because the type-$1$ constraint binds, its multiplier makes the leakage term in \eqref{eq:adjprice} decisive, and the optimum accepts a worse exchange rate on report $T_2$ in return for a smaller transfer to the mimic. Disagreement that is large but misaligned therefore does not translate into a usable own-type rebate: what the operator can exploit is a state that is cheap and quiet at the same time, and Section~\ref{sec:projection} shows that projection delivers one.

Misalignment is one way disagreement fails to pay; its direction is another. Take $n=2$ and suppose each type overweights the \emph{opposite} type, so that $\pi_{12}>q_2$ and $\pi_{21}>q_1$, or equivalently $\pi_{11}<q_1$ and $\pi_{22}<q_2$, since each row and $\mathbf{q}$ sum to one. Any multiplicative tilt toward the other type lies in this class, however strong. The cost and leakage rankings align at the opposite-type diagonals, so {\rm (C1)--(C2)} hold and Theorem~\ref{thm:compat} still applies.

\begin{cor}[Disagreement without premium]\label{cor:reverse}
Fix $m\geq2$, let $n=2$, and take any belief system with $\pi_{12}>q_2$ and $\pi_{21}>q_1$. The unique optimal mechanism in $\mathcal{A}_{EF}$ leaves $u_1^{\ast}(\mathbf{D}_2)=(T_1-T_2)/m$ and zero rent elsewhere. Its total objective rent is $q_1q_2^{\,m-1}(T_1-T_2)$, exactly the rational minimum under the same allocation, and its payment $x_j^{\ast}=T_jp_j-u_j^{\ast}$ is nonnegative in every state.
\end{cor}

The subjective weight has cancelled. Rent for report $T_1$ sits at $\mathbf{D}_2$, where one unit of perceived rent costs $1/\Lambda_1(\mathbf{D}_2)=(q_2/\pi_{12})^{m-1}$; but $\mathbf{D}_2$ is also the only state in which the deterred report wins, so the perceived rent required, $(T_1-T_2)\pi_{12}^{\,m-1}/m$, rises by exactly the same factor and $E(\boldsymbol\pi)=0$ however sharp the tilt. Projection separates the two states---rent sits on the buyer's own diagonal while a downward report wins where rivals are lower---so the cancellation fails there and the premium is real. Taken together with Example~\ref{exap:leakage}, the benchmark draws the line between a favorable likelihood ratio and a cheaper incentive system: only the second is worth money, so an operator should check that the two come apart before paying for a contingent architecture. This is a two-type class, not a theorem that every distortion other than own-type projection is valueless.

\section{From Projection to an Implementable Interface}\label{sec:projection}
In this section, we first identify a simple payment rule under projection in Section~\ref{sec:diagonal}, then implement it on a finite interface in Section~\ref{sec:bands}, and distinguish the receipt opportunity from the allocation and pricing consequences of that interface in Section~\ref{sec:thick}. 

\subsection{The Projection Specialization}\label{sec:diagonal}
Projection enters \eqref{eq:rent-min} only through the subjective weights $\phi_j(\mathbf{k})$, and by \eqref{eq:weights} those weights depend on the belief system only through per-rival ratios. Under \eqref{eq:projection-beliefs} the two families of ratios that Section~\ref{sec:compat} requires are
\begin{equation}\label{eq:per-rival-ratio}
\frac{\pi_{jr}}{q_r}=
\begin{cases}
A_j/q_j>1, & r=j,\\[2pt]
1-\alpha<1, & r\neq j,
\end{cases}
\qquad\qquad
\frac{\pi_{jr}}{\pi_{hr}}=
\begin{cases}
A_j/\bigl((1-\alpha)q_j\bigr)>1, & r=j,\\[2pt]
(1-\alpha)q_h/A_h<1, & r=h,\\[2pt]
1, & r\notin\{j,h\} .
\end{cases}
\end{equation}
These products are uniquely extremized at the relevant diagonals. Hence {\rm (C1)--(C2)} hold, with 
\begin{equation}\label{eq:proj-peak}
\mathbf{k}_j^{\ast}=\mathbf{D}_j,
\qquad
\Lambda_j=\Bigl(\frac{A_j}{q_j}\Bigr)^{m-1},
\qquad
\varrho_{jh}=\Bigl(\frac{(1-\alpha)q_h}{A_h}\Bigr)^{m-1}=:\varrho_h,\quad j\neq h .
\end{equation}
A type-$T_j$ buyer assigns her matching state the probability $A_j^{\,m-1}$ against its objective probability $q_j^{\,m-1}$, and the ratio compounds across rivals; unlike the opposite-type benchmark of Corollary~\ref{cor:reverse}, the peak is the buyer's \emph{own} diagonal. Since $A_h>(1-\alpha)q_h$, every $\varrho_h<1$, so $\bar\varrho<1$ and Theorem~\ref{thm:compat} applies in full.

Beyond locating the rent, projection does two things general beliefs do not. It \emph{collapses leakage}: the $n(n-1)$ factors $\varrho_{jh}$ reduce to the $n$ factors $\varrho_h$, so how visible a rent is depends on the imitated type alone and not on who is watching. And it \emph{orients} the constraints: every upward constraint turns out to be strictly slack, so \eqref{eq:rlp} ceases to be a two-directional program and becomes a ladder solved in one pass from the bottom.

\noindent\textbf{Diagonal payments and rent levels.} Concentrating rent at the diagonals means $u_j(\mathbf{k})=z_j\mathbf{1}(\mathbf{k}=\mathbf{D}_j)$, where $z_j$ is the matching rent, so that $\hat U_j=A_j^{\,m-1}z_j$ and the expected-payments are
\begin{equation}\label{eq:diagpay}
x_j(\mathbf{k})=T_j\,p_j(\mathbf{k})-z_j\,\mathbf{1}(\mathbf{k}=\mathbf{D}_j),\qquad j\in\mathcal{N} .
\end{equation}
We call \eqref{eq:diagpay} the \emph{diagonal payment rule} with rents $\mathbf{z}=(z_1,\ldots,z_n)$. It reads as a bet: a buyer surrenders her whole surplus in every configuration in which some rival differs from her, and is compensated only in the one configuration in which all of them match her report. Projection makes that configuration look disproportionately likely to its intended recipient and unusually unlikely to any mimic, which is exactly the pair of properties {\rm (C1)--(C2)} ask for. Robust-IR still bars charging above value; the gain comes from relocating the rent that incentive compatibility requires.

Write $Q_h:=\sum_{r\geq h}q_r$ with $Q_{n+1}=0$, and let
$\bar P_h:=\frac{Q_h^{\,m}-Q_{h+1}^{\,m}}{m\,q_h}$
be the \emph{objective} interim winning probability of a report $T_h$ under \eqref{eq:efficient-allocation}. The extent $\alpha$ enters the rent ladder through two separate channels, the denominator $A_j^{\,m-1}$ of the perceived level and the leakage factors $\varrho_h$. Rescaling by $\zeta_j:=\bigl(A_j/(1-\alpha)\bigr)^{m-1}z_j$ removes the first and leaves a clear recursion:
\begin{equation}\label{eq:zetarec}
\zeta_n=0,
\qquad
\zeta_j=\max_{h>j}\ \bigl[\varrho_h\,\zeta_h+(T_j-T_h)\,\bar P_h\bigr],
\qquad
z_j=\frac{(1-\alpha)^{m-1}}{A_j^{\,m-1}}\,\zeta_j .
\end{equation}
Each type is granted the least rent that deters her most attractive downward report, net of what that report's diagonal already leaks to her.

\begin{thm}[Optimal diagonal expected-payment rule]\label{thm:diagonal}
Fix the efficient allocation \eqref{eq:efficient-allocation}, and let $\boldsymbol\zeta$ and $\mathbf{z}$ be generated by \eqref{eq:zetarec}.

\noindent{\rm (i)} For every $\alpha\in(0,1)$, $\mathbf{z}\geq0$, the levels $\hat U_j^{\ast}=A_j^{\,m-1}z_j$ solve \eqref{eq:rlp}, every upward constraint is strictly slack, and $u_j^{\ast}(\mathbf{k})=z_j\mathbf{1}(\mathbf{k}=\mathbf{D}_j)$ is the unique optimal rent schedule for \eqref{eq:rent-min}, with objective rents $\overline U_j=q_j^{\,m-1}z_j$.

\noindent{\rm (ii)} The diagonal payment rule \eqref{eq:diagpay} is therefore the unique solution of \eqref{eq:rev-max}, and only the winner pays: $T_j-mz_j$ when all $m$ reports equal $T_j$, and $T_j$ otherwise.

\noindent{\rm (iii)} Revenue is $U_0^{\ast}(\alpha)=\overline T-m\sum_jq_j^{\,m}z_j$, continuous on $[0,1)$ and strictly increasing to $\overline T$ as $\alpha\uparrow1$. At $\alpha=0$ the recursion remains optimal and returns the rational rents $\overline U_j^{S}=\sum_{r>j}(T_{r-1}-T_r)\bar P_r$, but no state is then cheaper than another and rent \emph{location} is no longer unique---the classical revenue-equivalence boundary. Hence the behavioral increment satisfies
\begin{equation}\label{eq:recovery}
\Bigl(1-\max_{j<n}\varrho_j\Bigr)\bigl(\overline T-U_0^{S}\bigr)\ \leq\ E(\alpha)\ \leq\ \overline T-U_0^{S},
\end{equation}
with equality on the left when $n=2$, and is positive for every $\alpha\in(0,1)$.
\end{thm}

The mechanism changes the states in which surplus is conceded, not who receives the object: allocation stays efficient in every configuration, and part~{\rm (iii)} measures the gain against the \emph{same} allocation. By \eqref{eq:recovery}, projection recovers the rational floor rent up to a fraction $\max_{j<n}\varrho_j$, because a diagonal rent costs its leakage factor times its rescaled level, $\overline U_j=\varrho_j\zeta_j$, and rescaled levels never exceed rational rents.
Note that the matching price $T_j-mz_j$ can be negative, which is a subsidy rather than a violation of \eqref{eq:robust-ir}: an operator that forbids payments to bidders cannot post the rule unmodified, and Section~\ref{sec:extensions} solves the capped program instead.

\begin{exap}[A two-type payment illustration]\label{exap:main-two-type}
Let $m=2$, $(T_1,T_2)=(30,12)$, $\mathbf q=(0.20,0.80)$ and $\alpha=0.50$. A specialist believes her rival is a specialist with probability $A_1=0.60$ rather than $0.20$; a generalist puts $A_2=0.90$ on another generalist. Then $\Lambda_1=3.00$ and $\varrho_2=4/9$, and with $\bar P_2=0.40$ the recursion gives $\zeta_2=z_2=0$ and $\zeta_1=(T_1-T_2)\bar P_2=7.20$, hence $z_1=6.00$. A specialist who meets a generalist pays her full $30.00$ and keeps nothing; the matching configuration has objective probability $0.04$, and there the winner pays $T_1-2z_1=18.00$. Expected rent is $q_1^2z_1=0.24$ per bidder, 
so with efficient surplus $\overline T=18.48$ the revenue ladder is
$U_0^{S}=15.60\ <\ U_0^{\ast}=18.00\ <\ \overline T=18.48$,
of which $E=2.40=(1-\varrho_1)(\overline T-U_0^{S})$, with $\varrho_1=1/6$, is behavioral increment.
\end{exap}

The example illustrates the nominal truthful-equilibrium calculation. Its binding downward constraint supplies no strict incentive margin against belief misspecification.

\subsection{Exact Incentive Compatibility within Published Value Bands}\label{sec:bands}
Exact matching has objective probability zero under a continuous value distribution. Concentrating perceived rent on that event can therefore exploit a singular difference between objective and subjective probabilities, enough to extract the entire efficient surplus. A finite partition replaces an exact-value contingency with a positive-probability, report-verifiable event, at the cost of changing allocation and the prices that statewise participation permits.

Let values be independent draws from a continuous $G$ on a bounded support $[\underline s,\bar s]\subset(0,\infty)$. The operator publishes descending contiguous bands $\mathcal{I}=\{I_1,\ldots,I_n\}$, $I_j=[L_j,H_j]$ with $H_1=\bar s$, $L_n=\underline s$ and $L_j=H_{j+1}$ for $j<n$. Assign shared endpoints consistently. Let
$q_j=\Pr(\widetilde S_i\in I_j)>0$,
$\delta_j=H_j-L_j$, and
$\mu_j={\sf E}\bigl[\widetilde S_i\mid \widetilde S_i\in I_j\bigr]$.
Buyers know their exact values but report bands; allocation and payment use band reports only, with uniform selection among the highest-band reports.  Beliefs over rival bands are constant within the buyer's true band. Under the continuous analogue of \eqref{eq:projection-beliefs}:
$\widehat G(s'\mid s)=\alpha\mathbf{1}(s'\geq s)+(1-\alpha)G(s')$, a value-$s$ buyer assigns her own band probability $\alpha+(1-\alpha)q_j=A_j$ and other bands probability $(1-\alpha)q_h$. This law contains an atom at the buyer's exact value, analogous to \eqref{eq:projection-beliefs}. Band constancy and the fixed mass retained by shrinking own bands are substantive assumptions, not consequences of finite reporting or general evidence of similarity bias.

Because the mechanism cannot distinguish values inside a band, the floor $L_j$ is the highest price every member of $I_j$ can be made to pay. Define the \emph{floor rent} $u_j(\mathbf{k})=L_jp_j(\mathbf{k})-x_j(\mathbf{k})$. A truthful buyer of value $s\in I_j$ then realizes $u_j(\mathbf{k})+(s-L_j)p_j(\mathbf{k})$, so $u_j\geq0$ gives statewise participation at every value \eqref{eq:robust-ir}. Reporting band $h$ yields subjective payoff $\sum_{\mathbf{k}}\phi_j(\mathbf{k})u_h(\mathbf{k})+(s-L_h)P_{jh}$, affine in $s$ with coefficient $P_{jh}$. Consequently, the binding value is an endpoint of $I_j$ and which endpoint binds is decided by the sign of $P_{jh}-P_{jj}$. The worst endpoint yields
\begin{equation}\label{eq:bandd}
d^{B}_{jh}=(L_j-L_h)P_{jh}+\delta_j\max\{P_{jh}-P_{jj},0\} .
\end{equation}

\begin{thm}[Band implementation without representative-value substitution]\label{thm:band}
Fix a published partition $\mathcal{I}$ and band-constant beliefs $\boldsymbol\pi$.

\noindent{\rm (i)} Statewise participation at every value is equivalent to $\mathbf u\ge0$, and incentive compatibility at every value is \eqref{eq:biased-ic} on band primitives, with $T_j$ replaced by $L_j$ in floor rents and $(T_j-T_h)P_{jh}$ replaced by \eqref{eq:bandd}. Theorems~\ref{thm:compat}--\ref{thm:approx} and Proposition~\ref{prop:dual} apply to this continuous-value, finite-message problem.

\noindent{\rm (ii)} Under common projection \eqref{eq:projection-beliefs}, recursion \eqref{eq:zetarec} at $T_j=L_j$ solves the full band problem for every contiguous partition, without restricting band widths. Every upward constraint is strictly slack. 

\noindent{\rm (iii)} The rule is implemented by charging only the winner: the floor $L_j$ of the winning band in every profile in which some rival reports a different band, and the matching price $L_j-mz_j$ when all $m$ reports fall in $I_j$. Losers pay nothing. Truthful band reporting is a subjective equilibrium. 
\end{thm}

Part~{\rm (i)} is what makes the interface exact rather than approximate: incentives are checked at both endpoints of every band and therefore hold throughout it, so no representative value ever stands in for a continuum. The continuum changes only the deviation constant, and because \eqref{eq:bandd} enters the program exactly where $(T_j-T_h)P_{jh}$ did, the rent-placement theory of Section~\ref{sec:optimal}---compression, the loss bound, the shadow-price certificate---transfers without restatement.

Part~{\rm (ii)} is the substantive addition. The extra term in \eqref{eq:bandd} is the premium a buyer at the top of her band would earn by claiming the band above, and no a priori argument says the floor recursion covers it. It does, for every partition, because the slope bound behind Theorem~\ref{thm:diagonal} caps how fast rescaled rents fall along the ladder by exactly that amount. This is proved for projection beliefs, not assumed; other band-constant laws require the full endpoint program of part~{\rm (i)}.

Part~{\rm (iii)} turns the latent diagonal of Section~\ref{sec:diagonal} into an event an operator can announce and audit---\emph{all active bids fall in the same published band}---while bidders still submit ordinary numerical bids and only the payment rule changes. Two cautions follow. Allocation and payment must both use the disclosed bands, since breaking within-band ties by raw bids is a different mechanism to which none of the above applies. How finely to partition is then a design choice, not a question of approximation quality, Section~\ref{sec:thick} prices it. The matching price can be negative, so an operator that forbids subsidies must check $mz_j\leq L_j$, Section~\ref{sec:extensions} solves the capped program otherwise. 

\subsection{Pricing, Pooling, and Resolution Limits}\label{sec:thick}
A published partition buys contractibility at a price, and the price is not the rent. Let
$w_j:=Q_j^{\,m}-Q_{j+1}^{\,m}=mq_j\bar P_j$
be the probability that the highest occupied band is $I_j$, and let $\widehat S$ denote the value of the buyer who receives the object. Write $\widetilde S_{(1)}\geq\cdots\geq\widetilde S_{(m)}$ for descending order statistics, so that $\widetilde S_{(1)}$ is the highest and $\widetilde S_{(2)}$ the second highest. Since the allocation ignores positions inside the winning band, ${\sf E}[\widehat S]=\sum_jw_j\mu_j$, while $\overline T=\sum_jw_jL_j$ is the expected winning-band floor---all the mechanism can charge for. The efficient continuous surplus therefore is decomposed into:
\begin{equation}\label{eq:band-split}
{\sf E}\bigl[\widetilde S_{(1)}\bigr]
=\underbrace{U_0^{\ast}}_{\text{receipts}}
+\underbrace{U_{\min}^{\ast}}_{\text{floor rent}}
+\underbrace{\textstyle\sum_jw_j(\mu_j-L_j)}_{\text{within-band surplus}}
+\underbrace{\varepsilon_{\mathcal I}}_{\text{pooling loss}},
\qquad
\varepsilon_{\mathcal I}:={\sf E}\bigl[\widetilde S_{(1)}-\widehat S\bigr] ,
\end{equation}
the first two terms adding to $\overline T$ by Lemma~\ref{prop-revenue-rent-eq}. The second and third terms are what the interface costs. As shown in Section~\ref{sec:diagonal}, projection can reduce floor rent as it vanishes toward zero as projection sharpens, but it cannot remove the above-floor surplus of values pooled in one band: a buyer whose value sits above her band's floor keeps that difference whatever the operator does with rent.  Rent placement and resolution act on different terms. Pricing above the floor cannot recover within-band surplus while protecting every type.
\begin{cor}[Pricing bands at their means]\label{cor:mean}
Under common projection, let \eqref{eq:zetarec} run at $T_j=\mu_j$, producing $\mathbf z^{\mu}$, and charge the winner $\mu_j$, or $\mu_j-mz^{\mu}_j$ when all $m$ reports fall in $I_j$. Against truthful rivals, a buyer of value $s\in I_j$ never gains by reporting a higher band; if $s\geq\mu_j$, truthful reporting is optimal and leaves nonnegative realized utility; if $s<\mu_j$, an unmatched win leaves her $s-\mu_j<0$ and, for $j<n$, reporting the band $h$ that attains the maximum in \eqref{eq:zetarec} gains $(\mu_j-s)(P_{jj}-P_{jh})>0$. Each failure is at most $\mu_j-L_j\leq\delta_j$.
\end{cor}
Mean pricing thus fails at every value below the mean. In the base case ($m=4$, $\alpha=0.5$, ten equal-probability bands), pricing the top band at its \$34.12 mean charges a floor-value winner \$7.26 above her value and permits a \$2.61 gain from reporting the band below. Representative-value pricing is therefore not an exact implementation. The next result prices both instruments and the two benchmarks an operator would compare against.

\begin{prop}[Resolution and competition]\label{prop:resolution}
Fix a partition $\mathcal{I}$ and let $\delta:=\max_j\delta_j$.

\noindent{\rm (i)} \emph{Pooling loss bound.} $0\leq\varepsilon_{\mathcal I}\leq\delta$; if $G$ has density bounded by $\bar g$, then $\varepsilon_{\mathcal I}\leq\binom{m}{2}\bar g\,\delta^{2}$.

\noindent{\rm (ii)} \emph{Baseline.} $|U_0^{S}-U_0^{SPA}|\leq\delta$, where $U_0^{SPA}={\sf E}[\widetilde S_{(2)}]$ is the revenue of a second-price auction.

\noindent{\rm (iii)} \emph{Behavior increment.}  With partition $\mathcal{I}$ fixed, for every $\alpha\in(0,1)$,
\begin{equation}\label{eq:premium-bound}
0\ \leq\ E(\alpha)\ \leq\ m\,(L_1-L_n)(1-q_1)^{m-1}\ \xrightarrow{m\rightarrow\infty} \ 0.
\end{equation}

\noindent{\rm (iv)} \emph{Thick markets.} With partition $\mathcal{I}$ fixed, for every $\alpha\in(0,1)$,
\begin{equation}\label{eq:thicklimit}
U_0^{\ast}(\alpha)-U_0^{SPA}\ \xrightarrow{m\rightarrow\infty} \ L_1-H_1=-\delta_1,
\qquad
\varepsilon_{\mathcal I}\ \xrightarrow{m\rightarrow\infty} \ H_1-\mu_1 .
\end{equation}
\end{prop}

The two costs of a coarse interface behave very differently in $m$. Part~{\rm (iii)} says competition destroys the behavioral opportunity exponentially: with the partition held fixed, a top-band rival appears almost surely, and the rent that projection was saving disappears with it. Pooling, bounded by the mesh $\delta$ in part~{\rm (i)}, does not merely fail to improve with competition: by part~{\rm (iv)} it rises to $H_1-\mu_1$, because the winner is increasingly pooled with rivals she genuinely outvalues. The banded rule then ends up $\delta_1$ \emph{below} the second-price benchmark, and both losses are pure resolution costs that no amount of competition or projection removes. In thin markets the premium is worth having and coarse bands are tolerable; as the market thickens, the case for the rule turns entirely on whether the partition is fine enough that $\delta_1$ costs less than the interface is worth.

\begin{prop}[Refinement and the exact-matching limit]\label{prop:refine}
Fix $\alpha\in(0,1)$ and $m$.

\noindent{\rm (i)} \emph{Exact matching.} On exact-value reports, some efficient, statewise individually rational mechanism with truthful reporting as a subjective equilibrium under $\widehat G$ earns ${\sf E}[\widetilde S_{(1)}]$.  It concedes rent only when all reports coincide, an event of subjective probability $\alpha^{m-1}$ and objective probability zero.

\noindent{\rm (ii)} \emph{Refinement.} If $G$ has density bounded by $\bar g$, then for every partition
\[
0\ <\ {\sf E}\bigl[\widetilde S_{(1)}\bigr]-U_0^{\ast}(\alpha)\ \leq\ \delta+\binom{m}{2}\bar g\,\delta^{2}+\Bigl(\frac{(1-\alpha)\,\bar g\,\delta}{\alpha}\Bigr)^{m-1}\bigl(\overline T-U_0^{S}\bigr).
\]
Consequently, as $\delta\to0$ (i.e., the number of bands $n\to\infty$), $U_0^{\ast}(\alpha)\to{\sf E}[\widetilde S_{(1)}]$ and $E(\alpha)\to{\sf E}[\widetilde S_{(1)}-\widetilde S_{(2)}]$.
\end{prop}

Propositions~\ref{prop:resolution} and \ref{prop:refine} describe different comparative statics: increasing competition at fixed resolution, and refining resolution at fixed competition. The refinement conclusion specifically uses the atom in $\widehat G$ and unrestricted transfers: for every $\alpha\in(0,1)$ the premium tends to the whole information rent ${\sf E}[\widetilde S_{(1)}-\widetilde S_{(2)}]$ that second-price leaves to its winner. In the calibrated base case, the limiting information rent is \$7.94, compared with a \$4.37 premium at ten bands and \$6.89 at fifty. The all-matching probability $\sum_jq_j^m\le(\max_jq_j)^{m-1}$ vanishes; the calibrated recursion first requires a subsidy at 61 bands. Thus uncapped nominal receipts do not select a finite resolution, and the atomic limit is not evidence that arbitrarily fine interfaces are operationally desirable.


\section{Nonnegative Payments and Smooth Beliefs}\label{sec:extensions}
The baseline permits unrestricted transfers and band-constant beliefs. We examine these restrictions separately, rather than treating the following models as interchangeable implementations. 

\subsection{Nonnegative Payments}\label{sec:robust}
An operator that never pays a bidder requires, on top of \eqref{eq:robust-ir},
\begin{equation}\label{eq:cap}
0\ \leq\ u_j(\mathbf{k})\ \leq\ L_j\,p_j(\mathbf{k})
\end{equation}
for bands, or the same with $T_j$ for discrete types. If the rule of Theorem~\ref{thm:band} already satisfies \eqref{eq:cap}---that is, if $mz_j\leq L_j$ for every $j$---it remains optimal in the restricted class, and nothing changes. Otherwise the operator must solve the capped rent program, and two shortcuts fail. Clipping the discount to the cap generally destroys \eqref{eq:biased-ic}, because the binding constraints are those of the uncapped optimum. And while the cap limits the diagonal's capacity rather than its ranking, so that it is still filled first, it no longer absorbs all rent.
\begin{prop}[Diagonal saturation]\label{prop:sat}
Fix $\alpha\in(0,1)$ and a published partition. The capped program with constraints \eqref{eq:cap} attains its optimum, and every optimal schedule $\mathbf{u}^{\ast}$ satisfies
\[
u_j^{\ast}(\mathbf{k})>0\ \text{ for some }\ \mathbf{k}\neq\mathbf{D}_j
\qquad\Longrightarrow\qquad
u_j^{\ast}(\mathbf{D}_j)=\frac{L_j}{m},\qquad j\in\mathcal{N} .
\]
\end{prop}

The cap changes capacity, not the cost--leakage ranking. The cap restricts the diagonal's \emph{capacity}---at $\mathbf{D}_j$ a truthful buyer wins with probability $1/m$, so the state can carry at most $L_j/m$---but it leaves {\rm (C1)--(C2)} untouched, and those are properties of the belief system alone. The own diagonal therefore remains both the cheapest state for band $j$ and the least visible to its mimics, and is filled to the brim before any other state is used. What the cap does change is where the overflow goes: that is settled by the shadow prices of Proposition~\ref{prop:dual}, not by the ranking of Theorem~\ref{thm:compat}. The constraint binds when the required discount is large relative to the floor it must respect---weak projection, a wide gap to the band below, a rare top band---and where it binds the operator re-solves rather than clips.

\subsection{Smooth Beliefs and A Same-Belief Benchmark}\label{sec:smooth}
The atomic law $\widehat G(s'\mid s)=\alpha\mathbf{1}(s'\geq s)+(1-\alpha)G(s')$ underlying Theorem~\ref{thm:band} assigns probability $\alpha$ to a rival having \emph{exactly} the buyer's value, and the exact interface runs on that atom: the own-band ratio $A_j/q_j=1-\alpha+\alpha/q_j$ diverges as $q_j\downarrow0$, so the diagonal is cheap because a vanishing objective event keeps a fixed subjective mass, not because buyers think rivals are merely similar. The atom also makes a tie with all rivals at the buyer's own value a subjective event of probability $\alpha^{m-1}$, which monotone first-price analysis assumes away. We therefore study a \emph{separate} smooth law that concentrates near the buyer's own value:
\begin{equation}\label{eq:smoothlaw}
\widehat g(s'\mid s)=\frac{g(s')\exp\bigl[-(\log s'-\log s)^{2}/2\theta^{2}\bigr]}{Z(s)},
\qquad s'\in[\underline s,\bar s],
\end{equation}
here $Z(s)$ normalizes the density and $\theta>0$ is an exogenous \emph{belief-spread} parameter that measures how far subjective mass extends from the buyer's own log value---the spread of the neighborhood law of Appendix~\ref{app:caldetail}, read on raw values---and $\theta\to\infty$ returns rational beliefs. Unlike a band boundary, $\theta$ is not the seller's control. This exercise changes the kernel as well as the permitted payment dependence; it is not a smooth approximation validating the atomic mechanism at fixed prices.

Beliefs now vary inside each band, so incentives must hold at every value, and the endpoint reduction of Theorem~\ref{thm:band} is unavailable. Write $\widehat\phi_s(\mathbf{k})$ for the probability that a value-$s$ buyer assigns to rival configuration $\mathbf{k}$ under \eqref{eq:smoothlaw}, and, for $j<n$, let $\Gamma_j(\mathbf{k}):=u_j(\mathbf{k})-u_{j+1}(\mathbf{k})-\delta_{j+1}p_{j+1}(\mathbf{k})$ be the realized gain, at the shared boundary $L_j=H_{j+1}$, from reporting band $j$ rather than band $j+1$.

\begin{prop}[An exactly truthful class and an upper bound]\label{prop:exact}
Fix a published partition and the band allocation, and let beliefs follow \eqref{eq:smoothlaw}.

\noindent{\rm (i)} \emph{Exact truthfulness.} Suppose \eqref{eq:cap} holds and, for every $j<n$, $\sum_{\mathbf{k}\in\mathcal{K}}\widehat\phi_{L_j}(\mathbf{k})\Gamma_j(\mathbf{k})=0$ and
\begin{equation}\label{eq:triangular}
\Gamma_j(\mathbf{D}_j)\geq0,
\qquad
\Gamma_j(\mathbf{k})\leq0\ \text{ if }\textstyle\sum_{r\leq j}k_r=0,
\qquad
\Gamma_j(\mathbf{k})=0\ \text{ otherwise}.
\end{equation}
Truthful band reporting is then a subjective equilibrium, optimal at every value. These finite linear restrictions are satisfied, with $\Gamma_j\equiv0$, by the dominant-strategy rule~\eqref{eq:envelope}.

\noindent{\rm (ii)} \emph{Upper bound.} Write incentive compatibility constraints imposed at finitely many values as $\mathbf{u}\mathbf{A}\geq\mathbf{a}$. For every $\mathbf{y}\geq0$, every exactly truthful rule satisfying \eqref{eq:cap} earns at most
\[
\overline T-\mathbf{a}\mathbf{y}^{\prime}-\sum_{j\in\mathcal N}\sum_{\mathbf{k}\in\mathcal K}L_jp_j(\mathbf{k})\min\bigl\{mq_j\Phi(\mathbf{k})-(\mathbf{A}\mathbf{y}^{\prime})_{j\mathbf{k}},\,0\bigr\},
\]
where $(\mathbf{A}\mathbf{y}^{\prime})_{j\mathbf{k}}$ is the coefficient of $u_j(\mathbf{k})$ in $\mathbf{u}\mathbf{A}\mathbf{y}^{\prime}$.
\end{prop}

Part~{\rm (i)} replaces the endpoint reduction. Indifference pins each switch to its boundary, and since a higher value tilts \eqref{eq:smoothlaw} toward higher rival values, \eqref{eq:triangular} makes the tilt reinforce the switch rather than reverse it. The restrictions are triangular: a positive boundary gain can arise only at the all-matching configuration $\mathbf{D}_j$ and must be offset where every rival is below band $j$. The class is sufficient, not exhaustive, and its payments may vary with the whole configuration of rival bands. Part~{\rm (ii)} needs no optimality of $\mathbf{y}$, so the multipliers a solver returns for a grid program certify an upper bound on every exactly truthful, no-subsidy rule with the band allocation. An alternative that beats this bound beats every such rule on the partition, and a member of the class that beats the alternative settles the converse.

The natural alternative is a first-price auction under the \emph{same} beliefs, which uses the distortion without a matching discount, a published partition or pooling: a buyer who thinks rivals resemble her shades her bid less, and the seller collects the difference. Because \eqref{eq:smoothlaw} is ordered by the monotone likelihood ratio, with a reserve $s_0$ the auction has exactly one symmetric, strictly increasing, differentiable subjective equilibrium, in closed form and globally optimal. It earns $U_0^{FPA}(s_0)$, against $U_0^{SPA}(s_0)$ for a second-price auction with the same reserve, in which truthful bidding is dominant under any belief. 

\begin{prop}[A global first-price equilibrium]\label{prop:fpa}
Fix $s_0\in[\underline s,\bar s)$. Under \eqref{eq:smoothlaw} and the stated density regularity, a first-price auction with reserve $s_0$ and uniform tie-breaking has exactly one symmetric, strictly increasing, differentiable subjective equilibrium on $(s_0,\bar s]$ with $\mathsf{b}(s_0)=s_0$:
\begin{equation}\label{eq:fpabid}
\mathsf{b}(s)=s-\int_{s_0}^{s}\exp\Bigl[-(m-1)\int_{y}^{s}
\frac{\widehat g(z\mid z)}{\widehat G(z\mid z)}\,\mathrm{d}z\Bigr]\mathrm{d}y ,
\end{equation}
read at $s_0=\underline s$ as its continuous improper limit. Bidding according to \eqref{eq:fpabid} is global best response against itself, not merely a first-order solution. Objective receipts are
\begin{equation}\label{eq:fparev}
U_0^{FPA}(s_0)=m\int_{s_0}^{\bar s}\mathsf{b}(s)\,G(s)^{m-1}g(s)\,\mathrm{d}s.
\end{equation}
A second-price auction with the same reserve earns
\begin{equation}\label{eq:sparev}
U_0^{SPA}(s_0)=m\,s_0\bigl[1-G(s_0)\bigr]G(s_0)^{m-1}
+m(m-1)\int_{s_0}^{\bar s}s\bigl[1-G(s)\bigr]G(s)^{m-2}g(s)\,\mathrm{d}s ,
\end{equation}
with $U_0^{SPA}(\underline s)=U_0^{SPA}$. Both formats trade with probability $1-G(s_0)^{m}$.
\end{prop}

This is equilibrium uniqueness within the specified increasing differentiable class, not seller-optimality of first price among mechanisms. Second-price truthfulness is belief-independent in this private-value environment; strengthening its benchmark means optimizing its reserve, not correcting its beliefs. The smooth-belief comparison adds a first-price equilibrium under the same subjective kernel. Its allocation and possible reserve-induced exclusion differ from the band lottery, so the comparison evaluates whole designs rather than isolating a payment-only effect.

\section{Data-Scaled Calibration and Counterfactuals}\label{sec:evidence}
This section uses calibrated economies and numerical comparisons to quantify interface losses and compare candidate mechanisms at their specified equilibria. The historical records determine a price scale and discipline a maintained value model. They do not identify projection, establish the independent-private-values interpretation of the source market, or validate a proposed rule's operating receipts.

\subsection{Data, Calibration, and the Evaluated Economy}\label{sec:calibration_data}
\textbf{Data.} We use the \emph{ebaycoins} data of \citet{wegmann2011}: 1{,}000 eBay U.S. Mint proof-set auctions collected from 2007--2008, recording book value, reserve-to-book ratio, bidder count, closing price including shipping, and item and seller indicators. Of the 1{,}000 listings, 863 sold and 137 attracted no bidder. Among sold listings the median bidder count is four (mean 4.21; interquartile range 2--6), the median book value is \$22.50, and the median closing-price-to-book ratio is 0.91. We therefore set $m=4$ and $B=\$22.50$ as the baseline; Table~\ref{tab:cal-desc} in Appendix~\ref{app:caldetail} describes it. 


\noindent\textbf{Maintained value interpretation.} Let $X=\widetilde S_i/B$ denote value relative to book value, with $X\sim\mathrm{Lognormal}(\kappa,\sigma^{2})$. We retain auctions with a positive closing price, at least two bidders, a closing ratio at least $0.02$ above the reserve ratio and below $3$, and a reserve ratio below $1.8$, leaving 749 (Table~\ref{tab:cal-screen}); the $0.02$ screen removes sales effectively at the reserve, for which bid increments make a continuous-density likelihood inappropriate. The calibration reads closing price as the second-highest entered value, models entry as value exceeding the reserve, and normalizes the likelihood over the retained price window (Appendix~\ref{app:caldetail}). These are maintained assumptions for this calibration, not findings inherited from the source study, which analyzes common-value auctions. Book value is a referenced retail valuation, not an observed bidder value. The pooled fit does not reproduce that source study's structural common-value analysis.

The maximizer is $\widehat\kappa=-0.50$, $\widehat\sigma=0.53$, implying $\mathbb{E}[X]=0.70$. All 500 auction-level bootstrap refits converge, with percentile intervals $[-0.55,-0.45]$ and $[0.50,0.58]$. Relative fit favors the lognormal over the reported gamma and Weibull alternatives, but a 300-replication re-estimated parametric-bootstrap specification test gives $p=0.02$. The pooled specification is rejected; the fit therefore supplies a data-scaled scenario rather than a validated objective value law. The alternative-family and screening results in Appendix~\ref{app:caldetail} concern the projection-versus-second-price comparison, not the positive smooth-belief reversal.

\noindent\textbf{Counterfactual population and interface.} The evaluated economy fixes $m=4$, $B=\$22.50$, and values $\widetilde S_i=BX_i$ drawn independently from the central 99\% of the fitted distribution, renormalized to $[\underline s,\overline s]=[\$3.47,\$54.25]$. The base case uses equal-probability bands and $\alpha=0.5$. The count four is motivated by the median number of historical entrants, but the evaluated buyers are four fixed draws in a new bounded economy. Their population is not identified with reserve-selected historical entrants, and entry is not re-equilibrated after a change in mechanism.

Band means are conditional means of that evaluated law. Incentive constraints are imposed at both endpoints under band-constant beliefs, payments follow Theorem~\ref{thm:band}, and receipts, rents, buyer surplus, and pooling use \eqref{eq:band-split}. Each scenario redesigns the rule for its stated primitives. Neither the bootstrap nor the alternative specifications measure the performance of a previously posted schedule held fixed.


\subsection{Resolution Changes the Nominal Receipt Comparison}\label{sec:calibration_resolution}
In the continuous base economy, second-price auction earns $\$16.84$. The comparison decomposes as
\begin{equation}\label{eq:adopt}
U_0^{\ast}(\alpha)-U_0^{SPA}\;=\;\underbrace{E(\alpha)}_{\text{behavioral increment}}\;+\;\underbrace{U_0^{S}-U_0^{SPA}}_{\text{interface baseline}}.
\end{equation}
The second term is a signed consequence of the band interface, not a behavioral gain. Proposition~\ref{prop:resolution}{\rm (ii)} bounds it by the mesh $\delta$ and permits either sign; it is negative in the reported equal-probability scenarios and shrinks as resolution improves.

\begin{figure}[htbp]
\vspace{-0.1in}
\centering
\caption{Truthful-equilibrium receipts of the band rules ($m=4$, $\alpha=0.5$)}\label{fig:resolution}
\includegraphics[width=.72\textwidth]{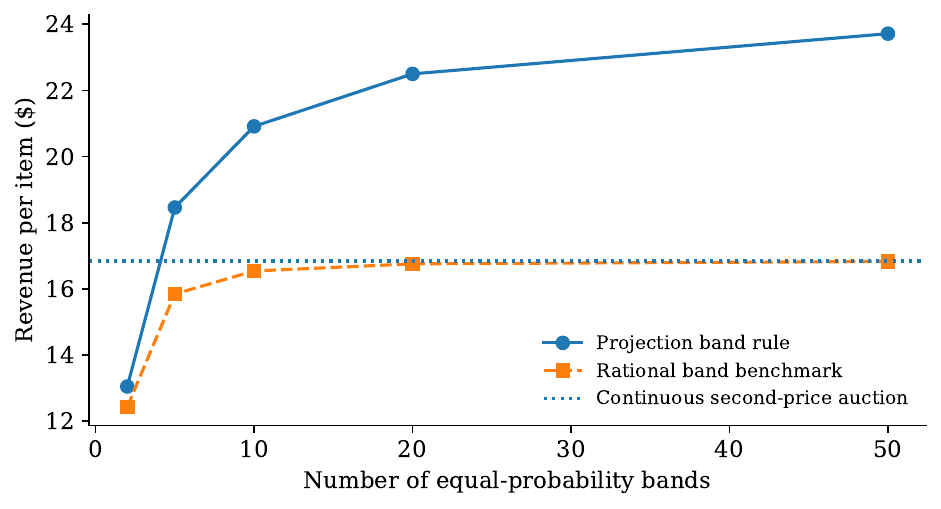}
\vspace{-0.2in}
\end{figure}

Figure~\ref{fig:resolution} compares mechanisms within the same continuous economy across different numbers of bands. With two bands, the \$0.62 behavioral increment does not offset a \$4.41 interface-baseline shortfall: the rule loses \$3.80 per item relative to second price and pools away \$3.75 of surplus. At ten bands, the increment is \$4.37, the shortfall \$0.30, and the net advantage \$4.07. Pooling loss is \$0.27, or 1.08\% of efficient surplus $\mathbb E[\widetilde S_{(1)}]=\$24.78$. Buyers retain \$3.60, essentially all within-band surplus rather than floor rent. Entries are rounded independently, so \eqref{eq:adopt} holds numerically up to one cent. The sign reversal is a model comparison, not an observed adoption outcome.

\noindent\textbf{The probabilities supporting the schedule.} The schedule of Theorem~\ref{thm:band}{\rm (iii)} makes the same point from the other side. At ten bands, the largest matching discount is \$11.17 (Table~\ref{tab:cal-sched}). The objective probability that all four reports match is $10^{-3}$; conditional on any own band, a buyer assigns all three rivals to it probability $A_j^3=0.55^3\approx0.17$. The top band spans \$26.86--\$54.25. At fifty bands, the largest discount is \$37.02. These magnitudes identify what an implementation would need to validate: beliefs about the relevant joint matching event, not merely the presence of projection in general. Finer partitions reduce pooling while attaching discounts to rarer objectively observed events. The calibration does not elicit those event probabilities.

\begin{figure}[htbp]
\vspace{-0.1in}
    \begin{minipage}{0.45\textwidth}
    {\footnotesize\centering\setlength{\tabcolsep}{2.5pt}
    \begin{tabular}{@{}lrrr@{}}\toprule
Ten-band & No-subsidy & Pooling & Uncapped\\
partition & revenue & loss & min.\ price\\\midrule
Equal probability & 20.91 & 0.27 & 3.47\\
Finer upper tail & 22.19 & 0.12 & 3.47\\
Equal dollar width & 22.24 & 0.24 & $-21.19$\\\bottomrule
\end{tabular}
        \captionof{table}{Boundary choices at ten messages}
        \label{tab:partition}
    }
    \end{minipage}\hfil
    \begin{minipage}{0.53\textwidth}
    {\centering
        \includegraphics[width=\textwidth]{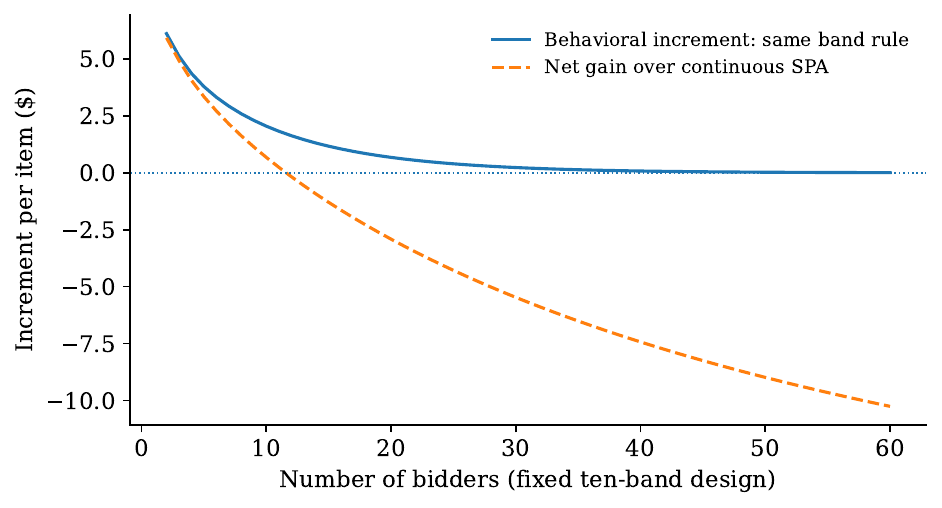}
        \captionof{figure}{Competition removes the behavioral premium but not fixed-resolution costs}
        \label{fig:thick}
    }
    \end{minipage}
\vspace{-0.2in}
\end{figure}

\noindent\textbf{Boundary placement at fixed message count.} Table~\ref{tab:partition} holds ten bands fixed and reoptimizes payments under \eqref{eq:cap}. Probability knots at $(i/n)^{1/2}$ refine the upper tail, raising no-subsidy receipts to \$22.19 and reducing pooling loss to \$0.12; equal dollar widths earn \$22.24. The uncapped equal-dollar recursion would require a minimum winner price of $-\$21.19$, whereas the finer-upper-tail design never prices below \$3.47. Enforcing the cap costs less than $10^{-5}$ per item in the former case because affected states are objectively rare. This small objective cost does not justify clipping the discount; the constrained rule reallocates rent.


Figure~\ref{fig:thick} and Table~\ref{tab:cal-thick} hold the ten-band partition fixed while raising the number of bidders $m$. The same-allocation increment is \$6.12 at two bidders and \$4.37 at four, and tends to zero. The comparison with second-price auction becomes negative at twelve bidders and reaches $-\$10.26$ at sixty, approaching $-\delta_1$, the theoretical limit indicated in \eqref{eq:thicklimit}. This fixed-interface result suggests reassessing resolution as the market thickens, though it does not address issues of endogenous participation or optimal redesign across different market sizes.

\subsection{Auction Choice under the Same Beliefs}\label{sec:calibration_auction}
We strengthen the comparison in two distinct ways: optimize the second-price reserve, then evaluate first price and band rules under the common smooth kernel~\eqref{eq:smoothlaw}. Private-value second-price bidding is truthful under either kernel. The stronger comparison comes from reserve optimization and the addition of first price, rather than a correction to second-price bidding incentives.

Truthful bidding is dominant in a second-price auction under any belief, so $U_0^{SPA}(s_0)$ can be optimized over the reserve $s_0$ without selecting an equilibrium. At four bidders, second price's best reserve is \$10.73, raising receipts from \$16.84 to \$16.87 with sale probability 0.99. The projection rule's receipt advantage is then \$4.04 rather than \$4.07. The correction is small here only because four bidders already do most of the reserve's work: at two bidders, reserve optimization raises second-price receipts from \$11.20 to \$11.71.

Under \eqref{eq:smoothlaw}, the endpoint reduction of Theorem~\ref{thm:band} is unavailable, so we bracket the best exactly truthful, no-subsidy rule on each of the three ten-band partitions of Table~\ref{tab:partition}. An explicit rule in the class of Proposition~\ref{prop:exact}{\rm (i)} supplies a feasible lower bound; multipliers of a grid relaxation supply an upper bound through part~{\rm (ii)}. First-price auction is evaluated over reserves. Appendix~\ref{app:caldetail} describes the exact and interval-arithmetic certificates. A ranking is established only when a feasible revenue exceeds the other design's upper bound.

\begin{figure}[htbp]
\vspace{-0.1in}
\centering
\caption{Nominal same-belief comparisons: boundary changes can reverse the ranking}\label{fig:smooth}
\includegraphics[width=.72\textwidth]{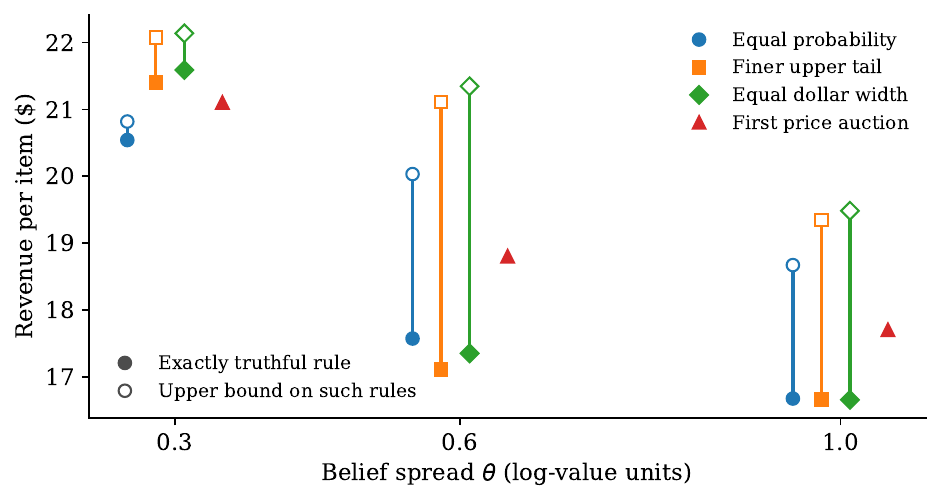}
\vspace{-0.2in}
\end{figure}

\begin{table}[tb]
\centering\caption{Decisive reported comparisons at $\theta=0.3$; four bidders and ten bands.}\label{tab:certified}
\begin{tabularx}{\linewidth}{l r X}\toprule
Band partition & Relevant band receipt & Supported comparison\\\midrule
Equal probability & Upper bound: 20.82 & Below first-price feasible lower bound 21.05.\\
Finer upper tail & Feasible rule: 21.40 & Above first-price upper bound 21.19.\\
Equal dollar width & Feasible rule: 21.59 & Above first-price upper bound 21.19.\\\bottomrule
\end{tabularx}
\end{table}
Table~\ref{tab:certified} and Figure~\ref{fig:smooth} show that band placement is a substantive auction-design lever: changing boundaries and redesigning payments can reverse revenue rankings under identical beliefs. At $\theta=0.3$, the equal-probability partition has a revenue upper bound of \$20.82---\$3.95 above reserve-optimized second-price revenue, but below the \$21.05 first-price lower bound at a feasible reserve. First price therefore dominates every exactly truthful, no-subsidy rule on that partition. In contrast, feasible rules on finer-upper-tail and equal-dollar bands earn \$21.40 and \$21.59, respectively, both exceeding the \$21.19 upper bound for reserve-optimized first price. This certified reversal highlights the value of evaluating band boundaries and contingent payments jointly, rather than judging band-based mechanisms from a single partition. At $\theta\in\{0.6,1\}$, however, feasible band revenues remain below first-price revenues while their upper bounds lie above them, leaving all six comparisons unresolved.


The positive reversal pertains to a richer contract rather than the straightforward matching-discount rule. Payments under this contract can depend on as many as 715 rival-configuration entries, while simple matching discounts---assuming smooth beliefs and no subsidies---yield at most \$0.09 above the rational band benchmark. The equal-dollar rule, however, results in a loss of \$0.24 in efficient surplus, leaving \$2.96 to buyers. Importantly, sampling error does not alter the negative result; recomputing the equal-probability comparison at $\theta=0.3$ using 100 bootstrap refits shows the gap remains negative in 98 cases, with a percentile interval of $[-0.53,-0.03]$. 

Taken together, the results show that the partition can change a nominal format ranking at a fixed message count, but that the economically successful smooth-belief rule also uses more contingent payments. They do not separate every incremental source of the gain through nested complexity-restricted optimizations, and do not establish robust superiority of a posted contract.

\section{Managerial Implications and Conclusion}\label{sec:conclusion}
The paper links a structural payment-design question to the constraints of a finite interface. Belief disagreement gives incentive rent a location as well as a level. Whether that improves a candidate auction depends on interacting deviations, band resolution, payment restrictions, and the conventional alternative. Four implications follow from the analysis.

\textit{Evaluate incentive cost, not probability distortion alone.} The right rent state must be cheap for its intended type and unattractive to mimics. Under exact alignment, location can be chosen before levels; approximate alignment admits a quantified architectural cost; otherwise shadow prices value the leakage tradeoff. Opposite-type overweighting in the two-type benchmark produces no premium. Measuring a bias without identifying its effect on the incentive system is therefore insufficient to justify a contingent payment design.

\textit{Treat the bidding interface as an allocation and pricing decision.} Bands determine who can win and the maximum price consistent with participation by every value in a band. The projection comparison moves from a \$3.80 loss at two bands to a \$4.07 gain at ten; at fixed ten-band resolution, the comparison turns negative from twelve bidders. Better boundary placement can improve receipts without adding messages. These are reasons to compare explicit interface candidates, not a recommendation to refine indefinitely: the full-extraction limit depends on atomic beliefs, rare contingencies, and unrestricted transfers.

\textit{Compare complete payment designs against a strong conventional auction.} A same-allocation behavioral increment is not the incremental receipt advantage over the available alternative. First price exploits distorted assessments of competition without a matching rebate, and can dominate all exactly truthful no-subsidy rules on a given partition. The positive smooth-belief reversal uses a substantially richer payment schedule. Its certified margin, rather than the larger projection-versus-second-price gap, is the relevant scenario budget for the added contract dependence. Nonnegative payments should be imposed inside the optimization; clipping an otherwise optimal discount is not an incentive-preserving shortcut.

\textit{Separate a nominal design benchmark from deployment validation.} The contemplated protocol requires a fixed participant set, published band boundaries, a disclosed highest-band lottery, and payments determined from recorded band reports. Winner selection, payment computation, and any rounding must be consistent with that rule. The model does not analyze withdrawals, duplicate identities, seller shills, collusion, or the cost of auditing them; it supplies no integrity tolerance or manipulation-proofness guarantee. Nor does statewise participation establish equilibrium selection or buyers' comprehension of configuration-contingent payments. Those requirements should be checked before treating truthful-equilibrium receipts as an operating forecast.

The analysis takes values and the participant population as primitives. The pooled calibration is rejected as a complete statistical specification, and beliefs and strategy conjectures are not identified. The positive smooth-belief comparison is numerically certified at its stated primitives but lacks the paired sampling and specification evidence needed for a broader superiority claim. Equilibrium selection, fixed-rule performance under belief error, endogenous entry, and payment-complexity costs remain outside the analysis. Within that scope, the framework identifies when incentive rent can be placed independently of its level and shows how interface and payment constraints can preserve, erode, or reverse the resulting nominal receipt advantage.

\vspace{.1in}
\theendnotes

\bibliographystyle{informs2014}
\bibliography{reference}

\newpage
\begin{APPENDICES}
\newpage
\setcounter{page}{1}
\setcounter{section}{0}

\begin{center}
{\large Online Appendix \\
\vspace{.1in}
\PaperTitle
}
\end{center}

\renewcommand{\thesection}{O\Alph{section}}
\renewcommand{\theequation}{OA-\arabic{equation}}
\renewcommand{\thefigure}{OA-\arabic{figure}}
\renewcommand{\thetable}{OA-\arabic{table}}
\renewcommand{\thethm}{OA-\arabic{thm}}
\renewcommand{\thelemma}{OA-\arabic{lemma}}
\renewcommand{\theprop}{OA-\arabic{prop}}
\renewcommand{\theexap}{OA-\arabic{exap}}
\renewcommand{\thecor}{OA-\arabic{cor}}
\renewcommand{\theremark}{OA-\arabic{remark}}
\renewcommand{\thedefi}{OA-\arabic{defi}}
\renewcommand{\theas}{OA-\arabic{as}}
\setcounter{equation}{0}
\setcounter{thm}{0}
\setcounter{lemma}{0}
\setcounter{prop}{0}
\setcounter{cor}{0}
\setcounter{figure}{0}
\setcounter{table}{0}
\setcounter{exap}{0}
\setcounter{remark}{0}
\setcounter{defi}{0}
\setcounter{as}{0}

This note is the online supplement to ``\PaperTitle.'' Appendix~\ref{app:proofs} proves the main-text results, and Appendix~\ref{app:caldetail} collects the additional calibration detail.

\section{Technical Proofs of Main Results}\label{app:proofs}
This appendix collects the proofs of the results stated in the main text.

\subsubsection*{Proof of Lemma~\ref{prop-revenue-rent-eq}.}
Hold the allocation at (\ref{eq:efficient-allocation}). The map $\mathbf{x}\mapsto\mathbf{u}$ with $u_j(\mathbf{k}):=T_j\,p_j(\mathbf{k})-x_j(\mathbf{k})$ is a bijection between payment rules and utility schedules that carries $\mathcal{A}_{EF}$ onto $\mathcal{U}_{EF}$: (\ref{eq:robust-ir}) is exactly $\mathbf{u}\geq0$, and (\ref{eq:biased-ic}) is already an inequality in $\mathbf{u}$. Substituting $x_j=T_jp_j-u_j$ turns the objective of (\ref{eq:rev-max}) into $\overline T-m\sum_jq_j\overline U_j$, whose first term is fixed by the allocation, so the two programs share their optimizers, paired by $x_j^{\ast}=T_jp_j-u_j^{\ast}$, and their values satisfy $U_0^{\ast}+U_{\min}^{\ast}=\overline T$. \qed

\subsubsection*{Proof of Theorem~\ref{thm:compat}.}
Throughout, $\hat U_j$ and $\overline{U}_j$ are the aggregates of a rent schedule $\mathbf{u}\geq0$, as in (\ref{eq:two-aggregates}) and (\ref{eq:rent-min}). As $\mathbf{q}$ and $\boldsymbol\pi$ have positive entries, $\Phi$ and every $\phi_j$ are positive on $\mathcal{K}$, so the ratios in (\ref{eq:leak-ratio}) are well defined and $\varrho_{jh}>0$.

\emph{Part {\rm (i)}: the compression boundary.} Fix a report $h$ and a state $\mathbf{k}^{\circ}\in\mathcal{K}$. Let $u_h\geq0$ be arbitrary with perceived rent $\hat U_h=\sum_{\mathbf{k}}\phi_h(\mathbf{k})u_h(\mathbf{k})$. If $\hat U_h=0$ then $u_h\equiv0$, as $\phi_h>0$, and there is nothing to prove. Suppose $\hat U_h>0$ and set $\omega_h(\mathbf{k}):=\phi_h(\mathbf{k})u_h(\mathbf{k})/\hat U_h$, a probability distribution on $\mathcal{K}$. Then
\[
\frac{1}{\hat U_h}\sum_{\mathbf{k}}\Phi(\mathbf{k})u_h(\mathbf{k})=\sum_{\mathbf{k}}\frac{\omega_h(\mathbf{k})}{\Lambda_h(\mathbf{k})},
\qquad
\frac{1}{\hat U_h}\sum_{\mathbf{k}}\phi_j(\mathbf{k})u_h(\mathbf{k})=\sum_{\mathbf{k}}\omega_h(\mathbf{k})\,\varrho_{jh}(\mathbf{k}),\quad j\neq h,
\]
so the objective cost of report $h$'s rent and every mimic's leakage from it, each per unit of $\hat U_h$, are the $\omega_h$-averages of the corresponding coordinates of the cost--leakage signature. The compressed schedule $u_h^{\circ}(\mathbf{k}):=\hat U_h\mathbf{1}(\mathbf{k}=\mathbf{k}^{\circ})/\phi_h(\mathbf{k}^{\circ})$ has the same perceived rent and puts $\omega_h$ at the point mass on $\mathbf{k}^{\circ}$, so its two per-unit values are $1/\Lambda_h(\mathbf{k}^{\circ})$ and $\varrho_{jh}(\mathbf{k}^{\circ})$.

If $\mathbf{k}^{\circ}$ satisfies (\ref{eq:compatibility}), each of these $n$ values is the minimum of the corresponding coordinate over $\mathcal{K}$ and therefore does not exceed its average under any $\omega_h$: replacement weakly lowers the objective cost and every mimic's leakage, and since the deviation term $(T_j-T_h)P_{jh}$ in (\ref{eq:biased-ic}) does not involve $u_h$, it weakly lowers every other type's payoff from reporting $T_h$. Conversely, suppose $\mathbf{k}^{\circ}$ permits universal incentive-safe compression, and apply the operation to the schedule carrying one unit of perceived rent at a single state $\mathbf{k}$, namely $u_h(\mathbf{k}')=\mathbf{1}(\mathbf{k}'=\mathbf{k})/\phi_h(\mathbf{k})$. Its two per-unit values are $1/\Lambda_h(\mathbf{k})$ and $\varrho_{jh}(\mathbf{k})$, so weak reduction of the objective cost gives $\Lambda_h(\mathbf{k}^{\circ})\geq\Lambda_h(\mathbf{k})$ and weak reduction of each mimic's payoff gives $\varrho_{jh}(\mathbf{k}^{\circ})\leq\varrho_{jh}(\mathbf{k})$. As $\mathbf{k}\in\mathcal{K}$ was arbitrary, $\mathbf{k}^{\circ}$ lies in the intersection (\ref{eq:compatibility}). Conditions~{\rm (C1)--(C2)} assert that each of these $n$ extrema is attained only at $\mathbf{k}_h^{\ast}$, so the intersection is then $\{\mathbf{k}_h^{\ast}\}$.

\emph{Part {\rm (ii)}--{\rm (iii)}.} Assume {\rm (C1)--(C2)} for the remainder. Note first that $\hat{\boldsymbol U}$ is feasible for (\ref{eq:rlp}) if and only if $\mathcal{R}(\hat{\boldsymbol U})\leq\hat{\boldsymbol U}$, with $\mathcal{R}$ as in (\ref{eq:fixed}).

\emph{Step 1 ($\bar\varrho<1$).} Fix $j\neq h$. By (\ref{eq:leak-ratio}), $\phi_j(\mathbf{k})\geq\varrho_{jh}\phi_h(\mathbf{k})$ for every $\mathbf{k}\in\mathcal{K}$; summing over $\mathcal{K}$, where both weights are probability distributions, gives $1\geq\varrho_{jh}$. If $\varrho_{jh}=1$, then $\sum_{\mathbf{k}}[\phi_j(\mathbf{k})-\phi_h(\mathbf{k})]=0$ with every summand nonnegative, so $\phi_j\equiv\phi_h$ and $\varrho_{jh}(\cdot)\equiv1$ attains its minimum at every configuration; since $m\geq2$ and $n\geq2$ give $|\mathcal{K}|\geq2$, this contradicts the uniqueness clause of {\rm (C2)}. Hence $\varrho_{jh}<1$ for every $j\neq h$, and $\bar\varrho<1$.

\emph{Step 2 (a smallest feasible vector).} Since $|\max\{0,a\}-\max\{0,b\}|\leq|a-b|$ and $|\max_ra_r-\max_rb_r|\leq\max_r|a_r-b_r|$, Step 1 gives $\|\mathcal{R}(\hat{\boldsymbol U})-\mathcal{R}(\hat{\boldsymbol U}')\|_\infty\leq\bar\varrho\,\|\hat{\boldsymbol U}-\hat{\boldsymbol U}'\|_\infty$: the map $\mathcal{R}$ is a contraction on $\mathbb{R}^{n}$ and has a unique fixed point $\hat{\boldsymbol U}^{\ast}$, which satisfies $\mathcal{R}(\hat{\boldsymbol U}^{\ast})\leq\hat{\boldsymbol U}^{\ast}$ and is therefore feasible for (\ref{eq:rlp}), nonnegativity included since $\mathcal{R}$ takes values in $\mathbb{R}^n_{\geq0}$. It is also the smallest feasible vector: $\mathcal{R}$ is monotone, each $\varrho_{jr}$ being nonnegative, and $\mathcal{R}(\hat{\boldsymbol y})\leq\hat{\boldsymbol y}$ for any feasible $\hat{\boldsymbol y}$, so the iterates from $\hat{\boldsymbol U}^{(0)}=\mathbf{0}\leq\hat{\boldsymbol y}$ satisfy $\mathbf{0}\leq\hat{\boldsymbol U}^{(t)}\leq\hat{\boldsymbol y}$ for every $t$, increase in $t$, and converge to $\hat{\boldsymbol U}^{\ast}$. Every objective coefficient $q_j/\Lambda_j$ being strictly positive, any feasible $\hat{\boldsymbol y}\neq\hat{\boldsymbol U}^{\ast}$ has strictly larger objective; $\hat{\boldsymbol U}^{\ast}$ is therefore the componentwise-minimal---and hence the unique---solution of (\ref{eq:rlp}). This proves {\rm (iii)}.

\emph{Step 3 (concentrated schedules).} For $\hat{\boldsymbol U}\geq0$ let $u_j(\mathbf{k}):=\hat U_j\mathbf{1}(\mathbf{k}=\mathbf{k}_j^{\ast})/\phi_j(\mathbf{k}_j^{\ast})$, a schedule whose perceived rent is indeed $\hat U_j$. By {\rm (C2)} the ratio $\varrho_{jh}(\cdot)$ equals $\varrho_{jh}$ at $\mathbf{k}_h^{\ast}$, so for $h\neq j$ the perceived leakage from a report of $T_h$ is $\sum_{\mathbf{k}}\phi_j(\mathbf{k})u_h(\mathbf{k})=\varrho_{jh}(\mathbf{k}_h^{\ast})\hat U_h=\varrho_{jh}\hat U_h$, while $\mathbf{k}_j^{\ast}$ being the peak of $\Lambda_j(\cdot)$ by {\rm (C1)} makes (\ref{eq:exchange}) an equality, $\overline{U}_j=\hat U_j/\Lambda_j$. Such a schedule therefore satisfies (\ref{eq:robust-ir}) and (\ref{eq:biased-ic}) exactly when $\hat{\boldsymbol U}$ is feasible for (\ref{eq:rlp}), and its objective in (\ref{eq:rent-min}) is then $m\sum_j(q_j/\Lambda_j)\hat U_j$. In particular the schedule concentrated at $\hat{\boldsymbol U}^{\ast}$ is feasible for (\ref{eq:rent-min}).

\emph{Step 4 (lower bound).} Let $\mathbf{u}$ be feasible for (\ref{eq:rent-min}); its perceived rents satisfy $\hat{\boldsymbol U}\geq0$. Recording the gap in (\ref{eq:exchange}) for later use,
\begin{equation}\label{eq:compat-cost}
\overline{U}_j-\frac{\hat U_j}{\Lambda_j}=\sum_{\mathbf{k}}\Phi(\mathbf{k})u_j(\mathbf{k})\Bigl(1-\frac{\Lambda_j(\mathbf{k})}{\Lambda_j}\Bigr)\ \geq\ 0 ,
\end{equation}
a sum of nonnegative terms because $\Lambda_j(\mathbf{k})\leq\Lambda_j$ by (\ref{equ:peak-ratio}). By {\rm (C2)}, $\sum_{\mathbf{k}}\phi_j(\mathbf{k})u_h(\mathbf{k})\geq\varrho_{jh}\hat U_h$, so (\ref{eq:biased-ic}) gives $\hat U_j\geq\varrho_{jh}\hat U_h+(T_j-T_h)P_{jh}$ for every $h\neq j$: the induced $\hat{\boldsymbol U}$ is feasible for (\ref{eq:rlp}), hence $\hat{\boldsymbol U}\geq\hat{\boldsymbol U}^{\ast}$ by Step 2. Therefore $m\sum_jq_j\overline{U}_j\geq m\sum_j(q_j/\Lambda_j)\hat U_j\geq m\sum_j(q_j/\Lambda_j)\hat U_j^{\ast}$, and by Step 3 the right-hand side is the value attained by the concentrated schedule (\ref{eq:compat-support}) at $\hat{\boldsymbol U}^{\ast}$, which is therefore optimal for (\ref{eq:rent-min}); in particular the two programs share their optimal value.

\emph{Step 5 (uniqueness).} Let $\mathbf{u}$ be optimal for (\ref{eq:rent-min}). Both inequalities of Step 4 are then equalities, and since $q_j>0$ and $q_j/\Lambda_j>0$ they hold coordinate by coordinate. Equality in (\ref{eq:compat-cost}), with $u_j(\cdot)\geq0$ and the uniqueness in {\rm (C1)}, forces $u_j(\mathbf{k})=0$ for every $\mathbf{k}\neq\mathbf{k}_j^{\ast}$, so that $\hat U_j=\phi_j(\mathbf{k}_j^{\ast})u_j(\mathbf{k}_j^{\ast})$; equality in the second gives $\hat U_j=\hat U_j^{\ast}$, whence $u_j(\mathbf{k}_j^{\ast})=\hat U_j^{\ast}/\phi_j(\mathbf{k}_j^{\ast})$. This is (\ref{eq:compat-support}) and completes {\rm (ii)}. \qed

\subsubsection*{Proof of Theorem~\ref{thm:approx}.}
Write $\mathcal{R}$ and $\mathcal{R}^{\rm D}$ for the map \eqref{eq:fixed} with coefficient matrices $(\varrho_{jh})$ and $(\varrho^{\rm D}_{jh})$ respectively. Both matrices are nonnegative, and $\varrho_{jh}\leq\varrho^{\rm D}_{jh}$ because $\varrho_{jh}$ minimizes $\varrho_{jh}(\cdot)$ over $\mathcal{K}$, so $\bar\varrho\leq\bar\varrho^{\rm D}<1$. Step~2 of the proof of Theorem~\ref{thm:compat}, which uses neither {\rm (C1)} nor {\rm (C2)}, applies to either matrix: each map is a contraction of the stated modulus, and its fixed point is the componentwise-minimal, hence unique, solution of the corresponding rent-level program. Denote the two fixed points $\hat{\boldsymbol U}^{\ast}$ and $\hat{\boldsymbol U}^{\rm D}$.

\emph{Step 1 (the restricted program).} Fix $\hat{\boldsymbol U}\geq0$ and let $u_j(\mathbf{k}):=\hat U_j\mathbf{1}(\mathbf{k}=\mathbf{k}_j^{\ast})/\phi_j(\mathbf{k}_j^{\ast})$ be the schedule that places each report's rent at its peak diagonal. Its perceived rent is $\hat U_j$, its objective cost is $\overline{U}_j=\Phi(\mathbf{k}_j^{\ast})u_j(\mathbf{k}_j^{\ast})=\hat U_j/\Lambda_j(\mathbf{k}_j^{\ast})=\hat U_j/\Lambda_j$ by \eqref{equ:peak-ratio}, and the perceived leakage from a report of $T_h$ is $\phi_j(\mathbf{k}_h^{\ast})u_h(\mathbf{k}_h^{\ast})=\varrho^{\rm D}_{jh}\hat U_h$, neither requiring {\rm (C1)--(C2)}. Consequently \eqref{eq:robust-ir} and \eqref{eq:biased-ic} hold for this schedule exactly when $\hat{\boldsymbol U}$ is feasible for \eqref{eq:rlp} with $\varrho^{\rm D}$ in place of $\varrho$, and its objective in \eqref{eq:rent-min} is $m\sum_h(q_h/\Lambda_h)\hat U_h$. Minimizing over this class therefore gives the value $U_{\min}^{\rm D}$ of Section~\ref{sec:approx}. Since every such schedule is feasible for \eqref{eq:rent-min}, $U_{\min}^{\ast}\leq U_{\min}^{\rm D}$.

\emph{Step 2 (the lower bound).} Let $\mathbf{u}\geq0$ be feasible for \eqref{eq:rent-min} and $\hat{\boldsymbol U}$ its perceived rents. Then $\overline{U}_j\geq\hat U_j/\Lambda_j$ by \eqref{eq:exchange} and $\sum_{\mathbf{k}}\phi_j(\mathbf{k})u_h(\mathbf{k})\geq\varrho_{jh}\hat U_h$ by \eqref{eq:leak-ratio}, so \eqref{eq:biased-ic} forces $\hat{\boldsymbol U}$ to be feasible for \eqref{eq:rlp} and hence $\hat{\boldsymbol U}\geq\hat{\boldsymbol U}^{\ast}$. Therefore $m\sum_jq_j\overline{U}_j\geq m\sum_j(q_j/\Lambda_j)\hat U_j^{\ast}=U_{\min}^{\rm L}$, which is the left-hand inequality of $U_{\min}^{\rm L}\ \leq\ U_{\min}^{\ast}\ \leq\ U_{\min}^{\rm D}$. This proves {\rm (i)}.

\emph{Step 3 (comparison of the fixed points).} Both maps are monotone in $\hat{\boldsymbol U}$ and $\mathcal{R}(\hat{\boldsymbol U})\leq\mathcal{R}^{\rm D}(\hat{\boldsymbol U})$ for $\hat{\boldsymbol U}\geq0$, because $\varrho\leq\varrho^{\rm D}$ entrywise. Starting both iterations at $\mathbf 0$ and inducting on $t$ gives $\mathcal{R}^{t}(\mathbf 0)\leq(\mathcal{R}^{\rm D})^{t}(\mathbf 0)$ for every $t$, and both sides are nonnegative; letting $t\to\infty$ yields $0\leq\hat{\boldsymbol U}^{\ast}\leq\hat{\boldsymbol U}^{\rm D}$.

\emph{Step 4 (the loss bound).} For $\hat{\boldsymbol U}\geq0$ and each $j$, the two maps differ only in their coefficients, so
\[
\bigl|\mathcal{R}^{\rm D}(\hat{\boldsymbol U})_j-\mathcal{R}(\hat{\boldsymbol U})_j\bigr|
\ \leq\ \max_{r\neq j}\bigl(\varrho^{\rm D}_{jr}-\varrho_{jr}\bigr)\hat U_r
\ \leq\ \Delta\varrho\,\|\hat{\boldsymbol U}\|_\infty ,
\]
using $|\max\{0,a\}-\max\{0,b\}|\leq|a-b|$ and $|\max_ra_r-\max_rb_r|\leq\max_r|a_r-b_r|$ twice. Both vectors being fixed points, the triangle inequality and the contraction property of $\mathcal{R}^{\rm D}$ give $\|\hat{\boldsymbol U}^{\rm D}-\hat{\boldsymbol U}^{\ast}\|_\infty\leq\|\mathcal{R}^{\rm D}(\hat{\boldsymbol U}^{\rm D})-\mathcal{R}^{\rm D}(\hat{\boldsymbol U}^{\ast})\|_\infty+\|\mathcal{R}^{\rm D}(\hat{\boldsymbol U}^{\ast})-\mathcal{R}(\hat{\boldsymbol U}^{\ast})\|_\infty\leq\bar\varrho^{\rm D}\|\hat{\boldsymbol U}^{\rm D}-\hat{\boldsymbol U}^{\ast}\|_\infty+\Delta\varrho\|\hat{\boldsymbol U}^{\ast}\|_\infty$, and rearranging gives $\|\hat{\boldsymbol U}^{\rm D}-\hat{\boldsymbol U}^{\ast}\|_\infty\leq\Delta\varrho\,\|\hat{\boldsymbol U}^{\ast}\|_\infty/(1-\bar\varrho^{\rm D})$. Combining the two ends of $U_{\min}^{\rm L}\ \leq\ U_{\min}^{\ast}\ \leq\ U_{\min}^{\rm D}$ with Step~3,
\[
0\ \leq\ U_{\min}^{\rm D}-U_{\min}^{\ast}
\ \leq\ U_{\min}^{\rm D}-U_{\min}^{\rm L}
\ =\ m\sum_{h}\frac{q_h}{\Lambda_h}\bigl(\hat U_h^{\rm D}-\hat U_h^{\ast}\bigr)
\ \leq\ m\Bigl(\sum_{h}\frac{q_h}{\Lambda_h}\Bigr)\|\hat{\boldsymbol U}^{\rm D}-\hat{\boldsymbol U}^{\ast}\|_\infty ,
\]
which is the first display of \eqref{eq:approx}. For the auxiliary estimate, write $d_+:=\max_{j<h}(T_j-T_h)P_{jh}$, which is positive because $T_j>T_h$ for $j<h$ and $P_{jh}>0$ under full support. A deviation constant $(T_j-T_r)P_{jr}$ is positive only when $r>j$, so every one of them is at most $d_+$. Each coordinate of $\hat{\boldsymbol U}^{\ast}=\mathcal{R}(\hat{\boldsymbol U}^{\ast})$ therefore satisfies $\hat U_j^{\ast}\leq\max\{0,\bar\varrho\|\hat{\boldsymbol U}^{\ast}\|_\infty+d_+\}=\bar\varrho\|\hat{\boldsymbol U}^{\ast}\|_\infty+d_+$, and taking the maximum over $j$ and rearranging yields the second display of \eqref{eq:approx}.

\emph{Step 5 (the compatible case).} Under {\rm (C1)--(C2)} the peak diagonal $\mathbf{k}_h^{\ast}$ minimizes $\varrho_{jh}(\cdot)$ for every mimic, so $\varrho^{\rm D}_{jh}=\varrho_{jh}$, $\Delta\varrho=0$, $\bar\varrho^{\rm D}=\bar\varrho$, and the two maps coincide. Then $\hat{\boldsymbol U}^{\rm D}=\hat{\boldsymbol U}^{\ast}$, so $U_{\min}^{\rm L}=U_{\min}^{\rm D}$ and the bracket $U_{\min}^{\rm L}\ \leq\ U_{\min}^{\ast}\ \leq\ U_{\min}^{\rm D}$ closes at $U_{\min}^{\ast}$, which is the conclusion of Theorem~\ref{thm:compat}. \qed

\subsubsection*{Proof of Proposition~\ref{prop:dual}.}
Throughout write $d_{jh}:=(T_j-T_h)P_{jh}$ for the deviation constants of \eqref{eq:biased-ic}.

\emph{Feasibility and attainment.} If $\bar\varrho^{\rm D}<1$, take the peak-concentrated schedule of Step~1 of the proof of Theorem~\ref{thm:approx} with the constant perceived rents $\hat U_j\equiv t$, where $t:=\max\{0,\max_{j\neq h}d_{jh}/(1-\varrho^{\rm D}_{jh})\}$; every denominator is positive and each constraint $t\geq\varrho^{\rm D}_{jh}t+d_{jh}$ holds by construction, so \eqref{eq:rent-min} is feasible. Given feasibility, the linear objective is bounded below by zero on $\mathbf{u}\geq0$, so the program attains its minimum.

\emph{Part {\rm (i)}.} Attach $\eta_{jh}\geq0$ to the constraint of \eqref{eq:biased-ic} deterring $T_j$ from reporting $T_h$ and form
\[
L(\mathbf{u},\boldsymbol\eta)
=m\sum_{h}q_h\sum_{\mathbf{k}}\Phi(\mathbf{k})u_h(\mathbf{k})
-\sum_{j\neq h}\eta_{jh}\Bigl[\sum_{\mathbf{k}}\phi_j(\mathbf{k})\bigl(u_j(\mathbf{k})-u_h(\mathbf{k})\bigr)-d_{jh}\Bigr].
\]
Collecting the terms in $u_h(\mathbf{k})$---from the objective, from report $h$'s own constraints and from those of the types that may imitate it---its coefficient is
\begin{equation}\label{eq:dualcoeff}
mq_h\Phi(\mathbf{k})-\phi_h(\mathbf{k})\sum_{r\neq h}\eta_{hr}+\sum_{j\neq h}\eta_{jh}\phi_j(\mathbf{k}).
\end{equation}
The infimum of $L(\cdot,\boldsymbol\eta)$ over $\mathbf{u}\geq0$ is $-\infty$ unless every coefficient in \eqref{eq:dualcoeff} is nonnegative, in which case it equals $\sum_{j\neq h}\eta_{jh}d_{jh}$ and is attained at $\mathbf{u}=\mathbf{0}$. Dividing \eqref{eq:dualcoeff} by $\phi_h(\mathbf{k})>0$ and using \eqref{eq:weights} and \eqref{eq:leak-ratio} turns the sign restriction into \eqref{eq:dual}. The dual program is therefore to maximize $\sum_{j\neq h}\eta_{jh}d_{jh}$ over $\boldsymbol\eta\geq0$ subject to \eqref{eq:dual}. The primal is a feasible linear program with an attained finite optimum, so finite-dimensional linear-programming duality gives equality of the two values, and a feasible pair $(\mathbf{u},\boldsymbol\eta)$ is optimal for both programs if and only if complementary slackness holds in both directions: equality in \eqref{eq:dual} at every $(h,\mathbf{k})$ with $u_h(\mathbf{k})>0$, and equality in \eqref{eq:biased-ic} for every $(j,h)$ with $\eta_{jh}>0$. This is the stated certificate, and evaluating the dual objective at such an $\boldsymbol\eta$ gives $U_{\min}^{\ast}$.

\emph{Part {\rm (ii)}.} If $\hat U_h=0$, full support and $u_h\geq0$ force $u_h\equiv0$ and there is nothing to prove, so fix an optimal $\mathbf{u}$ and a report $h$ with $\hat U_h>0$; hold every other report's schedule and the scalar $\hat U_h$ fixed, and re-optimize $u_h$ alone. In the variables $y_{\mathbf{k}}:=\phi_h(\mathbf{k})u_h(\mathbf{k})\geq0$ the objective contributed by report $h$ is $\sum_{\mathbf{k}}y_{\mathbf{k}}/\Lambda_h(\mathbf{k})$, and $u_h$ enters the constraint system in exactly two ways. The $n-1$ \emph{incoming} constraints, which deter each $j\neq h$ from reporting $h$, read $\sum_{\mathbf{k}}\varrho_{jh}(\mathbf{k})y_{\mathbf{k}}\leq\hat U_j-d_{jh}$; the $n-1$ \emph{outgoing} constraints, which deter $h$ from reporting some $r$, involve $u_h$ only through the fixed aggregate $\hat U_h$; all remaining constraints do not involve $u_h$ at all. Adding the definition $\sum_{\mathbf{k}}y_{\mathbf{k}}=\hat U_h$, the local feasible set is the intersection of a scaled simplex with $n-1$ half-spaces: nonempty, since the current $y$ belongs to it, and compact, so the linear objective attains its minimum at an extreme point. Writing the $n-1$ inequalities with slacks gives $n$ equalities, so a basic solution has at most $n$ strictly positive variables and hence at most $n$ positive $y_{\mathbf{k}}$. Substituting this solution preserves every constraint and cannot raise the objective; since $\mathbf{u}$ was optimal, the cost is unchanged and the new schedule is again optimal. Applying the substitution once to each report leaves earlier reports' schedules untouched, so the final schedule is optimal with at most $n$ states supporting each report's rent.

\emph{Part {\rm (iii)}.} If $\mathbf{k}^{\circ}$ satisfies \eqref{eq:compatibility}, then it maximizes $\Lambda_h(\cdot)$ and minimizes every $\varrho_{jh}(\cdot)$, so each of the $n$ terms of \eqref{eq:adjprice} is minimized at $\mathbf{k}^{\circ}$ for any $\boldsymbol\eta\geq0$, and so is their sum. Conversely, suppose $\mathbf{k}^{\circ}$ minimizes $c_h(\cdot;\boldsymbol\eta)$ over $\mathcal{K}$ for every $\boldsymbol\eta\geq0$. Taking $\boldsymbol\eta=\mathbf{0}$ gives $mq_h/\Lambda_h(\mathbf{k}^{\circ})\leq mq_h/\Lambda_h(\mathbf{k})$ for all $\mathbf{k}$, that is, $\mathbf{k}^{\circ}\in\arg\max_{\mathbf{k}}\Lambda_h(\mathbf{k})$. Now fix $j_0\neq h$, set $\eta_{j_0h}=t>0$ and all other multipliers to zero, so that $mq_h/\Lambda_h(\mathbf{k}^{\circ})+t\varrho_{j_0h}(\mathbf{k}^{\circ})\leq mq_h/\Lambda_h(\mathbf{k})+t\varrho_{j_0h}(\mathbf{k})$ for every $\mathbf{k}$. Dividing by $t$ and letting $t\to\infty$ gives $\varrho_{j_0h}(\mathbf{k}^{\circ})\leq\varrho_{j_0h}(\mathbf{k})$. As $j_0$ and $\mathbf{k}$ were arbitrary, $\mathbf{k}^{\circ}$ lies in the intersection \eqref{eq:compatibility}. \qed

\noindent\textbf{Verification of Example~\ref{exap:leakage}.} With $m=2$ a configuration is the rival's type, so $\Phi(\mathbf{D}_r)=q_r$ and $\phi_j(\mathbf{D}_r)=\pi_{jr}$. The ratios $\Lambda_j(\mathbf{D}_r)=\pi_{jr}/q_r$ are $(17/10,3/4,1/10)$, $(3/8,3/2,5/4)$ and $(1/4,1/4,4)$ for $j=1,2,3$, so each type's own diagonal is its unique peak and {\rm (C1)} holds. The leakage ratios $\varrho_{12}(\mathbf{D}_r)=\pi_{1r}/\pi_{2r}$ are $68/15$, $1/2$ and $2/25$, minimized at $\mathbf{D}_3$ rather than $\mathbf{D}_2$, so {\rm (C2)} fails for report $T_2$ and, by Theorem~\ref{thm:compat}{\rm (i)}, no state permits universal incentive-safe compression for it. With $P_{jh}=\sum_r\pi_{jr}p_h(\mathbf{D}_r)$ under \eqref{eq:efficient-allocation}, the deviation constants are $d_{12}=1.70$, $d_{13}=0.20$, $d_{23}=1.25$, $d_{21}=-9.25$, $d_{31}=-19.00$ and $d_{32}=-8.50$. The schedule $u_1^{\ast}(\mathbf{D}_1)=45/17$, $u_2^{\ast}(\mathbf{D}_3)=5$, zero elsewhere, induces perceived rents $(9/5,5/4,0)$; the constraints $1\to2$ and $2\to3$ bind, the other four have slacks $8/5$, $687/68$, $637/34$ and $9/2$, and the objective cost is $2[q_1^2u_1^{\ast}(\mathbf{D}_1)+q_2q_3u_2^{\ast}(\mathbf{D}_3)]=28/17$. With $\eta_{12}=8/17$, $\eta_{23}=288/425$ and all other multipliers zero, the coefficients \eqref{eq:dualcoeff}, with rows indexed by reports and columns by the rival's type, are
\[
\begin{pmatrix}
0&76/425&64/425\\
1144/2125&116/2125&0\\
556/2125&1204/2125&106/425
\end{pmatrix}.
\]
They are nonnegative, so \eqref{eq:dual} holds; they vanish exactly where rent is positive, and the two positively priced constraints bind, so complementary slackness holds in both directions and the dual value $(8/17)(1.70)+(288/425)(1.25)=28/17$ confirms optimality by Proposition~\ref{prop:dual}{\rm (i)}. Confining each report to its own peak, $u_j(\mathbf{D}_r)=z_j\mathbf{1}(r=j)$, turns \eqref{eq:biased-ic} into $\pi_{jj}z_j\geq\pi_{jh}z_h+d_{jh}$, so every such schedule has $z_2\geq d_{23}/\pi_{22}=25/12$ and $z_1\geq(\pi_{12}z_2+d_{12})/\pi_{11}\geq465/136$. These bounds with $z_3=0$ satisfy all six constraints, so the restricted optimum costs $2(q_1^2z_1+q_2^2z_2)=449/255$, exceeding $28/17$ by $29/255$.

\subsubsection*{Proof of Corollary~\ref{cor:reverse}.}
With $n=2$, \eqref{eq:efficient-allocation} reads $p_1(\mathbf{k})=1/(k_1+1)$ and $p_2(\mathbf{k})=\mathbf{1}(\mathbf{k}=\mathbf{D}_2)/m$. By \eqref{eq:weights} and \eqref{eq:leak-ratio}, with $k_1+k_2=m-1$, $\Lambda_1(\mathbf{k})=(\pi_{11}/q_1)^{k_1}(\pi_{12}/q_2)^{k_2}$ and $\varrho_{12}(\mathbf{k})=(\pi_{11}/\pi_{21})^{k_1}(\pi_{12}/\pi_{22})^{k_2}$, and symmetrically for $\Lambda_2(\cdot)$ and $\varrho_{21}(\cdot)$. The hypothesis gives $\pi_{11}<q_1<\pi_{21}$ and $\pi_{22}<q_2<\pi_{12}$, so $\Lambda_1(\cdot)$ is uniquely maximized and $\varrho_{21}(\cdot)$ uniquely minimized at $\mathbf{D}_2$, while $\Lambda_2(\cdot)$ is uniquely maximized and $\varrho_{12}(\cdot)$ uniquely minimized at $\mathbf{D}_1$. Hence {\rm (C1)--(C2)} hold with $\mathbf{k}_1^{\ast}=\mathbf{D}_2$, $\mathbf{k}_2^{\ast}=\mathbf{D}_1$, $\varrho_{12}=(\pi_{11}/\pi_{21})^{m-1}$ and $\varrho_{21}=(\pi_{22}/\pi_{12})^{m-1}$, and Theorem~\ref{thm:compat} applies.

In \eqref{eq:rlp}, $P_{12}=\phi_1(\mathbf{D}_2)/m=\pi_{12}^{\,m-1}/m$ and $P_{21}=\sum_{\mathbf{k}}\phi_2(\mathbf{k})/(k_1+1)\geq\phi_2(\mathbf{D}_2)=\pi_{22}^{\,m-1}$. The vector $\hat U_1^{\ast}=(T_1-T_2)\pi_{12}^{\,m-1}/m$, $\hat U_2^{\ast}=0$ meets $\hat U_1\geq\varrho_{12}\hat U_2+(T_1-T_2)P_{12}$ with equality and $\hat U_2\geq\varrho_{21}\hat U_1-(T_1-T_2)P_{21}$ strictly, since $\varrho_{21}\hat U_1^{\ast}=(T_1-T_2)\pi_{22}^{\,m-1}/m<(T_1-T_2)P_{21}$. Every feasible vector has $\hat U_2\geq0$ and hence $\hat U_1\geq\hat U_1^{\ast}$, so $\hat{\boldsymbol U}^{\ast}$ is the componentwise-minimal solution, and \eqref{eq:compat-support} gives the unique optimal schedule $u_1^{\ast}(\mathbf{k})=(T_1-T_2)\mathbf{1}(\mathbf{k}=\mathbf{D}_2)/m$, $u_2^{\ast}\equiv0$, whose payments Lemma~\ref{prop-revenue-rent-eq} pins down. Its objective rent is $mq_1\Phi(\mathbf{D}_2)u_1^{\ast}(\mathbf{D}_2)=q_1q_2^{\,m-1}(T_1-T_2)$. At $\pi_{jh}=q_h$, Theorem~\ref{thm:diagonal}{\rm (iii)} with $n=2$ gives the rational rents $\overline U_1^{S}=(T_1-T_2)\bar P_2=(T_1-T_2)q_2^{\,m-1}/m$ and $\overline U_2^{S}=0$, so the rational minimum is the same and $E(\boldsymbol\pi)=0$. Finally, $x_2^{\ast}=T_2p_2\geq0$, $x_1^{\ast}(\mathbf{k})=T_1p_1(\mathbf{k})\geq0$ for $\mathbf{k}\neq\mathbf{D}_2$, and $x_1^{\ast}(\mathbf{D}_2)=T_1-(T_1-T_2)/m>0$ because a high report wins for certain at $\mathbf{D}_2$. \qed

\subsubsection*{Proof of Theorem~\ref{thm:diagonal}.}
Fix $\alpha\in(0,1)$ and write $\hat Q_h:=\alpha+(1-\alpha)Q_h$, so that $\hat Q_1=1$, together with
\[
\widetilde W_h:=\frac1{q_h}\int_{Q_{h+1}}^{Q_h}\Bigl(s+\frac{\alpha}{1-\alpha}\Bigr)^{m-1}\mathrm{d}s ,
\]
the counterpart of $\bar P_h$ after the projection shift induced by an upward report.

\emph{Step 1 ({\rm (C1)--(C2)}).} Section~\ref{sec:diagonal} reads the two conditions off the per-rival ratios \eqref{eq:per-rival-ratio}, with the constants \eqref{eq:proj-peak} and $\bar\varrho<1$, so Theorem~\ref{thm:compat} applies in full and only the levels $\hat{\boldsymbol U}^{\ast}$ remain to be identified.

\emph{Step 2 (deviation-winning probabilities).} A type-$T_j$ buyer treats her rivals as $m-1$ independent draws from $\boldsymbol\pi_j$. Write $\chi:=\pi_{jh}$ for the mass she assigns a rival at $T_h$ and $\xi:=\sum_{r>h}\pi_{jr}$ for the mass strictly below it. She assigns probability $\binom{m-1}{b}\chi^b\xi^{\,m-1-b}$ to the event that no rival reports above $T_h$ while exactly $b$ report $T_h$, an event in which \eqref{eq:efficient-allocation} awards a report of $T_h$ the object with probability $1/(b+1)$. Using $\binom{m-1}{b}/(b+1)=\binom{m}{b+1}/m$,
\begin{equation}\label{eq:Pinterval}
P_{jh}=\sum_{b=0}^{m-1}\binom{m-1}{b}\chi^b\xi^{\,m-1-b}\frac1{b+1}
=\frac{(\chi+\xi)^m-\xi^m}{m\,\chi},
\end{equation}
which is the average of $t^{\,m-1}$ over $[\xi,\chi+\xi]$. The endpoints depend on where the atom of $\boldsymbol\pi_j$ at $T_j$ sits. For $h>j$ the atom lies above $T_h$ and so enters neither $\chi$ nor $\xi$: $\chi=(1-\alpha)q_h$, $\xi=(1-\alpha)Q_{h+1}$ and $\chi+\xi=(1-\alpha)Q_h$. For $h<j$ the atom lies below $T_h$ and is part of $\xi$: $\chi=(1-\alpha)q_h$ and $\xi=A_j+(1-\alpha)(Q_{h+1}-q_j)=\hat Q_{h+1}$, whence $\chi+\xi=\hat Q_h$, endpoints free of $j$. For $h=j$, $\chi=A_j$ and $\xi=(1-\alpha)Q_{j+1}$, whence $\chi+\xi=\hat Q_j$. Substituting $t=(1-\alpha)s$ in the first case and $t=\alpha+(1-\alpha)s$ in the second turns each average into $(1-\alpha)^{m-1}$ times an average over $[Q_{h+1},Q_h]$, of $s^{\,m-1}$ in the first and of $\bigl(s+\alpha/(1-\alpha)\bigr)^{m-1}$ in the second:
\begin{equation}\label{eq:Ptwo}
P_{jh}=(1-\alpha)^{m-1}\bar P_h\ \ (h>j),
\qquad
P_{jh}=(1-\alpha)^{m-1}\widetilde W_h\ \ (h<j).
\end{equation}
The intervals $[Q_{h+1},Q_h]$ tile $[0,1]$ and move left as $h$ increases, and $t\mapsto t^{\,m-1}$ is strictly increasing for $m\geq2$, so $\bar P_h$ is strictly decreasing in $h$; and $s+\alpha/(1-\alpha)\geq s$ gives $\widetilde W_h\geq\bar P_h$, strictly for $\alpha>0$. For fixed $j$ the intervals $[\xi,\chi+\xi]$ read off above are contiguous, running from $\hat Q_1=1$ down to $(1-\alpha)Q_{n+1}=0$ as $h$ increases, so $P_{jh}$ is strictly decreasing in $h$ as well. In particular
\begin{equation}\label{eq:Porder}
P_{jh}>P_{jj}\ \ (h<j),
\qquad
P_{jh}<P_{jj}\ \ (h>j),
\qquad
P_{jj}>(1-\alpha)^{m-1}\bar P_j ,
\end{equation}
the last because the interval for $h=j$ shares its left endpoint with $[(1-\alpha)Q_{j+1},(1-\alpha)Q_j]$ and, for $\alpha>0$, extends strictly further right.

\emph{Step 3 (the recursion).} On a diagonal schedule the weights \eqref{eq:weights} reduce to $\phi_j(\mathbf{D}_r)=\pi_{jr}^{\,m-1}$, so $\phi_j(\mathbf{D}_j)=A_j^{\,m-1}$ and $\phi_j(\mathbf{D}_h)=\bigl((1-\alpha)q_h\bigr)^{m-1}$ for $h\neq j$. Hence \eqref{eq:robust-ir} reads $\mathbf{z}\geq0$, the perceived level is $\hat U_j=A_j^{\,m-1}z_j$, and \eqref{eq:biased-ic} deterring $T_j$ from a report of $T_h$ becomes
\begin{equation}\label{eq:GSC}
A_j^{\,m-1}z_j\ \geq\ \bigl((1-\alpha)q_h\bigr)^{m-1}z_h+(T_j-T_h)\,P_{jh},
\qquad h\neq j .
\tag{GSC}
\end{equation}
For $h>j$, substituting \eqref{eq:Ptwo}, dividing by $(1-\alpha)^{m-1}$ and using
$q_h^{\,m-1}z_h=\bigl(\tfrac{(1-\alpha)q_h}{A_h}\bigr)^{m-1}\bigl(\tfrac{A_h}{1-\alpha}\bigr)^{m-1}z_h=\varrho_h\zeta_h$
turns \eqref{eq:GSC} into $\zeta_j\geq\varrho_h\zeta_h+(T_j-T_h)\bar P_h$. The downward subsystem is therefore exactly the family maximized in \eqref{eq:zetarec}, which assigns each $\zeta_j$ its smallest admissible value. Taking $h=n$ there gives $\zeta_j\geq(T_j-T_n)\bar P_n>0$ for $j<n$, so $\boldsymbol\zeta\geq0$ and $\mathbf{z}\geq0$. The same substitution with $h$ replaced by $j$ gives $\overline U_j=\Phi(\mathbf{D}_j)z_j=q_j^{\,m-1}z_j=\varrho_j\zeta_j$.

\emph{Step 4 (rent slopes).} We claim $\zeta_h-\zeta_{h+1}\leq(T_h-T_{h+1})\bar P_{h+1}$ for every $h<n$. Let $l$ attain the maximum defining $\zeta_h$, and recall $\varrho_r\leq1$ and $\zeta_r\geq0$. If $l=h+1$ the claim is immediate from $\varrho_{h+1}\zeta_{h+1}\leq\zeta_{h+1}$. If $l>h+1$ then $l$ is also admissible in the maximum defining $\zeta_{h+1}$, so $\zeta_{h+1}\geq\varrho_l\zeta_l+(T_{h+1}-T_l)\bar P_l$; subtracting this from $\zeta_h=\varrho_l\zeta_l+(T_h-T_l)\bar P_l$ gives $\zeta_h-\zeta_{h+1}\leq(T_h-T_{h+1})\bar P_l\leq(T_h-T_{h+1})\bar P_{h+1}$, the last step by Step~2. Summing the one-step bound over $r=h,\ldots,j-1$ and using $\bar P_{r+1}\leq\bar P_{h+1}$,
\begin{equation}\label{eq:zetatele}
\zeta_h-\zeta_j\ \leq\ (T_h-T_j)\,\bar P_{h+1},\qquad h<j .
\end{equation}

\emph{Step 5 (upward constraints are strictly slack).} Fix $h<j$. By $\zeta_j=(A_j/(1-\alpha))^{m-1}z_j$, the identity in Step~3 and \eqref{eq:Ptwo}, the slack in \eqref{eq:GSC} equals
\[
(1-\alpha)^{m-1}\bigl[\zeta_j-\varrho_h\zeta_h+(T_h-T_j)\widetilde W_h\bigr]
\ \geq\ (1-\alpha)^{m-1}\bigl[\zeta_j-\zeta_h+(T_h-T_j)\widetilde W_h\bigr]
\ \geq\ (1-\alpha)^{m-1}(T_h-T_j)\bigl[\widetilde W_h-\bar P_{h+1}\bigr],
\]
using $\varrho_h\leq1$ and $\zeta_h\geq0$ in the first inequality and \eqref{eq:zetatele} in the second. Since $h<j\leq n$ the index $h+1$ is admissible, and $\widetilde W_h\geq\bar P_h>\bar P_{h+1}$ by Step~2, so the bound is strictly positive.

\emph{Step 6 (optimality and uniqueness).} By Steps~3 and~5 the vector $\hat U_j^{\ast}=A_j^{\,m-1}z_j$ is feasible for \eqref{eq:rlp}. It is also the smallest feasible vector, by downward induction on $j$: $\hat y_n\geq0=\hat U_n^{\ast}$, and if $\hat y_h\geq\hat U_h^{\ast}$ for every $h>j$ then each downward constraint gives $\hat y_j\geq\varrho_h\hat y_h+(T_j-T_h)P_{jh}\geq\varrho_h\hat U_h^{\ast}+(T_j-T_h)P_{jh}$, whence $\hat y_j\geq\hat U_j^{\ast}$ on taking the maximum. Step~2 of the proof of Theorem~\ref{thm:compat} makes the smallest feasible vector the unique solution of \eqref{eq:rlp}, and Steps~4 and~5 of that proof make \eqref{eq:compat-support} at these levels the unique optimum of \eqref{eq:rent-min}; since $\phi_j(\mathbf{D}_j)=A_j^{\,m-1}$, that optimum is $u_j^{\ast}(\mathbf{k})=z_j\mathbf{1}(\mathbf{k}=\mathbf{D}_j)$. Lemma~\ref{prop-revenue-rent-eq} then makes \eqref{eq:diagpay} the unique solution of \eqref{eq:rev-max}.

\emph{Step 7 (the payment rule).} Outside $\mathbf{D}_j$ the rent is zero, so \eqref{eq:diagpay} gives $x_j(\mathbf{k})=T_jp_j(\mathbf{k})\geq0$. At $\mathbf{D}_j$ a truthful buyer wins with probability $p_j(\mathbf{D}_j)=1/m$, so $x_j(\mathbf{D}_j)=T_j/m-z_j$; charging the selected winner $T_j-mz_j$ there, $T_j$ elsewhere, and nothing to losers reproduces both.

\emph{Step 8 (comparative statics and the rational benchmark).} Lemma~\ref{prop-revenue-rent-eq} and Step~3 give $U_0^{\ast}(\alpha)=\overline T-m\sum_jq_j\overline U_j=\overline T-m\sum_jq_j^{\,m}z_j$. The factor $(1-\alpha)q_h/A_h$ is strictly decreasing in $\alpha$, so each $\varrho_h$ is, and backward induction through \eqref{eq:zetarec} makes every $\zeta_j$ weakly decreasing. The rescaling factor $\bigl((1-\alpha)/A_j\bigr)^{m-1}$ is strictly decreasing, and $\zeta_j>0$ for $j<n$ by Step~3, so each such $z_j$ strictly decreases; since \eqref{eq:zetarec} bounds $\zeta_j\leq T_j-T_n$ while the rescaling factor vanishes as $\alpha\uparrow1$, $z_j\to0$. Hence $U_0^{\ast}(\alpha)$ strictly increases to $\overline T$; each $\varrho_h$ and $A_j$ is continuous in $\alpha$ and \eqref{eq:zetarec} composes maxima of continuous functions, so $\mathbf{z}$ and $U_0^{\ast}(\cdot)$ are continuous on $[0,1)$. At $\alpha=0$ every $\varrho_h=1$ and $\overline U_j=\zeta_j$, so \eqref{eq:zetarec} reads $\zeta_j=\max_{h>j}[\zeta_h+(T_j-T_h)\bar P_h]$. Put $V_n:=0$ and $V_j:=\sum_{r>j}(T_{r-1}-T_r)\bar P_r$. Writing $T_j-T_h=\sum_{r=j+1}^{h}(T_{r-1}-T_r)$ for $h>j$ gives $V_j-[V_h+(T_j-T_h)\bar P_h]=\sum_{r=j+1}^{h}(T_{r-1}-T_r)(\bar P_r-\bar P_h)\geq0$, with equality exactly at $h=j+1$, since $\bar P$ is strictly decreasing by Step~2. Thus $\mathbf{V}$ satisfies the recursion, which determines its solution bottom-up, so $\overline U_j^{S}=V_j$. For \eqref{eq:recovery}, summing the one-step bound of Step~4 from $j$ to $n$ gives $\zeta_j\leq V_j=\overline U_j^{S}$, so by Step~3 and $\overline U_n^{S}=0$, $U_{\min}^{\ast}=m\sum_jq_j\varrho_j\zeta_j\leq\max_{j<n}\varrho_j\,m\sum_jq_j\overline U_j^{S}=\max_{j<n}\varrho_j\,(\overline T-U_0^{S})$, with equality when $n=2$ because $\zeta_1=(T_1-T_2)\bar P_2=\overline U_1^{S}$. Lemma~\ref{prop-revenue-rent-eq} gives $E(\alpha)=\overline T-U_0^{S}-U_{\min}^{\ast}$, and $E(\alpha)>0$ because $\max_{j<n}\varrho_j<1$ and $\overline U_j^{S}>0$ for $j<n$.

\emph{Step 9 (the boundary $\alpha=0$).} At $\alpha=0$ every ratio in \eqref{eq:per-rival-ratio} equals one, so the extrema in {\rm (C1)--(C2)} are attained at every configuration and Step~1 of the proof of Theorem~\ref{thm:compat} fails. Step~5 still gives strict slack, $(T_h-T_j)[\bar P_h-\bar P_{h+1}]>0$, and the induction of Step~6 uses only $\varrho_h\geq0$, so the levels remain the smallest feasible vector; Steps~3 and~4 of the proof of Theorem~\ref{thm:compat} use the extrema only through weak inequalities attained at the diagonals, so the diagonal rule stays optimal, and only the strict comparisons of Step~5 of that proof, hence uniqueness of rent location, are lost. \qed

\subsubsection*{Proof of Theorem~\ref{thm:band}.}
Throughout, $\phi_j$, $\Phi$, $\mathbf{D}_j$, $P_{jh}$, $\bar P_h$, $\widetilde W_h$, $\varrho_h$, $\zeta_j$ and $z_j$ are the objects of Sections~\ref{sec:model} and~\ref{sec:diagonal} read on the band primitives with $T_j=L_j$.

\emph{Step 1 (the endpoint reduction).} In rival configuration $\mathbf{k}$, a buyer of value $s\in I_j$ who reports band $h$ realizes $s\,p_h(\mathbf{k})-x_h(\mathbf{k})=u_h(\mathbf{k})+(s-L_h)p_h(\mathbf{k})$. Applied to $h=j$ with $p_j\geq0$, the worst value is $s=L_j$, so \eqref{eq:robust-ir} holds at every value of $I_j$ if and only if $u_j(\mathbf{k})\geq0$. Since beliefs are band-constant, that buyer's subjective payoff from reporting $h$ is $\Theta_h(s)=\sum_{\mathbf{k}}\phi_j(\mathbf{k})u_h(\mathbf{k})+(s-L_h)P_{jh}$, affine in $s$ with slope $P_{jh}$. Truthful reporting requires $\Theta_j(s)\geq\Theta_h(s)$ for every $s\in I_j$, that is $\hat U_j\geq\sum_{\mathbf{k}}\phi_j(\mathbf{k})u_h(\mathbf{k})+(L_j-L_h)P_{jh}+(s-L_j)(P_{jh}-P_{jj})$, whose right-hand side is affine in $s$, so its maximum over $[L_j,H_j]$ is attained at $s=L_j$ when $P_{jh}\leq P_{jj}$ and at $s=H_j$ otherwise; in either case the maximum is $\sum_{\mathbf{k}}\phi_j(\mathbf{k})u_h(\mathbf{k})+d^B_{jh}$ with $d^B_{jh}$ as in \eqref{eq:bandd}. The whole family of constraints is therefore \eqref{eq:biased-ic} with that constant. Finally, $x_j=L_jp_j-u_j$ gives
$m\sum_jq_j\sum_{\mathbf{k}}\Phi(\mathbf{k})x_j(\mathbf{k})=m\sum_jq_jL_j\sum_{\mathbf{k}}\Phi(\mathbf{k})p_j(\mathbf{k})-m\sum_jq_j\overline U_j$, whose first term is the same for every efficient band mechanism, so revenue maximization is again \eqref{eq:rent-min}. The proofs of Theorems~\ref{thm:compat} and~\ref{thm:approx} and of Proposition~\ref{prop:dual} treat the deviation constants as given numbers, and $d^B_{jh}>0$ exactly for $h>j$, since $d^B_{jh}=-(L_h-L_{j-1})P_{jh}-\delta_jP_{jj}<0$ when $h<j$; they therefore apply verbatim. This is {\rm (i)}.

\emph{Step 2 (downward constraints).} Assume a common projection intensity for the remainder. By \eqref{eq:Porder}, $P_{jh}<P_{jj}$ for $h>j$, so $d^B_{jh}=(L_j-L_h)P_{jh}$: every downward constraint is the discrete one at $T_j=L_j$, which \eqref{eq:zetarec} satisfies by construction, with equality along its maximizing branch.

\emph{Step 3 (upward constraints are strictly slack).} Fix $h<j$, so $P_{jh}>P_{jj}$ by \eqref{eq:Porder} and $d^B_{jh}=(L_j-L_h)P_{jh}+\delta_j(P_{jh}-P_{jj})$. On the diagonal schedule, $\hat U_j=A_j^{\,m-1}z_j=(1-\alpha)^{m-1}\zeta_j$ and $\sum_{\mathbf{k}}\phi_j(\mathbf{k})u_h(\mathbf{k})=\bigl((1-\alpha)q_h\bigr)^{m-1}z_h=(1-\alpha)^{m-1}\varrho_h\zeta_h$, so the slack in \eqref{eq:biased-ic} is, after collecting the terms in $P_{jh}$ through $L_j+\delta_j=H_j=L_{j-1}$, $(1-\alpha)^{m-1}(\zeta_j-\varrho_h\zeta_h)+(L_h-L_{j-1})P_{jh}+\delta_jP_{jj}$. Both $L_h-L_{j-1}$ and $\delta_j$ are nonnegative, and $P_{jh}=(1-\alpha)^{m-1}\widetilde W_h\geq(1-\alpha)^{m-1}\bar P_h$ while $P_{jj}>(1-\alpha)^{m-1}\bar P_j$ by \eqref{eq:Porder}; using also $\varrho_h\leq1$ and $\zeta_h\geq0$, the slack strictly exceeds $(1-\alpha)^{m-1}[\zeta_j-\zeta_h+(L_h-L_{j-1})\bar P_h+\delta_j\bar P_j]$. Telescoping the one-step bound of Step~4 of the proof of Theorem~\ref{thm:diagonal} and splitting off its last term,
\[
\zeta_h-\zeta_j\ \leq\ \sum_{r=h}^{j-1}(L_r-L_{r+1})\bar P_{r+1}
\ =\ \sum_{r=h}^{j-2}(L_r-L_{r+1})\bar P_{r+1}+\delta_j\bar P_j
\ \leq\ (L_h-L_{j-1})\bar P_h+\delta_j\bar P_j ,
\]
the last step because $\bar P_{r+1}\leq\bar P_h$ for every $r\geq h$ and $\sum_{r=h}^{j-2}(L_r-L_{r+1})=L_h-L_{j-1}$; for $h=j-1$ both sums are empty and the bound is the one-step bound itself. The bracket is therefore nonnegative and the slack strictly positive. No restriction on the widths $\delta_j$ is used.

\emph{Step 4 (optimality and uniqueness).} Replacing every $d^B_{jh}$ by $(L_j-L_h)P_{jh}$ relaxes the program, since the two agree for $h>j$ and $d^B_{jh}\geq(L_j-L_h)P_{jh}$ for $h<j$. That relaxation is \eqref{eq:rent-min} at the band primitives, whose unique optimum is by Theorem~\ref{thm:diagonal} the diagonal schedule generated by \eqref{eq:zetarec} at $T_j=L_j$. Steps~2 and~3 show that this schedule is feasible for the unrelaxed band program, so it attains the relaxed optimum there and is optimal; and any other band-feasible schedule of the same cost would be a second optimum of the relaxation, contradicting the uniqueness in Theorem~\ref{thm:diagonal}. This is {\rm (ii)}, and truthfulness at every value follows from Step~1.

\emph{Step 5 (implementation).} Outside $\mathbf{D}_j$ the floor rent is zero, so $x_j(\mathbf{k})=L_jp_j(\mathbf{k})$ and charging the selected winner $L_j$ reproduces it; at $\mathbf{D}_j$ the buyer wins with probability $1/m$, so $x_j(\mathbf{D}_j)=L_j/m-z_j$ and charging the selected winner $L_j-mz_j$ reproduces it. Losers pay zero in both cases. A winner of value $s\in I_j$ realizes $s-L_j\geq0$ in the first case and $s-L_j+mz_j\geq0$ in the second, so realized utility is nonnegative at every value of every band and no price exceeds the winner's value; and the prices are nonnegative exactly when $mz_j\leq L_j$. \qed

\subsubsection*{Proof of Corollary~\ref{cor:mean}.}
Replacing $L_j$ by $\mu_j$ changes neither bands nor beliefs, and $\mu_1>\cdots>\mu_n$, so the proofs of Theorems~\ref{thm:diagonal} and~\ref{thm:band} apply at $T=\boldsymbol\mu$. A winner of value $s\in I_j$ realizes $s-\mu_j$ off $\mathbf{D}_j$, where she wins with positive probability, and $s-\mu_j+mz^{\mu}_j$ at $\mathbf{D}_j$, which gives the claims on realized utility. As in Step~1 of the proof of Theorem~\ref{thm:band}, her payoff from reporting band $h$ is $\Theta_h(s)=\sum_{\mathbf{k}}\phi_j(\mathbf{k})u^{\mu}_h(\mathbf{k})+(s-\mu_h)P_{jh}$, with $u^{\mu}_h(\mathbf{k})=z^{\mu}_h\mathbf{1}(\mathbf{k}=\mathbf{D}_h)$. For $h>j$, $\Theta_j(\mu_j)\geq\Theta_h(\mu_j)$ is \eqref{eq:GSC} at $T=\boldsymbol\mu$, met with equality on the maximizing branch of \eqref{eq:zetarec}, so $\Theta_h(s)-\Theta_j(s)=[\Theta_h(\mu_j)-\Theta_j(\mu_j)]+(\mu_j-s)(P_{jj}-P_{jh})$ with $P_{jj}>P_{jh}$ by \eqref{eq:Porder}. For $h<j$ the slack $\Theta_j(s)-\Theta_h(s)$ decreases in $s$, as $P_{jh}>P_{jj}$, and at $s=H_j$ Step~3 of the proof of Theorem~\ref{thm:band}, with $L_h$ replaced by $\mu_h$ and $\delta_j$ by $H_j-\mu_j$, bounds it strictly below by $(1-\alpha)^{m-1}$ times $\zeta^{\mu}_j-\zeta^{\mu}_h+(\mu_h-H_j)\bar P_h+(H_j-\mu_j)\bar P_j$. Summing Step~4 of the proof of Theorem~\ref{thm:diagonal} gives $\zeta^{\mu}_h-\zeta^{\mu}_j\leq(\mu_h-\mu_{j-1})\bar P_h+(\mu_{j-1}-\mu_j)\bar P_j$, so this is at least $(\mu_{j-1}-H_j)(\bar P_h-\bar P_j)\geq0$, as $\mu_{j-1}>L_{j-1}=H_j$ and $\bar P_h>\bar P_j$. Each failure is at most $\mu_j-L_j$ because $P_{jj}-P_{jh}\leq1$. \qed

\subsubsection*{Proof of Proposition~\ref{prop:resolution}.}
\emph{Step 1 (the accounting identity).} A value lies in $I_j$ or below with probability $Q_j$, so the highest occupied band is $I_j$ with probability $Q_j^{\,m}-Q_{j+1}^{\,m}=w_j$, which equals $mq_j\bar P_j$ by the definition of $\bar P_j$. Conditional on the band profile, values inside a band are independent draws from $G$ restricted to it, and the winner is selected uniformly among the occupants of the highest occupied band without reference to her position in it; hence ${\sf E}[\widehat S\mid\text{highest occupied band }I_j]=\mu_j$ and ${\sf E}[\widehat S]=\sum_jw_j\mu_j$. Reading the definition of $\overline T$ in Section~\ref{sec:model} at the band primitives gives $\overline T=m\sum_jq_jL_j\bar P_j=\sum_jw_jL_j$. Subtracting, ${\sf E}[\widehat S]=\overline T+\sum_jw_j(\mu_j-L_j)$, and adding $\varepsilon_{\mathcal I}$ and Lemma~\ref{prop-revenue-rent-eq} gives \eqref{eq:band-split}.

\emph{Step 2 (pooling).} Both the winner and the highest bidder belong to the highest occupied band, so in every profile $0\leq\widetilde S_{(1)}-\widehat S\leq\delta$, which is the first bound. A strictly positive loss requires two bidders in that band, hence some band containing at least two bidders; a union bound over the $\binom{m}{2}$ pairs and the $n$ bands puts the probability of that event at $\binom{m}{2}\sum_jq_j^{2}$ at most. A bounded density gives $q_j\leq\bar g\delta_j\leq\bar g\delta$, so $\sum_jq_j^{2}\leq\bar g\delta\sum_jq_j=\bar g\delta$; multiplying by the maximum loss $\delta$ gives the quadratic bound.

\emph{Step 3 (baseline).} Fix bidder $i$ and her rivals' values, let $I_r$ be the highest band the rivals occupy and $k\geq1$ their number in it. The band allocation gives her the monotone step probability $p(v)=0$ for $v<L_r$, $p(v)=1/(k+1)$ for $v\in I_r$, and $p(v)=1$ for $v>H_r$. The dominant-strategy payment leaving the lowest value zero utility is
\begin{equation}\label{eq:envelope}
x(v)=v\,p(v)-\int_{\underline s}^{v}p(t)\,\mathrm{d}t,
\end{equation}
and integrating the step function gives $x(v)=L_r/(k+1)$ on $I_r$ and $x(v)=H_r-(H_r-L_r)/(k+1)$ above $H_r$. Both are constant in $v$, so \eqref{eq:envelope} is a band mechanism; being dominant-strategy incentive compatible with zero rent at $\underline s$, it leaves the minimal interim rents for this allocation and therefore attains $U_0^{S}$. Only its winner pays, at the price $x(v)/p(v)$, which is $L_r$ when at least two bidders occupy the highest band and $H_r-(H_r-L_r)/(k+1)$ when the winner is alone above $I_r$; in both cases the price lies in $I_r$. So does the second-highest value: in the first case the two highest values are both in $I_r$, and in the second the second-highest value is the highest rival value. The price and $\widetilde S_{(2)}$ therefore lie in a common band in every profile, so they differ by at most $\delta$, and taking expectations gives $|U_0^{S}-U_0^{SPA}|\leq\delta$. No regularity is used and the sign of the difference is not determined.

\emph{Step 4 (premium).} By Lemma~\ref{prop-revenue-rent-eq}, $E(\alpha)=m\sum_jq_j\bigl[\overline U_j^{S}-\overline U_j(\alpha)\bigr]$, which is nonnegative by Theorem~\ref{thm:diagonal}{\rm (iii)} and at most $m\sum_jq_j\overline U_j^{S}$ because rents are nonnegative. For $r\geq2$, the definition of $\bar P_r$ gives $\bar P_r\leq Q_r^{\,m-1}\leq Q_2^{\,m-1}=(1-q_1)^{m-1}$, so
\[
\overline U_j^{S}=\sum_{r>j}(L_{r-1}-L_r)\bar P_r\ \leq\ (1-q_1)^{m-1}\sum_{r>j}(L_{r-1}-L_r)=(1-q_1)^{m-1}(L_j-L_n),
\]
every index $r$ in the sum exceeding $j\geq1$. Since $\sum_jq_j=1$ and $L_j\leq L_1$, this gives \eqref{eq:premium-bound}; the bound vanishes as $m\to\infty$ because $q_1>0$ is held fixed.

\emph{Step 5 (thick markets).} As $m\to\infty$ with $\mathcal{I}$ fixed, $w_1=1-(1-q_1)^m\to1$, so Step~1 gives $\overline T\to L_1$ and ${\sf E}[\widehat S]\to\mu_1$, the other weights vanishing and $L_j,\mu_j$ being bounded. Step~4 and $\overline U_j\geq0$ bracket revenue by $\overline T-m(L_1-L_n)(1-q_1)^{m-1}\leq U_0^{S}\leq U_0^{\ast}(\alpha)\leq\overline T$, so $U_0^{\ast}(\alpha)\to L_1$. Since $H_1=\bar s$, for every $\epsilon>0$ the probability that fewer than two values exceed $H_1-\epsilon$ vanishes, so $\widetilde S_{(1)}$ and $\widetilde S_{(2)}$ converge to $H_1$ in probability and, being bounded, in expectation; hence $U_0^{\ast}(\alpha)-U_0^{SPA}\to L_1-H_1=-\delta_1$ and $\varepsilon_{\mathcal I}\to H_1-\mu_1$. \qed

\subsubsection*{Proof of Proposition~\ref{prop:refine}.}
\emph{(i)} Allocate efficiently on reports, charge an unmatched winner her report, and when all $m$ reports equal $s$ charge the selected winner $s-mz(s)$, with $z(s):=\bigl((1-\alpha)/\alpha\bigr)^{m-1}\max_{v\in[\underline s,s]}(s-v)G(v)^{m-1}\geq0$; losers pay nothing. Under $\widehat G(\cdot\mid s)$ a rival's value is $s$ with probability $\alpha$ and otherwise a draw from the atomless $G$. Reporting $s$ thus meets the all-match event with probability $\alpha^{m-1}$ and earns $\alpha^{m-1}z(s)$, since she pays her value elsewhere, whereas a report $v\neq s$ meets it with probability zero. A report $v<s$ then wins only if every rival lies below $v$, with probability $\bigl((1-\alpha)G(v)\bigr)^{m-1}$, at price $v$, earning at most $\alpha^{m-1}z(s)$ by definition of $z$, and a report $v>s$ wins only at a price above $s$. Realized utility is zero off the all-match event and $mz(s)$ for its winner, so \eqref{eq:robust-ir} holds statewise; the event is objectively null and the allocation efficient, so revenue is ${\sf E}[\widetilde S_{(1)}]$.

\emph{(ii)} By \eqref{eq:band-split}, ${\sf E}[\widetilde S_{(1)}]-U_0^{\ast}(\alpha)=U_{\min}^{\ast}+\sum_jw_j(\mu_j-L_j)+\varepsilon_{\mathcal I}$. The middle term lies in $(0,\delta]$, as $L_j<\mu_j\leq H_j$ and $\sum_jw_j=1$; the last is at most $\binom{m}{2}\bar g\delta^{2}$ by Proposition~\ref{prop:resolution}{\rm (i)}; and, the rule of Theorem~\ref{thm:band}{\rm (ii)} being the recursion at $T_j=L_j$, \eqref{eq:recovery} bounds the first by $\max_{j<n}\varrho_j(\overline T-U_0^{S})$, where $A_j\geq\alpha$ and $q_j\leq\bar g\delta$ give $\varrho_j\leq\bigl((1-\alpha)\bar g\delta/\alpha\bigr)^{m-1}$. Since $\overline T\leq{\sf E}[\widetilde S_{(1)}]$ and $|U_0^{S}-{\sf E}[\widetilde S_{(2)}]|\leq\delta$ by Proposition~\ref{prop:resolution}{\rm (ii)}, $\overline T-U_0^{S}$ stays bounded as $\delta\to0$, which gives both limits. \qed

\subsubsection*{Proof of Proposition~\ref{prop:sat}.}
\emph{Attainment.} The program minimizes a linear objective over a polyhedron and is bounded below by zero, so it suffices to exhibit a feasible point. Take the floor rents $u_j^{D}(\mathbf{k})=\sum_{r>j}\delta_rp_r(\mathbf{k})$ induced by the dominant-strategy rule \eqref{eq:envelope}; since $H_r=L_{r-1}$, $\sum_{r>j}\delta_r=L_j-L_n$. Under \eqref{eq:efficient-allocation} the allocation is nonincreasing in the band index: if $p_{h+1}(\mathbf{k})>0$ then $k_1=\cdots=k_h=0$ and $p_h(\mathbf{k})=1$. Hence $0\leq u_j^{D}(\mathbf{k})\leq(L_j-L_n)p_j(\mathbf{k})\leq L_jp_j(\mathbf{k})$, which is \eqref{eq:robust-ir} and \eqref{eq:cap}. In configuration $\mathbf{k}$ a buyer of value $s$ reporting band $h$ realizes $W_h(s,\mathbf{k})=(s-L_h)p_h(\mathbf{k})+\sum_{r>h}\delta_rp_r(\mathbf{k})$, and $L_{h+1}+\delta_{h+1}=H_{h+1}=L_h$ gives $W_h(s,\mathbf{k})-W_{h+1}(s,\mathbf{k})=(s-L_h)[p_h(\mathbf{k})-p_{h+1}(\mathbf{k})]$, whose sign is that of $s-L_h$. For $s\in I_j$ this difference is nonnegative at every $h\geq j$ and nonpositive at every $h<j$, so $W_j(s,\mathbf{k})\geq W_h(s,\mathbf{k})$ for every $h$: truthful band reporting is optimal configuration by configuration at every value of every band, hence under any belief, and the endpoint constraints of Theorem~\ref{thm:band}{\rm (i)} are satisfied.

\emph{The exchange.} Suppose at an optimum $\mathbf{u}^{\ast}$ some band $j$ has $u_j^{\ast}(\mathbf{k}_2)>0$ for a configuration $\mathbf{k}_2\neq\mathbf{D}_j$ while $u_j^{\ast}(\mathbf{D}_j)<L_j/m$. For small $\delta>0$ move rent onto the diagonal at the subjective exchange rate: lower $u_j(\mathbf{k}_2)$ by $\delta$, raise $u_j(\mathbf{D}_j)$ by $\delta\phi_j(\mathbf{k}_2)/\phi_j(\mathbf{D}_j)$, and leave every other entry unchanged, with $\delta$ small enough that both stay within $[0,L_jp_j(\cdot)]$, which is possible because $u_j^{\ast}(\mathbf{k}_2)>0$ and the diagonal is unsaturated. By construction $\hat U_j$ is unchanged.

\emph{Feasibility.} Band $j$'s own constraints are unaffected: their left-hand side is $\hat U_j$ and their right-hand sides involve only $u_h$, $h\neq j$, and the constant \eqref{eq:bandd}. Every other constraint carrying $u_j$ does so through the leakage term $\sum_{\mathbf{k}}\phi_h(\mathbf{k})u_j(\mathbf{k})$ on the right of the constraint that deters band $h$ from claiming $j$, and the exchange changes it by $\phi_j(\mathbf{k}_2)\delta[\phi_h(\mathbf{D}_j)/\phi_j(\mathbf{D}_j)-\phi_h(\mathbf{k}_2)/\phi_j(\mathbf{k}_2)]\leq0$, because by \eqref{eq:per-rival-ratio} the per-rival ratio $\pi_{hr}/\pi_{jr}$ is uniquely minimized at $r=j$, so its product over the $m-1$ rivals is uniquely minimized at $\mathbf{D}_j$; this is {\rm (C2)} with the roles of $j$ and $h$ exchanged. Leakage to every other band therefore weakly falls and every such constraint survives.

\emph{Strict improvement.} The objective of \eqref{eq:rent-min} changes by $mq_j\delta[\Phi(\mathbf{D}_j)\phi_j(\mathbf{k}_2)/\phi_j(\mathbf{D}_j)-\Phi(\mathbf{k}_2)]=mq_j\Phi(\mathbf{k}_2)\delta[\Lambda_j(\mathbf{k}_2)/\Lambda_j-1]<0$, strictly, since by \eqref{eq:per-rival-ratio} the ratio $\pi_{jr}/q_r$ is uniquely maximized at $r=j$, so $\Lambda_j(\mathbf{k}_2)<\Lambda_j$ for every $\mathbf{k}_2\neq\mathbf{D}_j$ when $\alpha\in(0,1)$. The exchange is feasible and strictly cheaper, contradicting optimality. Hence $u_j^{\ast}(\mathbf{D}_j)\geq L_j/m$ whenever band $j$ carries rent off its diagonal, and \eqref{eq:cap} at $\mathbf{D}_j$ makes that an equality. \qed

\subsubsection*{Proof of Proposition~\ref{prop:exact}.}
\emph{Part {\rm (i)}.} Fix $j<n$. For a value $s$, write ${\sf E}_s$ for expectation over the rivals' values $\mathbf{s}_{-i}$, drawn independently from \eqref{eq:smoothlaw} and reported truthfully, and $\mathbf{k}=\mathbf{k}(\mathbf{s}_{-i})$ for the induced configuration of rival bands, so that ${\sf E}_s[F(\mathbf{k})]=\sum_{\mathbf{k}\in\mathcal{K}}\widehat\phi_s(\mathbf{k})F(\mathbf{k})$; profiles with a value at a band endpoint have probability zero. Let
\[
Y:=\sum_{\ell\neq i}\log s_\ell-(m-1)\log L_j,
\qquad
\tau:=\frac{\log s-\log L_j}{\theta^{2}} .
\]
Expanding the squares in \eqref{eq:smoothlaw}, the joint density of $\mathbf{s}_{-i}$ at $s$ equals that at $L_j$ times $C(s)e^{\tau Y}$, where $C(s)>0$ collects the normalizers and the terms free of $\mathbf{s}_{-i}$. Hence ${\sf E}_s[F]=C(s)\,{\sf E}_{L_j}[F\,e^{\tau Y}]$ for every function $F$ of the rival profile.

A rival in band $j$ has value at least $L_j$, and a rival below band $j$ has value at most $H_{j+1}=L_j$. Thus $Y\geq0$ when $\mathbf{k}=\mathbf{D}_j$ and $Y\leq0$ when $\sum_{r\leq j}k_r=0$, and \eqref{eq:triangular} gives $\Gamma_j(\mathbf{k})\,Y\geq0$ at every profile. Since $e^{x}-1$ has the sign of $x$, the product $\Gamma_j(\mathbf{k})(e^{\tau Y}-1)$ has the sign of $\tau\,\Gamma_j(\mathbf{k})Y$, hence that of $\tau$. With indifference, ${\sf E}_{L_j}[\Gamma_j]=0$, this gives
\[
{\sf E}_s[\Gamma_j]=C(s)\,{\sf E}_{L_j}\bigl[\Gamma_j\,(e^{\tau Y}-1)\bigr],
\]
which is nonnegative for $s>L_j$ and nonpositive for $s<L_j$.

In configuration $\mathbf{k}$, a buyer of value $s$ gains $sp_j-x_j-(sp_{j+1}-x_{j+1})=(s-L_j)(p_j-p_{j+1})+\Gamma_j$ by reporting band $j$ rather than $j+1$, since $x_h=L_hp_h-u_h$ and $L_j-L_{j+1}=\delta_{j+1}$. Under the band allocation $p_j(\mathbf{k})\geq p_{j+1}(\mathbf{k})$: both vanish when a rival reports above band $j$, and otherwise $p_{j+1}(\mathbf{k})>0$ only when every rival is below band $j$, where $p_j(\mathbf{k})=1$. The expected gain $\gamma_j(s):={\sf E}_s\bigl[(s-L_j)(p_j-p_{j+1})+\Gamma_j\bigr]$ is therefore nonnegative for $s\geq L_j$ and nonpositive for $s\leq L_j$.

The same holds at every boundary $r<n$. Now let $s\in I_j$ and $h\neq j$. The expected gain from reporting band $j$ rather than $h$ telescopes over adjacent reports: it is $\sum_{r=j}^{h-1}\gamma_r(s)\geq0$ if $h>j$, because $s\geq L_j\geq L_r$ for $r\geq j$, and $-\sum_{r=h}^{j-1}\gamma_r(s)\geq0$ if $h<j$, because $s\leq H_j=L_{j-1}\leq L_r$ for $r\leq j-1$. Expected utility is linear in a report lottery, so truthful band reporting is optimal at every value against truthful rivals. The restrictions consist of \eqref{eq:cap}, one equality per boundary and one sign restriction per boundary and configuration, all linear in $\mathbf{u}$. The dominant-strategy rule has floor rents $u_j^{D}(\mathbf{k})=\sum_{r>j}\delta_rp_r(\mathbf{k})$ and satisfies \eqref{eq:cap} (proof of Proposition~\ref{prop:sat}), so $\Gamma_j=u_j^{D}-u_{j+1}^{D}-\delta_{j+1}p_{j+1}\equiv0$.

\emph{Part {\rm (ii)}.} An exactly truthful rule satisfies the incentive constraints at every value, hence $\mathbf{u}\mathbf{A}\geq\mathbf{a}$, and \eqref{eq:cap} gives $0\leq u_j(\mathbf{k})\leq L_jp_j(\mathbf{k})$. For $\mathbf{y}\geq0$, subtracting $(\mathbf{u}\mathbf{A}-\mathbf{a})\mathbf{y}^{\prime}\geq0$ from the objective of \eqref{eq:rent-min} and minimizing each remaining term over $[0,L_jp_j(\mathbf{k})]$ gives
\[
\begin{aligned}
m\sum_{j,\mathbf{k}}q_j\Phi(\mathbf{k})u_j(\mathbf{k})
&\geq\mathbf{a}\mathbf{y}^{\prime}+\sum_{j,\mathbf{k}}\bigl[mq_j\Phi(\mathbf{k})-(\mathbf{A}\mathbf{y}^{\prime})_{j\mathbf{k}}\bigr]u_j(\mathbf{k})\\
&\geq\mathbf{a}\mathbf{y}^{\prime}+\sum_{j,\mathbf{k}}L_jp_j(\mathbf{k})\min\bigl\{mq_j\Phi(\mathbf{k})-(\mathbf{A}\mathbf{y}^{\prime})_{j\mathbf{k}},\,0\bigr\}.
\end{aligned}
\]
Revenue is $\overline T$ minus the left-hand side, by Step~1 of the proof of Theorem~\ref{thm:band}. \qed

\subsubsection*{Proof of Proposition~\ref{prop:fpa}.}
\emph{Monotone likelihood ratio and the reverse hazard.} For $s>\tilde s$, \eqref{eq:smoothlaw} gives
\[
\frac{\widehat g(s'\mid s)}{\widehat g(s'\mid \tilde s)}
=C(s,\tilde s)\exp\Bigl[\frac{(\log s-\log\tilde s)\log s'}{\theta^{2}}\Bigr],
\]
where $C(s,\tilde s)$ collects the normalizers and the terms free of $s'$. The exponent is strictly increasing in $s'$, so the family is ordered by the monotone likelihood ratio. Fix $s'$ and write the reciprocal of the reverse hazard as $\widehat G(s'\mid s)/\widehat g(s'\mid s)=\int_{\underline s}^{s'}\bigl[\widehat g(z\mid s)/\widehat g(s'\mid s)\bigr]\mathrm{d}z$. Each integrand has $z<s'$ and is therefore decreasing in $s$ by the ratio property, so the reciprocal decreases and $\widehat g(s'\mid s)/\widehat G(s'\mid s)$ increases in $s$.

\emph{The candidate.} Under a strictly increasing $\mathsf{b}$, a buyer of value $s$ who bids $\mathsf{b}(y)$ for some $y\geq s_0$ wins exactly when every rival's value is below $y$, since lower values either bid less or abstain. Her subjective payoff is $[s-\mathsf{b}(y)]\widehat G(y\mid s)^{m-1}$. Differentiating at $y=s$ and setting the result to zero gives
\begin{equation}\label{eq:fpaode}
\mathsf{b}'(s)=(m-1)\frac{\widehat g(s\mid s)}{\widehat G(s\mid s)}\bigl[s-\mathsf{b}(s)\bigr] .
\end{equation}
Writing $\Psi(s)=s-\mathsf{b}(s)$ turns \eqref{eq:fpaode} into the linear equation $\Psi'=1-(m-1)\bigl[\widehat g(s\mid s)/\widehat G(s\mid s)\bigr]\Psi$ with $\Psi(s_0)=0$, whose unique solution is the integral in \eqref{eq:fpabid}. It is strictly positive for $s>s_0$, so $\mathsf{b}'>0$ by \eqref{eq:fpaode} and $\mathsf{b}(s)<s$. At $s_0=\underline s$ the positive density at the lower endpoint gives $(m-1)\widehat g(s\mid s)/\widehat G(s\mid s)\sim(m-1)/(s-\underline s)$ and hence $\Psi(s)\sim(s-\underline s)/m$, so the improper integral converges and the boundary condition holds.

\emph{Global optimality.} Write $\widehat\lambda(s'\mid s)=\widehat g(s'\mid s)/\widehat G(s'\mid s)$ and differentiate the payoff from bidding $\mathsf{b}(y)$:
\[
\frac{\mathrm{d}}{\mathrm{d}y}\Bigl\{[s-\mathsf{b}(y)]\widehat G(y\mid s)^{m-1}\Bigr\}
=(m-1)\widehat G(y\mid s)^{m-1}
\Bigl\{[s-\mathsf{b}(y)]\widehat\lambda(y\mid s)-[y-\mathsf{b}(y)]\widehat\lambda(y\mid y)\Bigr\},
\]
using \eqref{eq:fpaode} for $\mathsf{b}'(y)$. If $s>y$ then $s-\mathsf{b}(y)>y-\mathsf{b}(y)>0$ and $\widehat\lambda(y\mid s)>\widehat\lambda(y\mid y)$, so the brace is strictly positive. If $s<y$ and $s-\mathsf{b}(y)\leq0$ the brace is negative because the second product is strictly positive. If $s<y$ and $s-\mathsf{b}(y)>0$, both factors of the first product are smaller than their counterparts in the second and all four are positive, so the brace is again negative. The payoff therefore rises on $(s_0,s)$ and falls on $(s,\bar s]$, making $y=s$ the best bid in the range of $\mathsf{b}$. Bids above $\mathsf{b}(\bar s)$ win with the same probability at a higher price and bids below $s_0$ never trade, so no other bid can do better; a buyer with $s<s_0$ would pay at least $s_0>s$ whenever she won and so abstains.

\emph{Receipts.} Under a strictly increasing $\mathsf{b}$ the winner is the buyer with the highest value above $s_0$, of objective density $mG(s)^{m-1}g(s)$, and pays her bid, which gives \eqref{eq:fparev}. In the second-price auction the winner pays $s_0$ when exactly one value exceeds it, with objective probability $m[1-G(s_0)]G(s_0)^{m-1}$, and $\widetilde S_{(2)}$, of density $m(m-1)[1-G(s)]G(s)^{m-2}g(s)$ on $(s_0,\bar s]$, when at least two do; this is \eqref{eq:sparev}, which at $s_0=\underline s$ is ${\sf E}[\widetilde S_{(2)}]=U_0^{SPA}$. In both formats trade fails exactly when every value is below $s_0$, with objective probability $G(s_0)^{m}$. \qed

\section{Additional Calibration Detail}\label{app:caldetail}
This appendix records the sample construction, states the estimation likelihood, reports the specification checks and numerical tolerances behind Section~\ref{sec:evidence}, and extends its belief-law comparisons. Tables~\ref{tab:cal-screen}--\ref{tab:cal-smooth} collect the numbers underlying that section.

\noindent\textbf{Sample construction.} The public file records 1{,}000 listings, of which 863 report a closing price. Table~\ref{tab:cal-screen} applies the screens of Section~\ref{sec:calibration_data} in order and records the count each one retains; the final reserve-ratio screen removes no further observation. Table~\ref{tab:cal-desc} describes the sold and the retained samples. Book value is a referenced retail valuation, not an observed bidder valuation, and no synthetic record is added.

\begin{figure}[htbp]
\vspace{-0.1in}
    \begin{minipage}[t]{0.45\textwidth}
    {\footnotesize\centering\setlength{\tabcolsep}{3pt}
        \begin{tabular}{@{}lr@{}}\toprule
Screen applied & Retained\\\midrule
All listings & 1{,}000\\
Closing price recorded & 863\\
Closing price positive & 862\\
At least two bidders & 752\\
$y_a\geq r_a+0.02$ & 750\\
$y_a<3$ & 749\\
$r_a<1.8$ & 749\\\bottomrule
\end{tabular}
        \captionof{table}{Sample screens, in the order applied}
        \label{tab:cal-screen}
    }
    \end{minipage}\hfill
    \begin{minipage}[t]{0.53\textwidth}
    {\footnotesize\centering\setlength{\tabcolsep}{3pt}
        \begin{tabular}{@{}lrrrr@{}}\toprule
& \multicolumn{2}{c}{Sold (863)} & \multicolumn{2}{c}{Retained (749)}\\
\cmidrule(lr){2-3}\cmidrule(lr){4-5}
 & Median & Mean & Median & Mean\\\midrule
Bidders $m_a$ & 4.00 & 4.21 & 5.00 & 4.69\\
Book value $B_a$ (\$) & 22.50 & 31.38 & 24.50 & 33.10\\
Closing price (\$) & 20.25 & 28.01 & 22.06 & 29.42\\
Price-to-book $y_a$ & 0.91 & 0.97 & 0.91 & 0.96\\
Reserve-to-book $r_a$ & 0.41 & 0.51 & 0.35 & 0.44\\\bottomrule
\end{tabular}
        \captionof{table}{Descriptive statistics by sample}
        \label{tab:cal-desc}
    }
    \end{minipage}
\vspace{-0.2in}
\end{figure}

\noindent\textbf{Order-statistic likelihood.} Let $X=\widetilde S_i/B$ be value relative to book value, with cdf and density $G$ and $g$. For auction $a$, let $m_a$ be the number of entrants, $r_a$ the reserve-to-book ratio and $y_a$ the closing-price-to-book ratio. A bidder enters when her value exceeds the reserve, so conditional on entry $G_a(y)=[G(y)-G(r_a)]/[1-G(r_a)]$ and $g_a(y)=g(y)/[1-G(r_a)]$. Reading $y_a$ as the second-highest entered value gives the density $g^{(2)}_a$ below; because two screens condition on the realized price, estimation normalizes by the retained window $[r_a+0.02,3)$:
\begin{equation}\label{eq:cal-likelihood}
g^{(2)}_a(y)=m_a(m_a-1)\,G_a(y)^{m_a-2}\bigl[1-G_a(y)\bigr]g_a(y),
\qquad
(\widehat\kappa,\widehat\sigma)\in\arg\max\ \sum_a\log\frac{g^{(2)}_a(y_a)}{\int_{r_a+0.02}^{3}g^{(2)}_a(t)\,\mathrm{d}t} .
\end{equation}
This is a parsimonious pooled calibration, not a structural replication of the source study; Table~\ref{tab:cal-sens} reports what conditioning on the public covariates would change rather than assuming their effects away. Table~\ref{tab:cal-fit} reports the maximizer and the auction-level bootstrap intervals, and Table~\ref{tab:cal-boot} the intervals for the ten-band outcomes of Figure~\ref{fig:resolution}; each of the 500 replications refits $(\kappa,\sigma)$ and rebuilds the partition before the mechanism is resolved.

\begin{figure}[htbp]
\vspace{-0.1in}
    \begin{minipage}[t]{0.44\textwidth}
    {\footnotesize\centering\setlength{\tabcolsep}{3pt}
        \begin{tabular}{@{}lrr@{}}\toprule
 & Estimate & 95\% interval\\\midrule
$\kappa$ & $-0.50$ & $[-0.55,\,-0.45]$\\
$\sigma$ & 0.53 & $[0.50,\,0.58]$\\
${\sf E}[X]$ & 0.70 & $[0.68,\,0.72]$\\\bottomrule
\end{tabular}
        \captionof{table}{Fitted value law, 500 bootstraps}
        \label{tab:cal-fit}
    }
    \end{minipage}\hfill
    \begin{minipage}[t]{0.54\textwidth}
    {\footnotesize\centering\setlength{\tabcolsep}{3pt}
        \begin{tabular}{@{}lrr@{}}\toprule
 & Point & 95\% interval\\\midrule
Revenue $U_0^{\ast}$ & 20.91 & $[20.39,\,21.43]$\\
Behavioral increment $E$ & 4.37 & $[4.12,\,4.63]$\\
Second price $U_0^{SPA}$ & 16.84 & $[16.27,\,17.39]$\\
Net vs.\ SPA & 4.07 & $[3.84,\,4.31]$\\
Pooling loss $\varepsilon_{\mathcal I}$ & 0.27 & $[0.24,\,0.29]$\\\bottomrule
\end{tabular}
        \captionof{table}{Ten-band outcomes, bootstrap intervals}
        \label{tab:cal-boot}
    }
    \end{minipage}
\vspace{-0.2in}
\end{figure}

\noindent\textbf{Base-case schedule and competition.} Table~\ref{tab:cal-sched} gives the schedule of Theorem~\ref{thm:band}{\rm (iii)} at the base case in dollars, its last column the gap between the floor $L_j$ and the matching price. The discount is concentrated on the top bands, where the all-matching event is least likely, and vanishes at the bottom band, which earns no floor rent. Table~\ref{tab:cal-thick} lists the series behind Figure~\ref{fig:thick}: the published partition stays at ten bands while the bidder count rises, so the rational band revenue $U_0^{S}$ climbs toward the coarsened ceiling while the behavioral increment is competed away.

\begin{figure}[htbp]
\vspace{-0.1in}
    \begin{minipage}[t]{0.55\textwidth}
    {\footnotesize\centering\setlength{\tabcolsep}{3pt}
        \begin{tabular}{@{}rrrrrr@{}}\toprule
Band & Floor & Ceiling & Mean & Matching & \\
$j$ & $L_j$ & $H_j$ & $\mu_j$ & price & Discount\\\midrule
1 & 26.86 & 54.25 & 34.12 & 15.69 & 11.17\\
2 & 21.37 & 26.86 & 23.81 & 16.64 & 4.73\\
3 & 18.09 & 21.37 & 19.63 & 15.88 & 2.21\\
4 & 15.68 & 18.09 & 16.83 & 14.66 & 1.02\\
5 & 13.71 & 15.68 & 14.67 & 13.24 & 0.48\\
6 & 11.99 & 13.71 & 12.84 & 11.78 & 0.21\\
7 & 10.40 & 11.99 & 11.19 & 10.32 & 0.08\\
8 & 8.80 & 10.40 & 9.60 & 8.78 & 0.02\\
9 & 7.00 & 8.80 & 7.93 & 7.00 & 0.00\\
10 & 3.47 & 7.00 & 5.68 & 3.47 & 0.00\\\bottomrule
\end{tabular}
        \captionof{table}{Base-case payment schedule, ten bands}
        \label{tab:cal-sched}
    }
    \end{minipage}\hfill
    \begin{minipage}[t]{0.43\textwidth}
    {\footnotesize\centering\setlength{\tabcolsep}{3pt}
        \begin{tabular}{@{}rrrrr@{}}\toprule
Bidders & Rational & Projection & & Net vs.\\
$m$ & $U_0^{S}$ & $U_0^{\ast}$ & $E$ & SPA\\\midrule
2 & 11.01 & 17.13 & 6.12 & 5.93\\
4 & 16.54 & 20.91 & 4.37 & 4.07\\
6 & 19.35 & 22.67 & 3.32 & 2.72\\
8 & 21.17 & 23.76 & 2.60 & 1.63\\
10 & 22.46 & 24.51 & 2.05 & 0.68\\
11 & 22.97 & 24.80 & 1.83 & 0.25\\
12 & 23.41 & 25.04 & 1.63 & $-0.17$\\
20 & 25.47 & 26.15 & 0.68 & $-2.91$\\
40 & 26.69 & 26.78 & 0.08 & $-7.43$\\
60 & 26.84 & 26.85 & 0.01 & $-10.26$\\\bottomrule
\end{tabular}
        \captionof{table}{Competition, ten bands fixed}
        \label{tab:cal-thick}
    }
    \end{minipage}
\vspace{-0.2in}
\end{figure}

\noindent\textbf{Specification and distributional sensitivity.} Gamma and Weibull alternatives give AIC values $-264.13$ and $-248.55$ against $-274.39$ for the lognormal. Better relative AIC is not correct specification: the conditional probability-integral transform has Kolmogorov--Smirnov statistic $0.04$, and a 300-replication parametric bootstrap that re-estimates the model in every replication returns $p=0.02$, so the parsimonious specification is rejected. The calibration therefore anchors the counterfactuals to a real price scale; it does not certify the value model. What matters for the design conclusions is that they do not turn on it. Table~\ref{tab:cal-family} refits the value family and rebuilds the ten-band base case from it, comparing each mechanism to that family's own continuous second-price benchmark; every one of these schedules satisfies \eqref{eq:cap}. Table~\ref{tab:cal-tail} varies instead how much of each tail the evaluated economy discards, and Table~\ref{tab:cal-sens} varies the screen itself and adds a covariate-adjusted reference profile evaluated at the sample-mean item and seller indicators and at $B=\$22.50$. The net gain moves over \$3.52--\$4.32 across these specifications, a wider range than the sampling interval of Table~\ref{tab:cal-boot}: the qualitative conclusion is conditional but stable, the point return of any single design is not.

\begin{figure}[htbp]
\vspace{-0.1in}
    \begin{minipage}[t]{0.52\textwidth}
    {\footnotesize\centering\setlength{\tabcolsep}{3pt}
        \begin{tabular}{@{}lrrrrr@{}}\toprule
Value & & & & Second & Net vs.\\
family & AIC & $U_0^{\ast}$ & $E$ & price & SPA\\\midrule
Lognormal & $-274.39$ & 20.91 & 4.37 & 16.84 & 4.07\\
Gamma & $-264.13$ & 20.26 & 4.64 & 15.91 & 4.35\\
Weibull & $-248.55$ & 19.36 & 4.94 & 14.72 & 4.63\\\bottomrule
\end{tabular}
        \captionof{table}{Value-family sensitivity, ten bands}
        \label{tab:cal-family}
    }
    \end{minipage}\hfill
    \begin{minipage}[t]{0.46\textwidth}
    {\footnotesize\centering\setlength{\tabcolsep}{3pt}
        \begin{tabular}{@{}lrrrr@{}}\toprule
Each tail & & & Second & Net vs.\\
removed & $U_0^{\ast}$ & $E$ & price & SPA\\\midrule
$0.1\%$ & 21.03 & 4.45 & 16.91 & 4.13\\
$0.5\%$ & 20.91 & 4.37 & 16.84 & 4.07\\
$1.0\%$ & 20.77 & 4.28 & 16.77 & 3.99\\\bottomrule
\end{tabular}
        \captionof{table}{Tail truncation under the fitted law}
        \label{tab:cal-tail}
    }
    \end{minipage}
\vspace{-0.2in}
\end{figure}


\noindent\textbf{Numerical tolerances and certificates.} Across the central grid of Section~\ref{sec:evidence} the endpoint incentive constraints hold to within $1.1\times10^{-14}$. For the smooth-belief bounds of Section~\ref{sec:calibration_auction}, every band probability is, under the lognormal fit, a ratio of normal-CDF differences, enclosed in 100-digit interval arithmetic with an explicit series remainder. For a lower bound, the gains $\Gamma_j(\mathbf{k})$, $\mathbf{k}\neq\mathbf{D}_j$, are rationals obeying \eqref{eq:triangular}, rounded from a linear-programming solution; $\Gamma_j(\mathbf{D}_j)$ is then \emph{defined} by the indifference condition of Proposition~\ref{prop:exact}{\rm (i)}, payments are rebuilt from $x_n=L_np_n$ and $x_j=x_{j+1}+L_j(p_j-p_{j+1})-\Gamma_j$, and $\Gamma_j(\mathbf{D}_j)\geq0$ and \eqref{eq:cap} are verified, and revenue enclosed, in the same arithmetic. For an upper bound, part~{\rm (ii)}, valid for every $\mathbf{y}\geq0$, is evaluated with enclosed probabilities at rationalized solver multipliers for incentive compatibility at 17 log-spaced values per band. For first price, the shading $\Psi=s-\mathsf{b}$ solves $\Psi'=1-(m-1)\widehat\lambda(s\mid s)\Psi$ (proof of Proposition~\ref{prop:fpa}); interval bounds on $\widehat\lambda$ over 1{,}000 log-spaced cells bound $\Psi$ at every grid reserve, and since $\mathsf{b}$ increases, outward-rounded Stieltjes sums bracket \eqref{eq:fparev}. Differentiating \eqref{eq:fparev} through \eqref{eq:fpabid} gives $|\mathrm{d}U_0^{FPA}/\mathrm{d}s_0|\leq\max\{1,m\sup_ss\,g(s)\}$, covering the reserves between grid points.

\noindent\textbf{Broader belief laws.} Exact own-band concentration is not the only distortion worth pricing. Let $b_j=\tfrac12(\log L_j+\log H_j)$ be the log midpoint of band $j$ and let $\theta>0$ be a belief spread held fixed in log-value units as the partition changes. The \emph{neighborhood} law spreads the excess subjective mass over bands near the buyer's own, replacing \eqref{eq:projection-beliefs} by
\begin{equation}\label{eq:smooth}
\pi_{jr}=(1-\alpha)q_r+\alpha\,
\frac{q_r\exp\bigl[-(b_r-b_j)^{2}/2\theta^{2}\bigr]}
{\sum_{t\in\mathcal N}q_t\exp\bigl[-(b_t-b_j)^{2}/2\theta^{2}\bigr]} .
\end{equation}
These are scenario laws, not estimates. They are constant within a band, so Theorem~\ref{thm:band}{\rm (i)} applies and endpoint incentive compatibility remains exact; we solve the full no-subsidy program without imposing diagonal support. Table~\ref{tab:cal-smooth} reports the outcome. The gain survives a calculation that nowhere privileges the diagonal, but its size depends materially on the resolution of the interface and on how concentrated beliefs are. These are finite computations, not a convergence result for a continuous neighborhood model.

\begin{figure}[htbp]
\vspace{-0.1in}
    \begin{minipage}[t]{0.72\textwidth}
    {\footnotesize\centering\setlength{\tabcolsep}{3pt}
\begin{tabular}{@{}lrrrrr@{}}\toprule
Specification & Retained & $\widehat\kappa$ & $\widehat\sigma$ & $U_0^{\ast}$ & Net vs. SPA\\\midrule
Baseline screen & 749 & $-0.50$ & 0.53 & 20.91 & 4.07\\
Closing-ratio gap $0.01$ & 751 & $-0.50$ & 0.54 & 20.93 & 4.08\\
Closing-ratio gap $0.05$ & 721 & $-0.50$ & 0.54 & 20.93 & 4.09\\
Entry not conditioned on the reserve & 749 & $-0.44$ & 0.53 & 22.21 & 4.32\\
Excluding major blemishes & 721 & $-0.45$ & 0.51 & 21.36 & 3.99\\
Unsealed listings only & 656 & $-0.49$ & 0.49 & 20.19 & 3.67\\
Covariate-adjusted reference profile & 749 & $-0.34$ & 0.42 & 22.18 & 3.52\\\bottomrule
\end{tabular}
        \captionof{table}{Screening and covariate specifications (dollars per item)}
        \label{tab:cal-sens}
    }
    \end{minipage}\hfill
    \begin{minipage}[t]{0.28\textwidth}
    {\footnotesize\centering\setlength{\tabcolsep}{3pt}
        \begin{tabular}{@{}rrrrr@{}}\toprule
Bands & Spread & & & Net vs.\\
$n$ & $\theta$ & $U_0^{\ast}$ & $E$ & SPA\\\midrule
5 & 0.15 & 18.46 & 2.61 & 1.61\\
5 & 0.30 & 18.41 & 2.57 & 1.57\\
5 & 0.60 & 18.00 & 2.16 & 1.16\\
10 & 0.15 & 20.89 & 4.35 & 4.05\\
10 & 0.30 & 20.71 & 4.17 & 3.86\\
10 & 0.60 & 19.61 & 3.07 & 2.77\\
20 & 0.15 & 22.46 & 5.71 & 5.62\\
20 & 0.30 & 22.09 & 5.33 & 5.25\\
20 & 0.60 & 20.35 & 3.59 & 3.50\\\bottomrule
\end{tabular}
        \captionof{table}{Neighborhood beliefs}
        \label{tab:cal-smooth}
    }
    \end{minipage}
\vspace{-0.2in}
\end{figure}

\end{APPENDICES}

\end{document}